\documentclass[prd, amsfonts, twocolumn, nofootinbib, showpacs]{revtex4-1}
\usepackage{graphicx, epsfig}
\usepackage{color}
\usepackage{amsmath}
\usepackage{mathtools}
\newcommand{\be}{\begin{equation}}
\newcommand{\ee}{\end{equation}}
\newcommand{\bea}{\begin{eqnarray}}
\newcommand{\eea}{\end{eqnarray}}

\newcommand{\gapp}{\mathrel{\raise.3ex\hbox{$>$}\mkern-14mu \lower0.6ex\hbox{$\sim$}}}
\newcommand{\lapp}{\mathrel{\raise.3ex\hbox{$<$}\mkern-14mu \lower0.6ex\hbox{$\sim$}}}
\newcommand{\LSM}{L$\Sigma$M}

\newcommand{\GMLfull}{Gell-Mann-L\'evy~}
\newcommand{\GML}{GML~}

\newcommand{\HVEV}{\langle H\rangle}
\newcommand{\SVEV}{\langle S\rangle}
\newcommand{\half}{\frac{1}{2}}
\newcommand{\mpisq}{m_\pi^2}

\def\bbox{{\,\lower0.9pt\vbox{\hrule \hbox{\vrule height 0.2 cm
\hskip 0.2 cm \vrule  height 0.2 cm}\hrule}\,}}

\begin{document} 


\title{Hidden $U(1)_Y$ Ward-Takahashi identities in the spontaneously broken
Abelian Higgs model  and the
decoupling of certain heavy particles in its simple extensions
}

\author{Bryan W. Lynn$^{1,2,3}$, Glenn D. Starkman$^{1}$ 
and Raymond Stora$^{4,5,\dagger}$
}
\affiliation{$^1$ ISO/CERCA/Department of Physics, Case Western Reserve University, Cleveland, OH 44106-7079}
\affiliation{$^2$ University College London, London WC1E 6BT, UK}
\affiliation{$^3$ Department of Physics, University of Wisconsin, Madison, WI 53706-1390}
\affiliation{$^4$ Theory Division, Department of Physics, CERN, CH-1211 Geneva 23, Switzerland}
\affiliation{$^5$ Laboratoire d'Annecy-le-Vieux de Physique Th\'eorique (LAPTH), F-74941 Annecy-le-Vieux Cedex, France}
\email{bryan.lynn@cern.ch,  gds6@case.edu
, $\dagger$ deceased
}

\begin{abstract}


The spontaneously broken
$U(1)_Y$-hypercharge Abelian Higgs model (AHM) 
(i.e. the spontaneous symmetry breaking (SSB) gauge theory of a complex scalar 
$\phi = \frac{1}{\sqrt 2}(H+i\pi)= \frac{1}{\sqrt 2}{\tilde H}e^{i{\tilde \pi}/\HVEV}$
and a vector $A^\mu$) has, in Lorenz gauge, 
a massless pseudo-scalar $\pi$.  
Its {\em physical states} have a conserved $U(1)_Y$ global current 
(but no conserved charge), and a Goldstone Theorem (GT). 
$\tilde \pi$ becomes a Nambu-Goldstone boson (NGB), 
with only derivative couplings and a shift symmetry.

Since Slavnov-Taylor identities guarantee that 
on-shell T-matrix elements of physical states 
are independent  of local $U(1)_Y$ gauge transformations
(even though these break the Lagrangian's BRST symmetry), 
we observe that they are therefore also independent of  anomaly-free 
$U(1)_Y$ {\em rigid/global}  transformations. 
We derive
2 towers of $\phi$-sector $U(1)_Y$ Ward-Takahashi Identities (WTI) which give: 
relations among  Green's functions; 
relations among off-shell T-matrix elements; 
powerful constraints on the dynamics of the $\phi$-sector. 
We prove Adler self-consistency relations for the $U(1)_Y$ {\em gauge theory} (i.e. beyond those
usual for a global theory), which guarrantee infra-red finiteness for on-shell $\phi$-sector T-matrix elements in Lorenz gauge: one of these is the GT.

All ultra-violet quadratic divergences (UVQD) contribute only to $\mpisq$, 
a finite {\it pseudo}-NGB mass-squared, 
which appears in intermediate steps of the calculations.
The Goldstone theorem then enforces $\mpisq \equiv 0$ exactly, 
so that
all UVQD contributions (i.e. the only dangerous relevant AHM operators),  
originating in loops containing virtual $A^\mu;\phi$ and ghosts ${\bar \omega},\omega$
vanish. 
The NGB ${\tilde \pi}$ 
decouples from the {\em observable particle spectrum} in the usual way, 
when the {\em observable vector particle} 
$B_\mu \equiv A_\mu+\frac{1}{e\HVEV}\partial_\mu{\tilde \pi}$ absorbs it, 
as if it were a gauge transformation.
Our $U(1)_Y$ WTI are then ``hidden" from observable particle physics.

While the AHM has only one scale $m_{BEH}^2$, 
and so there are no ratios of scales to require fine-tuning,
our regularization-scheme-independent, WTI-driven results 
are unchanged by the addition of certain  
heavy $U(1)_Y$ matter representations ($M_{Heavy}^2 \gg |q^2|, \HVEV^2 \sim m_{Weak}^2$), 
because the extended rigid $U(1)_Y$ WTI and Goldstone theorem 
cause all UVQD, log-divergent  and  finite relevant operators
$(\sim M_{Heavy}^2 \ln [M_{Heavy}^2], 
M_{Heavy}^2 \ln [m_{Weak}^2], M_{Heavy}^2,m_{Weak}^2 \ln{M_{Heavy}^2})$
in the so-extended SSB AHM, to vanish. 
We prove 5 heavy-particle SSB decoupling theorems, 
illustrating them with two explicit examples:
a singlet $M^2_S \gg m_{Weak}^2$ real scalar field $S$ with 
discrete $Z_2$ symmetry and VEV $\SVEV=0$; 
and a  singlet right-handed Type I See-saw Majorana neutrino $\nu_R$ 
with $M_{\nu_R}^2\gg m_{Weak}^2$. 
Including all loops containing virtual gauge bosons, fermions, scalars and ghosts, 
we prove that certain heavy degrees of freedom decouple completely from the
$U(1)_Y$  low-energy effective SSB AHM Lagrangian,
contributing only irrelevant operators after quartic-coupling renormalization. 
We also display a non-decoupling exception: 
heavy type 1 See-saw $\nu_R^{Majorana}$ cannot {\em completely} decouple, 
but becomes invisible in practice.  

The NGB ${\tilde \pi}$ decouples as usual, hiding the WTI.
But our ``Hidden SSB $U(1)_Y$ WTI," 
and their embedded shift symmetry ${\tilde \pi}\to {\tilde \pi}+\HVEV \theta$,  
have protected the low-energy SSB AHM theory 
(i.e. its observable particle spectrum and dynamics) 
from loop contributions of heavy particles! 
Gauge-independent observable weak-scale 
$(m^2_{BEH;Pole} ,\HVEV^2 \sim m_{Weak}^2)$ 
are therefore ``Goldstone Exceptionally Natural",  not fine-tuned in the 
Abelian Higgs model with these judicious extensions.  
\end{abstract}
\maketitle

\section{Introduction}
\label{Introduction}

How can weak-scale $m^2_{Weak} \sim (100GeV)^2$ spontaneously broken  gauge theories protect themselves against quantum loop corrections 
involving very heavy matter particles $M^2_{Heavy} \gg m^2_{Weak}$?
The current consensus in the theoretical physics community is that, without the imposition of further new symmetry,  they cannot: the scale of  spontaneous symmetry breaking (SSB)
will ``naturally" rise from $m^2_{Weak}$ to $M^2_{Heavy}$ and only there will it be
quantum-mechanically stable.
Alternately, one is forced to fine-tune the theory, 
cancelling ${\cal O}(M^2_{Heavy})$ quantum loop contributions 
against similarly large bare counter-terms order-by-order
to obtain weak-scale physical quantities such as the 
mass $m_{BEH}\sim {\cal O}(100 GeV)$ of the Brout-Englert-Higgs (BEH) boson 
 \cite{Susskind1979,Wilson} and similar magnitude weak gauge boson masses.

We refer to theories requiring such cancellation among bare-Lagrangian terms
 as ``bare fine-tuned" (Bare-FT).\footnote{
	Our favorite fine-tuned theory is that of a real scalar ${\bar S}$, 
	with  (${\bar S}\to-{\bar S}$) $Z_2$ symmetry:
	\begin{eqnarray}
		\label{FineTunedScalarLagrangian}
		&&L_{\bar S}=
		\half(\partial_{\mu}{\bar S})^2 
		- \half \mu_{\bar S}^2 {\bar S}^2 
		- \frac {\lambda_{\bar S}^2}{4} {\bar S}^4 
	\end{eqnarray}
	Its symmetric $\mu_{\bar S}^2>0$ Wigner mode is Bare-FT \cite{AlvarezGaume2012}. 
	It is also Bare-FT in its $\mu_{\bar S}^2<0$ SSB mode \cite{AlvarezGaume2012}, 
	which spontaneously breaks the $Z_2$ symmetry, 
	because (crucially, as we see below)  it has no 
	Goldstone theorem or associated Nambu-Goldstone Bosons.
}
This indicates the theory's unsuitability 
as a UV-complete model of particle physics.\footnote{
	This BEH Fine tuning problem is separate from the Hierarchy problem -- 
	the philosophical quesiton of why the $\sim100$GeV scale of weak interactions
	is orders of magnitude smaller than that of gravity $\sim 2\times 10^{19}$GeV -- 
	but they are often confused or conflated.}
The Standard Model (SM) is regarded as having this ``BEH fine-tuning problem''
(hereafter FTP)
and an important goal of Beyond the Standard Model (BSM)
\cite{Raby2010,BaerTata2006,RandallReece2012,DimopoulosGeorgi,
LittleHiggs,Grinstein2007,DimopoulosArkaniHamed2004,Weinberg2005}. 
theories is to avoid the problem --
typically by adding new symmetries, such as supersymmetry.

This paper takes another  step forward 
\footnote{
	In June 2011 \cite{Lynn2011}  one of us introduced the idea of the ``Goldstone Exception"  
	(though not the term) for the SM, 
	and showed that the ultra-violet quadratic divergences (UVQD) of the SM 
	did not yield an  BEH-FT problem. 
	A December 2011 pedagogical companion paper \cite{Lynnetal2012} simplified UVQDs
	in the context of the {\em global} \GMLfull model \cite{GellMannLevy1960}, 
	and named that concept.
	We defined ``Goldstone Exceptional Naturalness" in \cite{LynnStarkman2013} 
	and showed that the Goldstone theorem protects the weak-scale {\em global} SSB $SO(2)$ Schwinger model \cite{Schwinger1957} (i.e.
	against 1-loop relevant operators $\sim M_{Heavy}^2 \gg m_{Weak}^2$ which arise from virtual heavy particles) by way of 2 explicit 1-loop examples:
	a real singlet scalar $S$ and a singlet Majorana neutrino $\nu_R$ with $M_S^2,M_{\nu_R}^2 \gg \vert q^2\vert, \HVEV^2$.
	Ref \cite{LSS-2} pushed those heavy-particle decoupling (and no-BEH-FT) results, for $\big< S \big> =0$,  to all-loop-orders, using 2 towers of recursive $SU(2)_L \times U(1)_Y$  WTI, while including the virtual effects of the lightest generation of SM quarks and leptons.
}
toward a non-BSM proposal, 
based on the Goldstone theorem \cite{Goldstone1961,Goldstone1962}, 
that may resolve the perceived crisis \cite{Lykken2014} 
due to tension between LHC8 data and simple BSM solutions 
of the FTP.   
We do so by explaining how 
spontaneously broken gauge theories can 
avoid the quantum instability responsible for the FTP,
and demonstrate that there is a wide class 
of heavy-particle $M_{Heavy}\gg m_{Weak}$ matter representations 
from which low-energy weak-scale physics is protected in 
such theories, fortified as they are by the Goldstone theorem. 


We show here that, in the SSB Abelian Higgs Model,
a tower of Ward-Takahashi Identities (WTI) relates all relevant-operator
contributions to AHM physical-scalar-sector physical observables to one
another, and the Goldstone Theorem  then enforces a strong version of no-FT 
by causing all such contributions  to vanish.
It does so through its insistence that the Nambu-Goldstone Boson (NGB) 
mass-squared vanishes exactly \cite{Nambu1959,Goldstone1961,Goldstone1962}.
This is regardless of the fact that the NGB is not a physical degree of freedom,
but is absorbed ("eaten") by the gauge boson.
The crucial advance over \cite{LSS-2}, which considered the 
global $SU(2)_L\times U(1)_Y$ Linear Sigma Model,
is a proof that the WTI remain in place in a SSB gauge theory, 
and the Goldstone Theorem continues to successfully play its protective role.

We call such \cite{LynnStarkman2013} theories, 
which are protected from quantum instability by the Goldstone Theorem,
Goldstone Exceptionally Natural (GEN). 
GEN is simply another (albeit unfamiliar) consequence of Ward-Takahashi Identities (WTI), 
spontaneous symmetry breaking (SSB) and the Goldstone theorem, 
and is a new exact property of the SSB vacuum and spectrum 
of the $U(1)_Y$ AHM \cite{ScarletBegonias}. 
It is the standard of naturalness/no-FT in this paper.
This suppression of FT is far more powerful than G. 't Hooft's
\cite{tHooft1980, AlvarezGaume2012} 
widely accepted criteria. 
\footnote{
	The literature provides various definitions/criteria for naturalness, 
	with increasing levels of suppression of fine-tuning.  
	 G.'t Hooft put forward a definition of what it means for parameters of 
	theories to be naturally small
	\cite{tHooft1980,AlvarezGaume2012}:
	``At any energy scale $\mu$, a [dimensionless] physical parameter or a set of physical parameters 
	$\alpha_i(\mu)$ is allowed to be very small only if the replacement $\alpha_i(\mu)=0$
	would  increase the symmetry of the system"  \cite{tHooft1980, AlvarezGaume2012}. 
}

Our GEN approach to the fine-tuning question differs markedly from myriad other proposals.
Leaving aside the standard supersymmetric or Technicolor approaches, 
some start from the perspective of  conformal invariance 
\cite{Tavares:2013dga,Meissner:2007xv}, or
suggest that the running of standard model parameters
returns the Higgs mass at high momentum-squared to its low momentum-squared value
\cite{Jegerlehner:2013cta,Jegerlehner:2013nna},
or aims to identify classes of heavy BSM fermion and scalar matter that don't
destabilize the Higgs mass
\cite{deGouvea:2014xba,Espinosa,Farina:2013mla}.  
None link ``naturalness" or heavy-matter decoupling to spontaneous symmetry breaking or the Goldstone Theorem.

The usual fine-tuning viewpoint prefers to cast FT in terms of bare parameters 
(eg. $\mu_{\phi;Bare}^2$, 
the coefficient of the quadratic self-coupling of the complex 
$U(1)_Y$ scalar $\phi=\frac{1}{\sqrt s}(H+i\pi)$)
instead of renormalized physical ones (e.g. $\mpisq$ and $\HVEV$),
and prefers the unitary Kibble representation of this AHM scalar 
as opposed to the linear one
(in which renormalization is straightforward,
and the symmetric $\HVEV =0$ Wigner mode makes sense).
That usual view thereby misses two crucial observations: 
that all ultra-violet quadratic divergences (UVQD) 
and finite relevant operator contributions 
are absorbed into the pseudo-NGB mass-squared $\mpisq$ 
which appears in intermediate steps of calculations; 
that a zero value for this $\mpisq$, 
and therefore for the only combination of UVQD and relevant operator contributions to
physical observables, is protected by the Goldstone theorem.


Our  no-FT results rely solely on the rigid $U(1)_Y$WTIs 
that govern the scalar-sector of the AHM
and of the extensions we consider in Section \ref{E-AHM}. 
They are therefore independent of regularization-scheme (assuming one exists).
Although not a gauge-independent procedure, 
it may help the reader to imagine that loop integrals 
are cut off at a short-distance  finite Euclidean UV scale, $\Lambda$, 
never taking the $\Lambda^2 \to \infty$ limit. 
Although that cut-off can be imagined to be near the Planck scale $\Lambda\simeq M_{Pl}$, 
quantum gravitational loops are not  included.

The reader should note that this paper concerns stability and protection 
of the Abelian Higgs model against UVQD 
(and, in Section \ref{E-AHM}, finite relevant) operators.
It does not address any of the other, more usual, stability issues  of the Standard Model 
(cf. the discussion in, for example, \cite{Ramond2004}, and references therein).

The structure of this paper is as follows:



Section \ref{AbelianHiggsModel} concerns the correct renormalization of the spontaneously broken AHM in 
Lorenz gauge.
We treat the AHM in isolation, 
as a stand-alone flat-space weak-scale quantum field theory,
not embedded or integrated into any higher-scale ``Beyond-AHM"  physics. 

\begin{itemize}

\item Subsection \ref{DefineAHM} defines the Abelian Higgs model

\item Subsection \ref{GlobalCurrents} builds a conserved rigid/global AHM current in 
Lorenz gauge, and reminds \cite{LSS-3Short} us that the rigid/global $U(1)_Y$ charge is {\it not} conserved, even for the physical states.

\item Sub-section \ref{GreensFunctionsAHM} constructs the $\phi$-sector effective Lagrangian from those WTIs which govern 1-$(h,\pi)$  Scalar-Particle-Irreducible (1-$\phi$-I) connected amputated Green's functions.

\item Subsection \ref{TMatrixAHM}  further contrains the $\phi$-sector effective Lagrangian with those WTIs which govern 1-$(h,\pi)$-{\em Reducible} (1-$\phi$-R) $\phi$-sector connected amputated {\it on-shell} T-Matrix elements, especially the Goldstone theorem.

\end{itemize}

Section \ref{E-AHM} extends our AHM
results to include the all-loop-orders virtual contributions of certain $M_{Heavy}^2 \gg m_{Weak}^2$ heavy $U(1)_Y$ matter representations (which might arise in certain Beyond-AHM models). 

\begin{itemize}

\item Subsection \ref{LagrangianE-AHM} constructs the effective Lagrangian for the extended-AHM, and proves 3 decoupling theorems.

\item Subsection \ref{HeavyScalar} gives an example of complete heavy-physics decoupling without fine-tuning: a virtual singlet $M^2_S \gg m_{Weak}^2$ real scalar field $S$ with 
discrete $Z_2$ symmetry and 
$\SVEV=0$

\item Sub-section \ref{SMQuarksLeptons} includes the virtual effects of the lightest generation of spin $S=\half$ Standard Model (SM) quarks and leptons, augmented by a right-handed neutrino $\nu_R$, with baryon and lepton-number conserving
 {\em Dirac} masses-squared $m_{Quark}^2,m_{Lepton}^2  \ll  m_{Weak}^2$, 
regarded here as $U(1)_Y$ Beyond-AHM matter representations. The SSB extended-AHM gauge theory remain anomaly-free because the fermion AHM quantum numbers are taken to be their SM hypercharges.

\item Sub-section \ref{HeavyNeutrino} gives an example of ``almost complete" heavy-physics decoupling (with a non-decoupling ``practical invisibility" subtlety) without a FTP: a  virtual singlet right-handed Type I See-saw Majorana neutrino $\nu_R$ with $M_{\nu_R}^2\gg m_{Weak}^2$.

\end{itemize}

Section \ref{ParticlePhysicsAHM} reminds \cite{LSS-3Short} us that the NGB $\tilde \pi$  disappears from the {\it observable} particle spectrum of the extended-AHM, carrying with it any fine-tuning problem due to heavy $M^2_{Heavy} \gg m_{Weak}^2$ Beyond-AHM particles.  
 \begin{itemize}

\item Subsection \ref{NGBDisappearsExtendeAHM} reminds \cite{LSS-3Short} us that the SSB extended-AHM's observable particle spectrum  {\em excludes} the NGB ${\tilde \pi}$, 
precisely because it is a spontaneously broken $U(1)_Y$ gauge theory. 

\item Sub-section \ref{DecouplingInGaugedE-AHM} derives our 4th and 5th decoupling theorems, and
shows complete decoupling, due to SSB, of heavy $M_{Heavy}^2 \gg m_{Weak}^2$ particles.

\end{itemize}

Section \ref{RigueurMathematiqueExigeante} discusses the exacting mathematical rigor which would have fully satisfied Raymond Stora.

Section \ref{Conclusions} reminds us that historically (with an important exception) the  decoupling of heavy particles is the usual experience of physics.

Appendix \ref{DerivationWTIAHM} gives detailed derivation, to be used here and in \cite{LSS-3Short}, of the $U(1)_Y$ WTIs governing the $\phi$-sector of the AHM. Our renormalized WTIs  include all contributions from virtual transverse gauge bosons; $\phi$-scalars; ghosts; $A^\mu;h,\pi;{\bar \omega},\omega;$ respectively. 

Appendix \ref{DerivationWTIE-AHM} gives detailed derivation, to be used here and in \cite{LSS-3Short}, of $U(1)_Y$ $(h,\pi)$-sector WTIs, which now  include the all-loop-orders contributions of certain 
additional $U(1)_Y$ matter representations: spin $S=0$ scalars $\Phi$, and $S=\half$ fermions $\psi$. They  include all contributions from virtual transverse gauge bosons; ghosts, scalars; and fermions: $A^\mu;h,\pi;{\bar \omega},\omega;\Phi;\psi;$.

\section{The Abelian Higgs model (AHM) in Lorenz gauge}
\label{AbelianHiggsModel}

\subsection{The Abelian Higgs model in Lorenz gauge}
\label{DefineAHM}

The BRST-invariant \cite{BecchiRouetStora,Tyutin1975,Tyutin1976} Lagrangian of the $U(1)_Y$ AHM gauge theory may be written, in Lorenz gauge, in terms of exact renormalized fields: a transverse vector $A_\mu$, a complex scalar $\phi$, a ghost $\omega$, and an anti-ghost $\bar \omega$ :
\begin{eqnarray}
	\label{LagrangianAHM}
	L_{AHM}^{Lorenz}&=&L_{AHM}^{GaugeInvariant} \\
	&+&L_{AHM}^{GaugeFix;Lorenz}  
	+L_{AHM}^{Ghost;Lorenz} \nonumber
\eea
where
\be
\label{LagrangianAHM-GI}
	L_{AHM}^{GaugeInvariant}=
	\vert D_\mu \phi \vert ^2 - \frac{1}{4}A_{\mu \nu}A^{\mu \nu} -V(\phi^\dagger \phi) 
\ee
with
\bea
D_\mu \phi &=&(\partial_\mu -ieY_\phi A_\mu)\phi \nonumber \\
A_{\mu \nu}&=&\partial_\mu A_\nu - \partial_\nu A_\mu \nonumber \\
V_{AHM}&=&\mu_\phi^2 \Big( \phi^\dagger \phi \Big)+ \lambda_\phi ^2 \Big(\phi^\dagger \phi\Big)^2 
\eea
and
\be
\label{LinearScalarRep}
\phi = \frac{1}{{\sqrt 2}}(H+i\pi); \quad H=\HVEV + h;  \quad Y_\phi =-1.
\ee
Also
\bea
L_{AHM}^{GaugeFix;Lorenz}&=& 
-\lim_{\xi \to0}\frac{1}{2\xi}\Big(\partial_\mu A^\mu  \Big) ^2   \nonumber \\
L_{AHM}^{Ghost;Lorenz}&=& {\bar \omega} (-\partial^2 ) \omega.
\end{eqnarray}
The complex scalar $\phi$ is manifestly renormalizable  
in the linear representation (\ref{LinearScalarRep}).
We shall see below that $m_A^2=e^2Y_\phi^2\HVEV^2$.

This paper distinguishes carefully between 
the local BRST-invariant $U(1)_Y$ Lagrangian (\ref{LagrangianAHM}), and its 3 physical modes \cite{Lee1970,Symanzik1970a,Symanzik1970b,Vassiliev1970,ItzyksonZuber}: symmetric Wigner mode, classically scale-invariant  point, and physical Goldstone mode.

\smallskip

{\bf 1) Symmetric Wigner mode} $\HVEV=0, m_A^2=0,\mpisq =m_{BEH}^2 = \mu^2_\phi \neq 0$: 

This is QED with massless photons and massive charged scalars. 
All UVQD $\sim \Lambda^2$
and finite relevant 
operators in the AHM are 
absorbed into
the {\it pseudo} Nambu-Goldstone boson mass-squared $\mpisq \neq0$. That pseudo-NGB mass, which survives in Wigner mode (but not in Goldstone mode), is the result of intermediate steps in the calculations, and has been a source of confusion and controversy surrounding  the BEH Fine-Tuning-Problem. 
The final step in Wigner mode calculations, i.e. setting $\HVEV=0$, would support such FT in $\mpisq$.
Since $m_{BEH}^2=\mpisq$ in the symmetric Wigner mode, any FTP would be passed onto the scalar mass. Wigner mode would be regarded as FT by the standards of GEN.

Thankfully, Nature is not in Wigner mode! 
Further analysis and renormalization of the Wigner mode  lies outside the scope of this paper.

{\bf 2) Classically scale-invariant point} $\HVEV=0, m_A^2=0,\mpisq =m_{BEH}^2 = 0$:
Analysis of the quantum scale-invariant point  is outside the scope of this paper.

{\bf 3) Spontaneously broken Goldstone mode} $\HVEV\neq0, m_A^2=e^2 \HVEV ^2 \neq0,\mpisq =0, m_{BEH}^2 \neq 0$:

 The ``famous" Abelian Higgs model, with its Nambu-Goldstone boson (NGB) ``eaten" by the Brout-Englert-Higgs mechanism 
(and, as we will see, WTI governed by the Goldstone theorem) 
 is actually only the SSB ``Goldstone mode" of the BRST-invariant local Lagrangian (\ref{LagrangianAHM}), 
and is the subject of this paper.

We work in  Lorenz gauge for many reasons: 
\begin{itemize}
\item The $U(1)_Y$  ghosts $({\bar \omega},\omega)$  decouple from the quantum loop dynamics, and can (and will) be benevolently ignored going forward.

\item After a subtlety concerning their mixing,
$\pi$ and $A^\mu$ are orthonormal species.
A term $\sim A_\mu \partial^\mu\pi$ arises from $\vert D_\mu \phi\vert^2$ after SSB in (\ref{LagrangianAHM}), 
A term $\sim \pi\partial^\mu A_\mu$ is shown to vanish {\em for physical states} in 
(\ref{GaugeConditions},\ref{GaugeConditionExtended}).
The resultant surface term $\partial^\mu \big( \pi A_\mu \big)$ vanishes (for physical states) because $A_\mu$ is massive,

\item Only in the SSB Goldstone mode of the BRST-invariant Lagrangian (\ref{LagrangianAHM}), and only after first renomalizing in the linear $\phi$ representation, does the renormalized Kibble $\phi$ unitary representation
\begin{eqnarray}
\label{Kibble}
\phi = \frac{1}{{\sqrt 2}}\big(H+i\pi\big) &\equiv&  \frac{1}{{\sqrt 2}} {\tilde H}e^{-i Y_\phi {\tilde \pi}/\HVEV} \nonumber \\
{H}=\HVEV + { h}; \quad {\tilde H} &=&\HVEV + {\tilde h} \nonumber \\
{\tilde \pi} &\equiv&  \HVEV \vartheta
\end{eqnarray}
 make sense. Here the $\phi$-hypercharge $Y_\phi =-1$.

\item We will prove an all-loop-orders  {\bf Goldstone theorem} (\ref{TMatrixGoldstoneTheoremPrime},\ref{TMatrixGoldstoneTheorem})
which forces the $\pi$ mass-squared $\mpisq =0$.

\item We use  {\bf ``pion-pole dominance"} (i.e. $\mpisq =0)$ arguments to derive $U(1)_Y$ SSB WTIs
 (\ref{AdlerSelfConsistencyPrime},\ref{AdlerSelfConsistency},\ref{InternalTMatrix}).

\item 
We prove \cite{LSS-3Short} with $U(1)_Y$ WTI
that, in SSB Goldstone mode, {\bf ${\tilde \pi}$ in (\ref{Kibble}) is a Nambu-Goldstone boson} (NGB), and that the resultant SSB gauge theory 
has a {\bf ``shift symmetry"} ${\tilde \pi} \to {\tilde \pi} +\HVEV \theta$ for constant $\theta$. 
\end{itemize}

Analysis is done in terms of the exact renormalized interacting fields, 
which asymptotically become the in/out states, i.e. free fields for physical S-Matrix elements.

The most important issue for fine-tuning (and heavy particle decoupling) is the classification and disposal of relevant operators, 
in this case the ${ \pi}$, $h$ and $A_\mu$ inverse propagators (together with  tadpoles).
 Define the exact renormalized pseudo-scalar propagator 
in terms of a massless $ \pi$, the K$\ddot a$ll$\acute e$n-Lehmann \cite{Bjorken1965,Lee1970} spectral density $\rho^{\pi}_{AHM}$, and  wavefunction renormalization $Z_{AHM}^\phi$. In Lorenz gauge: 
\begin{eqnarray}
\label{pNGBPropagator}
&&\Delta^{\pi}_{AHM}(q^2) = -i(2\pi)^2\langle 0\vert T\left[ \pi(y)\pi(0)\right]\vert 0\rangle\vert^{Fourier}_{Transform} \nonumber \\
&&\quad \quad = \frac{1}{q^2
+ i\epsilon} + \int dm^2 \frac{\rho^{\pi}_{AHM}(m^2)}{q^2-m^2 + i\epsilon} \nonumber \\
&&\Big[Z^{\phi}_{AHM}\Big]^{-1} = 1+ \int dm^2 \rho^{\pi}_{AHM}(m^2) 
\end{eqnarray} 

Define also the BEH scalar propagator in terms of a BEH scalar pole and the (subtracted) spectral density $\rho_{BEH}$, and the {\it same} wavefunction renormalization. We assume $h$ decays weakly, and resembles a resonance:
\begin{eqnarray}
\label{BEHPropagator}
&&\Delta^{BEH}_{AHM}(q^2) = -i (2\pi)^2\langle 0\vert T\left[ h(x) h(0)\right]\vert 0\rangle\vert^{Fourier}_{Transform} \nonumber \\
&& \quad \quad =\frac{1}{q^2-m_{BEH;Pole}^2 + i\epsilon}+ \int dm^2 \frac{\rho^{BEH}_{AHM}(m^2)}{q^2-m^2 + i\epsilon} \nonumber \\
&&\Big[ Z^{\phi}_{AHM}\Big]^{-1} = 1+ \int dm^2 \rho^{BEH}_{AHM}(m^2)  \nonumber \\
&& \int dm^2 \rho^{\pi}_{AHM}(m^2) = \int dm^2 \rho^{BEH}_{AHM}(m^2) 
\end{eqnarray}

Although 
$m^2_{BEH;pole}$ 
may (or may not!) be FT, the spectral density parts of the propagators 
\begin{eqnarray}
\label{SpectralDensityPropagators}
 \Delta^{\pi ;Spectral}_{AHM}(q^2) &=& \int dm^2 \frac{\rho^{\pi}_{AHM}(m^2)}{q^2-m^2 + i\epsilon}  \nonumber \\ 
\Delta^{BEH; Spectral}_{AHM}(q^2) &=& \int dm^2 \frac{\rho^{BEH}_{AHM}(m^2)}{q^2-m^2 + i\epsilon} \nonumber
\end{eqnarray}
are certainly {\bf not} fine-tuned. Dimensional analysis of the wavefunction renormalizations (\ref{pNGBPropagator},\ref{BEHPropagator}), shows that the contribution of a state  of mass/energy $\sim M_{Heavy}$ to the spectral densities 
$\rho_{AHM}^{\pi}(M_{Heavy}^2),\rho_{AHM}^{BEH}(M_{Heavy}^2)\sim \frac{1}{M_{Heavy}^2}$, and  to $\Delta_{AHM}^{\pi ;Spectral},\Delta_{AHM}^{BEH;Spectral}$ only irrelevant terms $\sim \frac{1}{M_{Heavy}^2}$. 
The finite Euclidean cut-off contributes only irrelevant terms $\sim \frac{1}{\Lambda^2}$.

\subsection{Rigid/global $U(1)_Y$ WTI and  conserved rigid/global current, for the {\em physical states} of the SSB AHM, in 
Lorenz gauge. Rigid/global $U(1)_Y$ Charge is {\it not} conserved!}
\label{GlobalCurrents}

In their seminal work,  E. Kraus and K. Sibold 
\cite{KrausSiboldAHM} 
 identified, in the Abelian Higgs model, 
 an anomaly-free ``deformed'' rigid/global $U(1)_Y$ symmetry 
 as a rigid  {subset} of that  anomaly-free deformed local/gauge symmetry. 

 The SSB case is tricky: neither the usual gauge/local symmetry, nor even its restriction to a rigid/global  symmetry
(where the gauge transformations are taken to be independent of position), 
commute with global BRST symmetry. Only deformed versions of them do. 

Kraus and Sibold then constructed deformed Ward-Takahashi Identities (WTI), allowing them to demonstrate
(with appropriate normalization conditions) 
proof of all-loop-orders renormalizability and unitarity for the SSB Abelian Higgs model. 
Because their renormalization relies only on  deformed $U(1)_Y$  WTI, 
Kraus and Sibold's results are independent of regularization scheme, for any acceptable scheme (i.e. if one exists).
\footnote{
E. Kraus and K. Sibold  also constructed, in terms of deformed WTI, all-loop-orders renormalized QED, QCD,and  the electro-weak Standard Model  \cite{KrausSiboldSM1996,KrausSiboldSM1997,KrausSM1997} to be independent of regularization  scheme. From this grew the powerful technology of ``Algebraic Renormalization", used by them, W. Hollik and others \cite{Hollik2002b}, to renormalize SUSY QED, SUSY QCD, and the MSSM.
}

Nevertheless, Slavnov-Taylor identities \cite{JCTaylor1976} prove that the
on-shell S-Matrix elements of {\em ``physical particles"} (i.e. spin $S=0$ scalars $h,\pi$, and $S=1$ transverse gauge bosons $A_\mu$, but not fermionic ghosts $({\bar \omega},\omega)$)
 are independent of  
the usual {\bf undeformed} anomaly-free 
$U(1)_Y$ local/gauge transformations, even though these break the Lagrangian's BRST symmetry. 

We therefore observed in \cite{LSS-3Short} that SSB S-Matrix elements are therefore 
also independent of  anomaly-free 
{\bf undeformed} 
$U(1)_Y$ global/rigid transformations, resulting in a ``new" global/rigid current and appropriate {\bf un-deformed} $U(1)_Y$ Ward-Takahashi identities.
All this is done without reference to the unbroken Wigner mode and scale-invariant point.

We are interested in rigid-symmetric relations among 1-$(h,\pi)$-Irreducible (1-$\phi$-I) connected amputated Green's functions $\Gamma_{N,M}$, and among 1-$(h,\pi)$-Reducible (1-$\phi$-R) connected amputated transition-matrix (T-Matrix) elements $T_{N,M}$,   with external $\phi$ scalars. It is convenient to use the powerful old tools (e.g. canonical quantization) from Vintage Quantum Field Theory (Vintage- QFT),
a name coined by Ergin Sezgin.  

We focus on the rigid/global AHM current
\footnote{
	This is related to the rigid/global hypercharge current of the Global Dirac Neutrino Standard Model ($\nu_D SM^G$) explored in \cite{LSS-2}: replace
	$ \pi \to \pi_3,\pi^2 \to {\vec \pi}^2$; un-gauge $A_\mu$;
	add a charged pion current $\pi_2 \partial^{\mu}\pi_1-\pi_1 \partial^{\mu}\pi_2$;
	add SM quarks (3 colors, 6 flavors) and leptons (3 charged flavors); 
	add three $\nu_R$ with SSB Dirac masses $m_{\nu}$;
	change the overall sign $ {J}^{\mu;SoModified}_{AHM} \to - {J}^{\mu}_{Y;\nu_D SM}$.
	}, 
which transforms as an axial-vector
\begin{eqnarray}
\label{AHMCurrentPrime}
 {J}^{\mu}_{AHM}&=& \pi \partial^\mu H-H\partial^\mu \pi-eA^\mu\Big(\pi^2 + H^2 \Big) \quad \quad
\end{eqnarray}

Rigid/global transformations of the fields are as usual: from the equal-time commutators (\ref{EqTimeCommAHM})
\begin{eqnarray}
\label{TransformationsAHMFields}
 \delta H(t,{\vec y})&=&-i\theta \int d^3 z\left[ {J}^0_{AHM}(t,{\vec z}),H(t,{\vec y})\right] \nonumber \\
&=& -\theta \int d^3 z \pi(t,{\vec z})\delta^3({\vec z}-{\vec y}) \nonumber \\
&=& - \pi (t,{\vec y}) \theta \\
 \delta \pi(t,{\vec y})&=&-i\theta \int d^3 z\left[ {J}^0_{AHM}(t,{\vec z}),\pi(t,{\vec y})\right] \nonumber \\
&=& \theta \int d^3 z H(t,{\vec z})\delta^3({\vec z}-{\vec y}) \nonumber \\
&=&  H (t,{\vec y})\theta \nonumber \\
 \delta A^\mu(t,{\vec y})&=&0 \nonumber \\
 \delta \omega(t,{\vec y})&=&0 \nonumber \\
 \delta {\bar \omega}(t,{\vec y})&=&0 \nonumber 
\end{eqnarray} 
so ${J}^0_{AHM}(t,{\vec z})$ serves as a ``proper" local current,  for commutator purposes. 

In contrast, \cite{LSS-3Short} showed that,  in  Lorenz gauge, $U(1)_Y$ AHM (and therefore also extended $U(1)_Y$ AHM) has no proper global charge $Q(t) \equiv\int d^3z {J}^0_{AHM}(t,{\vec z})$, because $\frac{d}{dt}Q(t)\neq0$. See Eqn. (\ref{ChargeAHMTimeOrderedProductsNotConserved}) below. 

The classical equations of motion reveal a crucial fact: due to gauge-fixing terms in the BRST-invariant Lagrangian (\ref{LagrangianAHM}), the 
classical current 
(\ref{AHMCurrentPrime}) is 
not conserved. In Lorenz gauge  
\begin{eqnarray}
\label{DivergenceAHMCurrentPrime}
\partial_{\mu} {J}^{\mu}_{AHM}&=& -H m_A F_A   \nonumber \\
m_A &=& e\HVEV \nonumber \\
F_A &=& \partial_{\beta}{A}^{\beta} 
\end{eqnarray}
with $F_A$ the gauge fixing condition.  

But the {\bf the global $U(1)_Y$ current (\ref{AHMCurrentPrime}) is conserved by the {\em physical states}} \cite{LSS-3Short}, and therefore still qualifies as a ``real current". Strict quantum constraints are imposed, which force the relativistically-covariant theory of gauge bosons to propagate {\it only} its true number of quantum spin $S=1$ degrees of freedom: these constraints are, in the modern literature, implemented by use of spin $S=0$ fermionic Fadeev-Popov ghosts $({\bar \omega},\omega)$.  The {\em physical states and their time-ordered products}, but not the BRST-invariant Lagrangian  (\ref{LagrangianAHM}),  then obey
G. 't Hooft's \cite{tHooft1971}
Lorenz gauge  gauge-fixing condition 
 interpreted as follows for the $\phi$-sector connected time-ordered product
\begin{eqnarray}
\label{GaugeConditionsPrime}
&&\big< 0\vert T\Big[ \Big( \partial_{\beta}{A}^{\beta}(z)\Big) \\
&&\quad \quad \times h(x_1)...h(x_N)\pi(y_1)...\pi(y_M)\Big]\vert 0\big>_{Connected} =0 \nonumber
\end{eqnarray}
Here we have N external renormalized scalars $h=H-\HVEV$ (coordinates x, momenta p), 
and M external ($CP=-1$) renormalized pseudo-scalars ${ \pi}$ (coordinates y, momenta q). 

Eqn. (\ref{GaugeConditionsPrime}) restores conservation of the rigid/global $U(1)_Y$ current for $\phi$-sector connected time-ordered products \cite{LSS-3Short}
\begin{eqnarray}
\label{PhysicalAHMCurrentConservation}
&&\Big< 0\vert T\Big[ \Big( \partial_{\mu}{J}^{\mu}_{AHM}(z) \Big) \\
&&\quad \quad \times h(x_1)...h(x_N) \pi(y_1)...\pi(y_M)\Big]\vert 0\Big>_{Connected} =0 \nonumber
\end{eqnarray}
It is in this ``physical" connected-time-ordered-product sense that the rigid global $U(1)_Y$ {\em ``physical current"} is conserved: the physical states, but not the BRST-invariant Lagrangian  (\ref{LagrangianAHM}), obey the physical current conservation equation 
(\ref{PhysicalAHMCurrentConservation}). 
It is this ``physical conserved current"  which generates our $U(1)_Y$ WTI \cite{LSS-3Short}. 


Appendix A derives 2 towers of quantum $U(1)_Y$ WTIs, which exhaust the information content of (\ref{PhysicalAHMCurrentConservation}), and severely constrain the dynamics (i.e. the connected time-ordered products) of the $\phi$-sector physical states 
of the SSB AHM.

We might have hoped to build a conserved charge
by restricting it to physical connected time-ordered products
\begin{eqnarray}
\label{ChargeAHMTimeOrderedProducts}
&&\Big< 0\vert T\Big[ \Big(  \frac{d}{dt}Q_{AHM}(t) \Big) \\
&&\quad \quad \times h(x_1)...h(x_N) \pi(y_1)...\pi(y_M)\Big]\vert 0\Big>_{Connected} \nonumber \\
&&\quad =\int d^3 z \Big< 0\vert T\Big[ \Big(  {\vec \nabla} \cdot {\vec J}_{AHM}(t,{\vec z})\Big) \nonumber \\
&&\quad \quad \times h(x_1)...h(x_N) \pi(y_1)...\pi(y_M)\Big]\vert 0\Big>_{Connected} \nonumber \\
&& \quad =\int_{2-surface}  d^2z \quad {\widehat {z}}^{2-surface} \cdot \Big< 0\vert T\Big[ \Big(  {\vec J}_{AHM}(t,{\vec z})\Big) \nonumber \\
&&\quad \quad \times h(x_1)...h(x_N) \pi(y_1)...\pi(y_M)\Big]\vert 0\Big>_{Connected} \nonumber 
\end{eqnarray}
where we have used Stokes theorem, and $ {\widehat {z}_\mu}^{2-surface}$ is a unit vector normal to the $2$-surface. The time-ordered-product constrains the $2$-surface to lie on-or-inside the light-cone.

At a given point on the surface of a large enough 3-volume $\int d^3z$ (i.e. the volume of all space), which lies on-or-inside the light cone, all fields {\em on the $z^{2-surface}$}: are asymptotic in-states and out-states; are properly quantized as free fields; with each field species orthogonal to the others; and evaluated at equal times, so that time-ordering is unnecessary.

But (\ref{ChargeAHMTimeOrderedProducts}) does not vanish \cite{LSS-3Short} because, after SSB, a specific term 
 in $J^\mu_{AHM}$ in (\ref{AHMCurrentPrime})
\begin{eqnarray}
\label{NGBSurfaceIntegralPrime}
&& \int_{LightCone\to\infty}  dz \quad {\widehat {z}}^{LightCone} \cdot \Big< 0\vert T\Big[  \nonumber \\
&&\quad \quad \times  \Big(-\HVEV {\vec \nabla} \pi (z) \Big)  \nonumber \\
&&\quad \quad \times h(x_1)...h(x_N) \pi_(y_1)...\pi(y_M)\Big]\vert 0\Big> \nonumber \\
&&\quad \quad \neq 0
\end{eqnarray}
does not vanish. $\pi$ is massless (in 
Lorenz gauge) in the SSB AHM, capable of carrying (along the light-cone) long-ranged pseudo-scalar forces out to the  very ends of the light-cone $(z^{LightCone}\to \infty)$, but not inside it.

Eqns. (\ref{ChargeAHMTimeOrderedProducts},\ref{NGBSurfaceIntegralPrime}) then show that the {\it spontaneously broken $U(1)_Y$ AHM charge is not conserved}, even for connected time-ordered products \cite{LSS-3Short}, in  
Lorenz gauge
\begin{eqnarray}
\label{ChargeAHMTimeOrderedProductsNotConserved}
&&\Big< 0\vert T\Big[ \Big(  \frac{d}{dt}Q_{U(1);AHM}(t) \Big) \nonumber \\
&&\quad \quad \times h(x_1)...h(x_N) \pi_{t_1}(y_1)...\pi_{t_M}(y_M)\Big]\vert 0\Big>_{Connected}  \nonumber \\
&&\quad \quad \neq 0 
\end{eqnarray}
dashing all further hope. 

The simplest, most powerful, and most general proof of the Goldstone theorem \cite{Goldstone1961,Goldstone1962,Kibble1967} requires a conserved charge $\frac{d}{dt}Q=0$, so that proof fails for spontaneously broken gauge theories.
This (of course) is a very famous result \cite{Higgs1964,Englert1964,Guralnik1964,Kibble1967}, and allows the spontaneously broken AHM to generate a mass-gap $m_A$ for the vector $A^\mu$, and avoid massless particles in its observable physical spectrum. 
This is true, even in 
Lorenz gauge, where there {\em is} a Goldstone theorem, $\pi$ is massless, and $\tilde \pi$ is a NGB \cite{Guralnik1964,Kibble1967}.

Massless $\pi$ is the basis \cite{LSS-3Short} of our pion-pole-dominance-based $U(1)_Y$ WTIs, derived in Appendix A, which give: relations among 1-$\phi$-R connected amputated  $\phi$-sector T-Matrix elements $T_{N,M}$ (\ref{AdlerSelfConsistencyPrime},  \ref{InternalTMatrix});  relations among 1-$\phi$-I connected amputated  $\phi$-sector Greens functions $\Gamma_{N,M}$ (\ref{GreensWTIPrime}, \ref{GreensFWTI}); 1-soft-pion theorems  (\ref{AdlerSelfConsistencyPrime}, \ref{AdlerSelfConsistency}, \ref{InternalTMatrix}); infra-red finiteness for $\mpisq =0$ (\ref{AdlerSelfConsistencyPrime}, \ref{AdlerSelfConsistency});  and a Goldstone theorem (\ref{TMatrixGoldstoneTheoremPrime}, \ref{TMatrixGoldstoneTheorem}).

\subsection{Construction of the scalar-sector effective Lagrangian from those $U(1)_Y$ WTIs which  govern connected amputated 1-$\phi$-I Greens functions}
\label{GreensFunctionsAHM}

In Appendix \ref{DerivationWTIAHM} we derive $U(1)_Y$ ``pion-pole-dominance"  1-$\phi$-R
connected amputated T-Matrix  WTI (\ref{InternalTMatrix})
for the SSB AHM. Their solution is a tower of recursive $U(1)_Y$ WTI (\ref{GreensFWTI}) which govern 1-$\phi$-I  $\phi$-sector connected amputated Greens functions $\Gamma_{N,M}$.
For $ \pi$ with $CP=-1$, the result
\begin{eqnarray}
	\label{GreensWTIPrime}
	&&\HVEV\Gamma_{N,M+1}(p_1 ...p_N;0q_1...q_M) \nonumber  \nonumber \\
	&&\quad \quad =\sum ^M_{m=1} \Gamma_{N+1,M-1}(q_mp_1...p_N;q_1...{\widehat {q_m}}...q_M) \nonumber \\
	&&\quad \quad -\sum ^N_{n=1}\Gamma_{N-1,M+1}(p_1 ...{\widehat {p_n}}...p_N;p_nq_1...q_M)
\end{eqnarray} 
is valid for $N,M \ge0$. 
On the left-hand-side of (\ref{GreensWTIPrime}) there are 
N renormalized $h=H-\HVEV$ external legs (coordinates x, momenta p), 
M renormalized ($CP=-1$) ${ \pi}$ external legs (coordinates y, momenta q), and 1 renormalized soft external $\pi(k_\mu=0)$  (coordinates z, momenta k).
``Hatted fields with momenta" $({\widehat {p_n}},{\widehat {q_m}})$ are omitted.

The rigid $U(1)_Y$ WTI  ``1-soft-pion theorems" (\ref{GreensWTIPrime})
relate a 1-$\phi$-I Green's function with $(N+M+1)$ external fields 
(which include an extra zero-momentum ${ \pi}$), 
to two 1-$\phi$-I  Green's functions with $(N+M)$ external fields. 
\footnote{
The rigid  $U(1)_Y$ WTI (\ref{GreensWTIPrime})  for the $U(1)_Y$ AHM {\em gauge theory} are a generalization
of the classic work of B.W. Lee \cite{Lee1970},  
who constructed 2 all-loop-orders renormalized towers of WTI's 
for the {\em global} $SU(2)_L\times SU(2)_R$ Gell-Mann Levy (GML) model \cite{GellMannLevy1960} with Partially Conserved Axial-vector Currents (PCAC).
We replace GML's strongly-interacting Linear Sigma Model (\LSM) with a weakly-interacting BEH \LSM, with explicit PCAC breaking $ =0$. Replace 
$\sigma \to H, {\vec \pi}\to { \pi}, m_{\sigma}\to m_{BEH}, f_{\pi} \to \HVEV$, and add {\bf local gauge group} $U(1)_Y$. This generates 
a set of {\bf global} $U(1)_Y$ WTI governing relations among {\bf weak-interaction} 1-$\phi$-R 
T-Matrix elements $T_{N,M}$. A solution-set of those $U(1)_Y$ WTI then govern relations among $U(1)_Y$ 1-$\phi$-I Green's functions $\Gamma_{N,M}$. As observed by Lee for \GML, one of those on-shell T-Matrix WTI is the Goldstone theorem. 
Appendix \ref{DerivationWTIAHM} includes, in Table 1,  translation between the WTI proofs in this paper (a gauge theory) and in  B.W. Lee (a global theory). 
}
The Green's functions $\Gamma_{N,M}(p_1...p_N;q_1...q_M)$ 
are not themselves gauge-independent. 
Furthermore, although 1-$\phi$-I, they are 1-$A^\mu$-Reducible (1-$A^\mu$-R) by cutting a transverse $A_\mu$ gauge boson line.

The 1-$\phi$-I ${ \pi}$  and $h$ inverse propagators are:
\begin{eqnarray}
\label{InversePropagators}
\Gamma_{0,2}(;q,-q) &\equiv& \left[ \Delta_{\pi}(q^2) \right]^{-1} \nonumber \\
\Gamma_{2,0}(q,-q;) &\equiv& \left[ \Delta_{BEH}(q^2) \right]^{-1} 
\end{eqnarray}

We can now form the $\phi$-sector effective {\em momentum space} Lagrangian in Lorenz gauge.
All perturbative quantum loop corrections, to all-loop-orders and including all UVQD, log-divergent and finite  contributions, 
are included in the $\phi$-sector effective Lagrangian:
1-$\phi$-I Green's functions $\Gamma_{N,M}(p_1...p_N;q_1...q_M)$; wavefunction renormalizations;  
renormalized $\phi$-scalar propagators (\ref{pNGBPropagator},\ref{BEHPropagator}); 
the BEH VEV $\HVEV$ (\ref{HVEV}); 
all gauge boson and ghost propagators. 
This includes the full all-loop-orders renormalization of the AHM $\phi$-sector,
originating in quantum loops containing transverse virtual gauge bosons, $\phi$-scalars and ghosts:
$A^\mu;h,\pi;{\bar \omega},\omega$ respectively. 
Because they arise entirely from global $U(1)_Y$ WTI, 
our results are independent of regularization-scheme \cite{KrausSiboldAHM}.

We want to classify operators arising in AHM loops,
and separate the finite operators  
\footnote{
In the Standard Model, there are finite operators that arise entirely from $SM$ degrees of freedom,   
which are crucially important for computing experimental observables. The most familiar are the  
	successful 1-loop high precision Standard Model  
	predictions for the top-quark from
	Z-pole physics \cite{LynnStuart1985,Kennedy1988,EXPOSTAR,
Levinthal1990,ALEPH1991,Levinthal1992a,Levinthal1992b,LEPWorkingGroup1993} in 1984
	and the $W^{\pm}$ mass \cite{Sirlin1980} in 1980, 
as well as the 2-loop BEH mass  from
	Z-pole physics \cite{LynnStuart1985,Verzegnassi1987,Kennedy1988,EXPOSTAR,Levinthal1992a,
LEPSLCWorkingGroup1995}
	and the $W^{\pm}$ mass \cite{Sirlin1980,Verzegnassi1987}: those precisely predicted the experimental discovery-masses of the top quark at FNAL \cite{Nobel1999}, and BEH scalar at CERN \cite{
LynnStuart1985,Kennedy1988,EXPOSTAR,
Levinthal1992a,LEPSLCWorkingGroup1995,
Nobel2013}.
But the $U(1)_Y$ analogy of such finite operators are not the point of this paper. 
}
from the divergent ones.
We focus on finite {\em relevant} operators, as well as quadratic and logarithmically divergent operators,
{\em which may cause a fine-tuning problem}.

There are 3 classes of {\it finite} operators, which cannot generate fine-tuning in the AHM: 
\begin{itemize}
\item Finite  ${\cal O}_{AHM}^{1/\Lambda^2;Irrelevant}$ vanish as $m_{Weak}^2/ \Lambda^2 \to 0$;

\item ${\cal O}_{AHM}^{Dim>4;Light}$ are finite dimension $Dim>4$ operators, where only the light degrees of freedom $A^\mu;h,\pi;{\bar \omega},\omega$ contribute to   all-loop-orders renormalization;  

\item ${\cal O}_{AHM}^{Dim\leq4;NonAnalytic}$ are finite dimension $Dim\leq4$ operators which are non-analytic in momenta or in a renormalization scale $\mu^2$  (e.g. finite renormalization-group logarithms). 
\end{itemize}

All such operators will be ignored.
\begin{eqnarray}
{\cal O}_{AHM}^{Ignore}&=&{\cal O}_{AHM}^{1/\Lambda^2;Irrelevant} + {\cal O}_{AHM}^{Dim>4;Light} \nonumber \\
&+&{\cal O}_{AHM}^{Dim\leq4;NonAnalytic}
\end{eqnarray}
Such finite operators appear throughout the $U(1)_Y$ Ward-Takahashi IDs (\ref{GreensWTIPrime}): 
\begin{itemize}
\item $N+M \geq 5$ is  ${\cal O}_{AHM}^{1/ \Lambda^2; Irrelevant}$ and ${\cal O}_{AHM}^{Dim>4;Light}$;
\item The left hand side of  (\ref{GreensWTIPrime}) for $N+M=4$ is also  
${\cal O}_{AHM}^{1/ \Lambda^2; Irrelevant}$ and ${\cal O}_{AHM}^{Dim>4;Light}$;
\item  $N+M\leq 4$ operators ${\cal O}_{AHM}^{Dim\leq 4;NonAnalytic}$  appear in (\ref{GreensWTIPrime}). 
\end{itemize}

Finally, there are  $N+M\leq 4$ operators that are analytic in momenta. 
We expand these in powers of momenta, 
count the resulting dimension of each term in the operator Taylor-series, 
and ignore ${\cal O}_{AHM}^{Dim>4;Light}$ and ${\cal O}_{AHM}^{1/ \Lambda^2; Irrelevant}$ terms in that series. 

The 
all-loop-orders renormalized scalar-sector effective Lagrangian is then formed
for ($h, \pi$) with CP=($1,-1$)
\begin{eqnarray}
\label{FormSchwingerPotential}
&& L^{Eff;Wigner,SI,Goldstone}_{AHM;\phi;Lorenz} \nonumber \\
&&\quad \quad =  \Gamma_{1,0}(0;)h +\frac{1}{2!} \Gamma_{2,0}(p,-p;)h^2 \nonumber \\
&&  \quad \quad + \frac{1}{2!} \Gamma_{0,2}(;q,-q)\pi^2 +\frac{1}{3!} \Gamma_{3,0}(000;)h^3  \nonumber \\ 
&& \quad \quad + \frac{1}{2!} \Gamma_{1,2}(0;00) h \pi^2  +\frac{1}{4!} \Gamma_{4,0}(0000;)h^4  \nonumber \\
&&  \quad \quad + \frac{1}{2!2!} \Gamma_{2,2}(00;00) h^2 \pi^2  \nonumber \\
&& \quad \quad +  \frac{1}{4!}\Gamma_{0,4}(;0000)\pi^4 + {\cal O}^{AHM}_{Ignore}   
\end{eqnarray}

The Ward-Takahashi IDs (\ref{GreensWTIPrime}) for Greens functions severely constrain the effective Lagrangian (\ref{FormSchwingerPotential}). 
\begin{itemize}


\item WTI $N=0, M=1$
\begin{eqnarray}
\label{NM01}
\Gamma_{1,0}(0;) &=& \HVEV \Gamma_{0,2}(;00) 
\end{eqnarray}
since no momentum can run into the tadpoles.

\item WTI $N=1, M=1$\footnote{
In previous papers on  $SU(2)_L\times SU(2)_R$ \GMLfull\cite{GellMannLevy1960}, we have written this $N=1,M=1$  WTI 
as a mass-relation between the BEH $h$ scalar and the {\em pseudo}-Nambu-Goldstone boson ${\pi}$ pseudo-scalar. In the K$\ddot a$ll$\acute e$n-Lehmann representation
\begin{eqnarray}
\label{MassRelation}
&&m_{BEH}^2 = \mpisq + 2\lambda_{\phi}^2 \HVEV ^2 \\
&&\mpisq  =  \Bigg[ \frac{1}{m_{\pi ;Pole}^2}+ \int dm^2 \frac{\rho_{\pi}(m^2)}{m^2 } \Bigg]^{-1} \nonumber \\
&&m_{BEH}^2 =  \Bigg[ \frac{1}{m_{BEH;Pole}^2}+ \int dm^2 \frac{\rho_{BEH}(m^2)}{m^2 } \Bigg]^{-1} \nonumber 
\end{eqnarray}
}
\begin{eqnarray}
\label{NM11}
\Gamma_{2,0}(-q,q;) &-& \Gamma_{0,2}(;q,-q)  \nonumber \\
&=&\HVEV \Gamma_{1,2}(-q;q0)  \nonumber \\
&=&\HVEV \Gamma_{1,2}(0;00)  + {\cal O}^{AHM}_{Ignore} \nonumber \\ \nonumber
\Gamma_{2,0}(00;) &=& \Gamma_{0,2}(;00) + \HVEV \Gamma_{1,2}(0;00) \\
&+&  {\cal O}^{AHM}_{Ignore}
\end{eqnarray}

\item WTI $N=2, M=1$
\begin{eqnarray}
\label{NM21}
\HVEV\Gamma_{2,2}(00;00) &=& \Gamma_{3,0}(000;) -2\Gamma_{1,2}(0;00) 
\end{eqnarray}

\item WTI $N=0, M=3$
\begin{eqnarray}
\label{NM03}
\HVEV\Gamma_{0,4}(;0000) &=& 3 \Gamma_{1,2}(0;00)
\end{eqnarray}

\item WTI $N=1, M=3$
\begin{eqnarray}
\label{NM13}
0&=&3\Gamma_{2,2}(00;00)-\Gamma_{0,4}(;0000)
\end{eqnarray}

\item WTI $N=3, M=1$
\begin{eqnarray}
\label{NM31}
0&=&\Gamma_{4,0}(0000;)-3\Gamma_{2,2}(00;00) 
\end{eqnarray}

\item The quartic coupling constant is defined in terms of a 4-point 1-$\phi$-I Green's function
\begin{eqnarray}
\label{SchwingerWignerDataA}
 \Gamma_{0,4}(;0000)  &\equiv& -6 \lambda_{\phi}^2 
\end{eqnarray} 

\end{itemize}

The  all-loop-orders renormalized $\phi$-sector momentum-space effective Lagrangian (\ref{FormSchwingerPotential}) - constrained 
{\bf only} by those $U(1)_Y$ WTI governing Greens functions (\ref{GreensWTIPrime}) - may be written
\begin{eqnarray}
\label{LEffectiveSM}
&&L^{Eff;Wigner,SI,Goldstone}_{AHM;\phi;Lorenz} = L^{Kinetic;Eff;Wigner,SI,Goldstone}_{AHM;\phi;Lorenz} \nonumber \\
&& \quad \quad \qquad \qquad -V^{Eff;Wigner,SI,Goldstone}_{AHM;\phi;Lorenz} + {\cal O}_{Ignore}^{AHM} \nonumber \\
&&L^{Kinetic;Eff;Wigner,SI,Goldstone}_{AHM;\phi;Lorenz} \nonumber \\
&&\quad \quad \qquad \qquad =\half \Big( \Gamma_{0,2}(;p,-p) - \Gamma_{0,2}(;00) \Big) h^2 \nonumber \\
&&\quad \quad \qquad \qquad +\half \Big( \Gamma_{0,2}(;q,-q)- \Gamma_{0,2}(;00)  \Big) { \pi}^2 \nonumber \\
&&V^{Eff;Wigner,SI,Goldstone}_{AHM;\phi;Lorenz}= \mpisq \Big[ \frac{h^2 + { \pi}^2}{2} +\HVEV h \Big] \nonumber \\
&&\quad \quad \qquad \qquad + \lambda_{\phi}^2 \Big[ \frac{h^2 + { \pi}^2}{2} +\HVEV h \Big]^2
\end{eqnarray} 
with
finite non-trivial wavefunction renormalization 
\begin{eqnarray}
\label{WavefunctionA}
\Gamma_{0,2}(;q,-q)-\Gamma_{0,2}(;00) \sim q^2 
\end{eqnarray}

The $\phi$-sector effective Lagrangian (\ref{LEffectiveSM}) has insufficient boundary conditions 
to distinguish among 
the 3 modes \cite{Lee1970,Symanzik1970a,Symanzik1970b,Vassiliev1970}  of the 
BRST-invariant Lagrangian $L_{AHM}$ in (\ref{LagrangianAHM}). 
For example, the effective potential $V^{Eff;Wigner;SI;Goldstone}_{AHM;\phi;Lorenz} $ becomes in various limits: 
\footnote{
	The inclusive Gell-Mann L${\acute e}$vy \cite{GellMannLevy1960} effective potential derived \cite{Lynnetal2012} from B.W. Lee's WTI \cite{Lee1970}, reduces to the three different effective potentials of the global $SU(2)_L \times SU(2)_R$ Schwinger model \cite{Schwinger1957}: 
	Schwinger Wigner mode $(\HVEV =0,\mpisq =m_{BEH}^2 \neq 0)$; 
	Schwinger Scale-Invariant point $(\HVEV =0,\mpisq = m_{BEH}^2 =0)$; 
	or Schwinger Goldstone mode $(\HVEV \neq 0,\mpisq = 0;m_{BEH}^2\neq 0)$.
	} 
AHM Wigner mode $(m_A^2=0;\HVEV =0;\mpisq =m_{BEH}^2\neq 0)$; 
AHM  ``Scale-Invariant" (SI) point $(m_A^2=0;\HVEV =0;\mpisq = m_{BEH}^2= 0)$; 
or AHM Goldstone mode $(m_A^2\neq0;\HVEV \neq 0;\mpisq = 0; m_{BEH}^2\neq 0)$;
\begin{eqnarray}
\label{WignerSIGoldstonePotentials}
V^{Eff;Wigner}_{AHM;\phi;Lorenz}&=& \mpisq \Big[ \frac{h^2 + { \pi}^2}{2} \Big] + \lambda_{\phi}^2 \Big[ \frac{h^2 + { \pi}^2}{2}  \Big]^2 \nonumber \\
V^{Eff;ScaleInvariant}_{AHM;\phi;Lorenz}&=& \lambda_{\phi}^2 \Big[ \frac{h^2 + { \pi}^2}{2}  \Big]^2 \nonumber \\
V^{Eff;Goldstone}_{AHM;\phi;Lorenz}&=& \lambda_{\phi}^2 \Big[ \frac{h^2 + { \pi}^2}{2} +\HVEV h \Big]^2
\end{eqnarray}

But (\ref{LEffectiveSM})  has exhausted the constraints (i.e. on the allowed terms in the $\phi$-sector effective Lagrangian) due to those $U(1)_Y$ WTIs which govern 1-$\phi$-I $\phi$-sector Green's functions $\Gamma_{N,M}$ (\ref{GreensWTIPrime}, \ref{GreensFWTI}). In order to provide boundary conditions  which distinguish between the effective potentials in
(\ref{WignerSIGoldstonePotentials}), we must turn to those $U(1)_Y$ WTIs which govern $\phi$-sector  1-$\phi$-R 
 T-Matrix elements  $T_{N,M}$.

\subsection{Further constraints on the $\phi$-sector effective Lagrangian:
IR finiteness; Goldstone theorem; automatic tadpole renormalization;\\ 
$\tilde \pi$ is a NGB;
gauge-independent observable $m_{BEH}^2 = 2 \lambda _\phi ^2 \HVEV ^2$
 is GEN not-FT
}
\label{TMatrixAHM}

\begin{quote}
{\it ``Whether you like it or not, 
you have to include in the Lagrangian all possible terms consistent with locality and power counting, 
{\bf unless otherwise constrained by Ward identities}."}
Kurt Symanzik, in a private letter to Raymond Stora \cite{SymanzikPC} 
\end{quote}

In strict obedience to K. Symanzik's edict, we now further constrain the allowed terms in the $\phi$-sector effective Lagrangian with {\bf those $U(1)_Y$ Ward-Takahashi identities which govern 1-$\phi$-R T-Matrix elements $T_{N,M}$}.

In Appendix A, we extend Adler's self-consistency condition  (originally written for 
the $SU(2)_L\times SU(2)_R$ \GMLfull model with PCAC\cite{Adler1965,AdlerDashen1968}), 
to AHM in Lorenz 
gauge (\ref{AdlerSelfConsistency})
\begin{eqnarray}
\label{AdlerSelfConsistencyPrime} 
&&\HVEV T_{N,M+1}(p_1...p_N;0q_1...q_M)\nonumber \\
&& \quad \quad \times (2\pi)^4\delta^4 \Big(\sum_{n=1}^N p_n +\sum_{m=1}^M q_m \Big) \Big\vert^{p_1^2 =p_2^2...=p_N^2=m_{BEH}^2}_{q_1^2 =q_2^2...=q_M^2=0}  \nonumber \\
&& \quad \quad =0 
\end{eqnarray}
The T-matrix vanishes as one of the pion momenta goes to zero (i.e. 1-soft-pion theorems), provided all other physical scalar particles are on mass-shell. 
Eqn. (\ref{AdlerSelfConsistencyPrime}) also ``asserts the absence of infrared (IR) divergences in the scalar-sector (of AHM) Goldstone mode 
(in Lorenz gauge). Although individual Feynman diagrams are IR divergent, those IR divergent parts cancel exactly in each order of perturbation theory. Furthermore, the Goldstone mode amplitude must vanish in the soft-pion limit" \cite{Lee1970}.

The $N=0,M=1$ case of (\ref{AdlerSelfConsistencyPrime}) is the Goldstone theorem (\ref{TMatrixGoldstoneTheorem}) itself \cite{Lee1970}: 
\begin{eqnarray}
\label{TMatrixGoldstoneTheoremPrime}
\HVEV T_{0,2}(;00)=0
\end{eqnarray}
Since the 2-point $T_{0,2}$ is already 1-$\phi$-I, we may write the Goldstone theorem as a further constraint on the 1-$\phi$-I Greens function
\begin{eqnarray}
\label{GFGoldstoneTheoremPrime}
\HVEV \Gamma_{0,2}\left(;00\right) =\HVEV \big[ \Delta_\pi (0) \big]^{-1} =0
\end{eqnarray}

Another crucial effect of the Goldstone theorem,
together with the  $N=0, M=1$ $U(1)_Y$ Ward-Takahashi Greens function identity (\ref{GreensWTIPrime}), is to automatically eliminate tadpoles in (\ref{FormSchwingerPotential})
\begin{eqnarray}
\label{ZeroTadpoles}
\Gamma_{1,0}(0;) &=& \HVEV \Gamma_{0,2}(;00) =0
\end{eqnarray}
so that separate tadpole renormalization is un-necessary.

We re-write the effective potential (\ref{LEffectiveSM}) but now including the constraint from the Goldstone theorem
 (\ref{TMatrixGoldstoneTheoremPrime}, \ref{GFGoldstoneTheoremPrime}):
\begin{eqnarray}
\label{LEffectiveGoldstoneTheorem}
L^{Eff;Goldstone}_{AHM;\phi;Lorenz} \nonumber &=& L^{Kinetic;Eff;Goldstone}_{AHM;\phi;Lorenz} \nonumber \\
 &-&V^{Eff;Goldstone}_{AHM;\phi;Lorenz} \nonumber \\
&+&{\cal O}_{Ignore}^{AHM} \nonumber \\
V^{Eff;Goldstone}_{AHM;\phi;Lorenz} &=& \lambda_{\phi}^2  
\left[ {\frac{h^2+{\vec \pi}^2}{2}} +\HVEV h\right] ^2 
\end{eqnarray} 
and wavefunction renormalization
\begin{eqnarray}
\label{GoldstoneWavefunction}
\Gamma_{0,2}(;q,-q)-\Gamma_{0,2}(;00) = q^2 +{\cal O}_{Ignore}^{AHM}
\end{eqnarray}
so the $\phi$-sector Goldstone mode effective  {\bf coordinate space} Lagrangian becomes
\begin{eqnarray}
\label{GoldstoneLagrangian}
L^{Eff;Goldstone}_{AHM;\phi;Lorenz} &=& \vert \partial_{\mu}\phi \vert ^2 -\lambda_{\phi}^2  \Big[ {\frac{h^2+{\pi}^2}{2}} +\HVEV h\Big] ^2 \nonumber \\
&+&{\cal O}_{Ignore}^{AHM}
\end{eqnarray} 

Eqn. (\ref{GoldstoneLagrangian}) is the $\phi$-sector effective Lagrangian of the  {\em spontaneously broken} Abelian Higgs model, in 
Lorenz gauge, contrained by the Goldstone theorem:
\footnote{
Imagine we suspected that $\pi$ {\em is not} all-loop-orders massless in 
Lorenz gauge SSB AHM, and simply/naively wrote a mass-squared $m_{\pi:Pole}^2$ (which we even imagined to be fine-tuned!)
into the  $\pi$ inverse-propagator
\begin{eqnarray}
\label{GoldstoneTheoremViolation}
&& \left[ \Delta_{\pi}(0)  \right]^{-1}  \equiv -\mpisq =-m_{\pi ;Pole}^2\Bigg[ 1+ m_{\pi ;Pole}^2\int dm^2 \frac{\rho_{\pi}(m^2)}{m^2 } \Bigg]^{-1} \quad
\end{eqnarray}
But the Goldstone theorem (\ref{GFGoldstoneTheoremPrime}) insists instead that
\begin{eqnarray}
\label{GoldstoneTheoremNGBMass}
&& \HVEV \left[ \Delta_{\pi}(0) \right]^{-1} \equiv -\HVEV \mpisq = \HVEV \Gamma_{0,2}(;00)=0 \quad \quad
\end{eqnarray} 
The $\pi$-pole-mass vanishes {\em exactly}, and is GEN not FT.
\begin{eqnarray}
\label{PionPoleMass}
m_{\pi;Pole}^2 &=& \mpisq\Bigg[ 1- \mpisq\int dm^2 \frac{\rho_{\pi}(m^2)}{m^2 } \Bigg]^{-1} =0 \quad \quad 
\end{eqnarray}
}
\begin{itemize}
\item It includes all divergent
${\cal O}(\Lambda^2),{\cal O}(\ln \Lambda^2)$ and finite terms which arise, 
to all perturbative loop-orders in the full 
$U(1)_Y$ gauge theory, due to virtual transverse gauge bosons, $\phi$ scalars and ghosts; $A^\mu;h,\pi;{\bar \omega},\omega$ respectively.
\item It obeys the Goldstone theorem  (\ref{TMatrixGoldstoneTheoremPrime},\ref{GFGoldstoneTheoremPrime}) and all other $U(1)_Y$ Ward-Takahashi Green's function and T-Matrix identities; 
\item It is minimized at $(H=\HVEV, {\pi}=0)$; and obeys stationarity  \cite{ItzyksonZuber} of that true minimum; 
\item It preserves the theory's renormalizability and unitarity, which require that wavefunction  renormalization, 
$\HVEV_{Bare}=\big[ Z^{\phi}_{AHM}\big]^{1/2}\HVEV$ \cite{LSS-2,Bjorken1965,ItzyksonZuber}, forbid UVQD, relevant, or any other dimension-2 operator corrections to $\HVEV$;


\item {\bf The Goldstone theorem (\ref{TMatrixGoldstoneTheoremPrime}) has caused all relevant operators in the spontaneously broken Abelian Higgs model to vanish!}

\end{itemize}

In order to make manifest  that $\tilde  \pi$ is a true NGB \cite{JCTaylor1976,Georgi2009} in 
Lorenz gauge,
re-write (\ref{GoldstoneLagrangian})  in the unitary Kibble representation. 
\cite{Ramond2004,Georgi2009} with $Y_\phi =-1$ the $\phi$ hypercharge. In coordinate space
\begin{eqnarray}
\label{GoldstoneKibbleAHMLagrangian}
\phi&=& \frac{1}{\sqrt{2}} {\tilde H} e^{-iY_\phi {\tilde \pi}/\HVEV} \nonumber \\
L_{AHM;\phi;Lorenz}^{Eff;Goldstone} &=& \frac{1}{2} \left( \partial_{\mu} {\tilde H} \right)^2 + \frac{1}{2} 
\frac {{\tilde H}^2}{\HVEV ^2}  \left( \partial_{\mu} {\tilde \pi} \right)^2  \nonumber  \\
&-&\frac{\lambda_{\phi}^2}{4}  \Big[ {\tilde H}^2 -\HVEV ^2 \Big]^2+{\cal O}_{Ignore}^{AHM} \nonumber \\
&=& \frac{1}{2} \left( \partial_{\mu} {\tilde h} \right)^2 + \frac{1}{2} \Big( 1+\frac {\tilde h}{\HVEV} \Big)^2
 \left( \partial_{\mu} {\tilde \pi} \right)^2  \nonumber  \\
&-&\frac{\lambda_{\phi}^2}{4}  \Big[ {\tilde h}^2 +\HVEV {\tilde h} \Big]^2+{\cal O}_{Ignore}^{AHM} 
\end{eqnarray}
shows that $\tilde \pi$ has only derivative couplings and, for constant $\theta$, a shift symmetry
\begin{eqnarray}
\label{ShiftSymmetryAHM}
{\tilde \pi} &\to& {\tilde \pi} + \HVEV \theta
\end{eqnarray}

The  Green's function Ward-Takahashi ID (\ref{GreensWTIPrime}) for $N=1,M=1$, constrained by the Goldstone theorem (\ref{GFGoldstoneTheoremPrime}), relates the BEH mass  to the coefficient of the $h{ \pi}^2$ vertex 
\begin{eqnarray}
\label{WTIHiggsMass}
\Gamma_{2,0}(00;)=\HVEV\Gamma_{1,2}(0;00) 
\end{eqnarray} 
Therefore, the BEH mass-squared in (\ref{GoldstoneKibbleAHMLagrangian}) 
\begin{eqnarray}
\label{BEHMassAHMS}
m_{BEH}^2 = 2\lambda_{\phi}^2\HVEV^2
\end{eqnarray} 
{\bf arises entirely from SSB}, as does (together with its AHM decays) the gauge-independent observable resonance pole-mass-squared
\begin{eqnarray}
\label{BEHPoleMassPrime}
m^2_{BEH;Pole}  &=& 2\lambda_\phi^2 \HVEV^2\Big[ 1- 2\lambda_\phi^2 \HVEV^2 \int dm^2 \frac{{\tilde \rho}^{BEH}_{AHM}(m^2)}{m^2 - i\epsilon} \Big]^{-1} \nonumber \\
&+& {\cal O}^{Ignore}_{AHM;\phi}
\end{eqnarray}
where the spectral density ${\tilde \rho}^{BEH}_{AHM}$ is displayed in the Kibble representation.

Since weak scale $\HVEV=\big[Z_{AHM}^{\phi}\big]^{-\half}\HVEV_{Bare}$ and dimensionless $\lambda_\phi^2$, absorb no relevant operators, and are therefore not FT, 
the  $\phi$-sector of the SSB AHM gauge theory is Goldstone Exceptionally Natural, with far more powerful suppression of fine-tuning than G. 't Hooft's naturalness criteria \cite{tHooft1980} would demand.

\section{extended-AHM: $U(1)_Y$ WTI  cause certain heavy matter representations to decouple from the low-energy  $\phi$-sector effective Lagrangian}
\label{E-AHM}

If the Euclidean cutoff $\Lambda^2$ were a true proxy for very heavy $M_{Heavy}^2 \gg m_{Weak}^2$   spin $S=0$ scalars $\Phi$, and $S=\half$ fermions $\psi$, 
we would already be in a position to comment on their de-coupling.
Unfortunately, although the literature seems to cite such proxy, it is simply not true. In order to prove theorems which reveal symmetry-driven results in gauge theories,
one must keep {\it all} of the terms arising from {\em all} Feynman graphs; i.e.  not just ``a selection of interesting terms from a representative subset of Feynman graphs" (Ergin Sezgin's dictum).

\subsection{$\phi$-sector effective Lagrangian for the extended-AHM}
\label{LagrangianE-AHM}

{\bf i) 1-$\phi$-I connected amputated $\phi$-sector Green's functions $\Gamma^{Extended}_{N,M}$:} In Appendix \ref{DerivationWTIE-AHM} we derive a tower of recursive $U(1)_Y$ WTI (\ref{ExtendedGreensFWTI}) which govern connected amputated 1-$\phi$-I Green's functions  for the extended-AHM:
\begin{eqnarray}
	\label{GreensWTIPrimeExtended}
	&&\HVEV\Gamma_{N,M+1}^{Extended}(p_1 ...p_N;0q_1...q_M) \nonumber  \nonumber \\
	&&\quad \quad =\sum ^M_{m=1} \Gamma^{Extended}_{N+1,M-1}(q_mp_1...p_N;q_1...{\widehat {q_m}}...q_M) \nonumber \\
	&&\quad \quad -\sum ^N_{n=1}\Gamma^{Extended}_{N-1,M+1}(p_1 ...{\widehat {p_n}}...p_N;p_nq_1...q_M)
\end{eqnarray} 
valid for $N,M \ge0$.

$\Gamma_{N,M}^{Extended}$ includes the all-loop-orders renorma;ization of the $\phi$-sector SSB extended-AHM, 
including virtual transverse gauge bosons, $\phi$-scalars, ghosts new scalars, and new fermions: $A^\mu;h, \pi;{\bar \omega},\omega;\Phi;\psi;$ respectively.

There are 4 classes of {\it finite} operators in the full SSB extended-AHM gauge theory, which can generate neither fine-tuning nor ``non-decoupling" of heavy particles 
\begin{itemize}

\item Finite ${\cal O}_{E-AHM;\phi}^{1/\Lambda^2;Irrelevant} $  vanish as $m_{Weak}^2/ \Lambda^2 \to 0$ 
or $M_{Heavy}^2/ \Lambda^2 \to 0$;

\item Finite ${\cal O}_{E-AHM;\phi}^{Dim>4;Light} $ are
 dimension $Dim>4$ operators,  where only the light degrees of freedom, 
including ghosts $({\bar \omega},\omega$) and $(\Phi_{Light} ,\psi_{Lght})$,  
contribute to   all-loop-orders renormalization. 

\item
${\cal O}_{E-AHM;\phi}^{Dim \leq 4;NonAnalytic;Light} $ are finite dimension $Dim \leq4$ operators, 
which are non-analytic in momenta or in a renormalization scale $\mu^2$, where only the light degrees of freedom $A^\mu;h,\pi;{\bar \omega},\omega;\Phi_{Light};\psi_{Light}$
contribute to   all-loop-orders renormalization.

\item ${\cal O}_{E-AHM;\phi}^{1/M_{Heavy}^2;Irrelevant} $  vanish as $m_{Weak}^2/ M_{Heavy}^2 \to 0$;

\end{itemize}

In addition
${\cal O}_{E-AHM;\phi}^{Dim \leq 4;NonAnalytic;Heavy}$ are finite dimension $Dim \leq4$ operators, 
which are non-analytic in momenta or in a renormalization scale $\mu^2$, where the heavy degrees of freedom $\Phi_{Heavy};\psi_{Heavy}$
contribute to   all-loop-orders renormalization. Analysis of these operators lies outside the scope of this paper.

All such operators will be ignored
\begin{eqnarray}
\label{IgnoreE-AHMOperators}
&&{\cal O}_{E-AHM;\phi}^{Ignore} \nonumber \\
&&\qquad ={\cal O}_{E-AHM;\phi}^{1/\Lambda^2;Irrelevant}+{\cal O}_{E-AHM;\phi}^{Dim>4;Light} \nonumber \\
&&\qquad +{\cal O}_{E-AHM\phi}^{Dim \leq 4;NonAnalytic;Light} \nonumber \\
&&\qquad +{\cal O}_{E-AHM\phi}^{Dim \leq 4;NonAnalytic;Heavy} \nonumber \\
&&\qquad +{\cal O}_{E-AHM;\phi}^{1/M_{Heavy}^2;Irrelevant} 
\end{eqnarray}

Such finite operators appear throughout the extended $U(1)_Y$ Ward-Takahashi IDs (\ref{GreensWTIPrimeExtended}) 
\begin{itemize}
\item $N+M \geq 5$ is  ${\cal O}_{E-AHM;\phi}^{1/ \Lambda^2; Irrelevant}$, ${\cal O}_{E-AHM;\phi}^{Dim>4;Light}$ 
and ${\cal O}_{E-AHM;\phi}^{1/M_{Heavy}^2;Irrelevant}$;
\item The left hand side of  (\ref{GreensWTIPrimeExtended}) for $N+M=4$ is also  
${\cal O}_{E-AHM;\phi}^{1/ \Lambda^2; Irrelevant}$, ${\cal O}_{E-AHM;\phi}^{Dim>4;Light}$ 
and ${\cal O}_{E-AHM;\phi}^{1/M_{Heavy}^2;Irrelevant}$;
\item  $N+M\leq 4$ operators ${\cal O}_{E-AHM;\phi}^{Dim\leq 4;NonAnalytic;Light}$  also appear in (\ref{GreensWTIPrimeExtended}). 
\end{itemize}

Finally, there are  $N+M\leq 4$ operators that are analytic in momenta. 
We expand these in powers of momenta, 
count the resulting dimension of each term in the operator Taylor-series, 
and ignore ${\cal O}_{E-AHM;\phi}^{Dim>4;Light}$, ${\cal O}_{E-AHM;\phi}^{1/ \Lambda^2; Irrelevant}$ 
and ${\cal O}_{E-AHM;\phi}^{1/M_{Heavy}^2;Irrelevant}$ in that series.

The 
all-loop-orders renormalized $\phi$-sector effective momentum-space Lagrangian for extended-AHM is then formed
for ($h,\vec \pi$) external particles with CP=($1,-1$)
\begin{eqnarray}
\label{FormSchwingerPotentialExtended}
&& L^{Eff;Wigner,SI,Goldstone}_{E-AHM;\phi} =  \Gamma_{1,0}^{Extended}(0;)h \nonumber \\
&& \quad \quad +\frac{1}{2!} \Gamma^{Extended}_{2,0}(p,-p;)h^2 \nonumber \\
&&  \quad \quad + \frac{1}{2!} \Gamma^{Extended}_{0,2}(;q,-q)\pi^2 +\frac{1}{3!} \Gamma^{Extended}_{3,0}(000;)h^3  \nonumber \\ 
&& \quad \quad + \frac{1}{2!} \Gamma^{Extended}_{1,2}(0;00) h \pi^2  +\frac{1}{4!} \Gamma^{Extended}_{4,0}(0000;)h^4  \nonumber \\
&&  \quad \quad + \frac{1}{2!2!} \Gamma^{Extended}_{2,2}(00;00) h^2 \pi^2  \\
&& \quad \quad +  \frac{1}{4!}\Gamma^{Extended}_{0,4}(;0000)\pi^4 + {\cal O}^{E-AHM}_{Ignore}  \nonumber 
\end{eqnarray}

The $U(1)_Y$ Ward-Takahashi IDs (\ref{GreensWTIPrimeExtended})  severely constrain the effective Lagrangian of the extended-AHM. 
\begin{itemize}

\item WTI $N=0, M=1$
\begin{eqnarray}
\label{NM01E-AHM}
\Gamma^{Extended}_{1,0}(0;) &=& \HVEV \Gamma^{Extended}_{0,2}(;00) 
\end{eqnarray}
since no momentum can run into the tadpoles.

\item WTI $N=1, M=1$
\begin{eqnarray}
\label{NM11E-AHM}
&&\Gamma^{Extended}_{2,0}(-q,q;) - \Gamma^{Extended}_{0,2}(;q,-q)  \nonumber \\
&&\quad \quad =\HVEV \Gamma^{Extended}_{1,2}(-q;q0)  \nonumber \\
&&\quad \quad =\HVEV \Gamma^{Extended}_{1,2}(0;00)  + {\cal O}^{E-AHM}_{Ignore} \nonumber \\ \nonumber
&&\Gamma^{Extended}_{2,0}(00;) = \Gamma^{Extended}_{0,2}(;00) + \HVEV \Gamma^{Extended}_{1,2}(0;00) \\
&&\quad \quad +  {\cal O}_{E-AHM;\phi}^{Ignore}
\end{eqnarray}

\item WTI $N=2, M=1$
\begin{eqnarray}
\label{NM21E-AHM}
\HVEV\Gamma^{Extended}_{2,2}(00;00) &=& \Gamma^{Extended}_{3,0}(000;) \nonumber \\
&-&2\Gamma^{Extended}_{1,2}(0;00) 
\end{eqnarray}

\item WTI $N=0, M=3$
\begin{eqnarray}
\label{NM03E-AHM}
\HVEV\Gamma^{Extended}_{0,4}(;0000) &=&3 \Gamma^{Extended}_{1,2}(0;00)
\end{eqnarray}

\item WTI $N=1, M=3$
\begin{eqnarray}
\label{NM13E-AHM}
0&=&3\Gamma^{Extended}_{2,2}(00;00)-\Gamma^{Extended}_{0,4}(;0000)
\end{eqnarray}

\item WTI $N=3, M=1$
\begin{eqnarray}
\label{NM31E-AHM}
0&=&\Gamma^{Extended}_{4,0}(0000;)-3\Gamma^{Extended}_{2,2}(00;00) 
\end{eqnarray}

\item The quartic coupling constant is defined in terms of a 4-point 1-SPI connected amputated GF
\begin{eqnarray}
\label{SchwingerWignerDataB}
 \Gamma^{Extended}_{0,4}(;0000)  &\equiv& -6 \lambda_{\phi}^2 
\end{eqnarray}

\end{itemize}

The  all-loop-orders renormalized $\phi$-sector effective Lagrangian (\ref{FormSchwingerPotentialExtended}), severely constrained only by the $U(1)_Y$ WTI
 governing connected amputated Greens functions (\ref{GreensWTIPrimeExtended}),  may be written
\begin{eqnarray}
\label{LEffectiveE-AHM}
&&L^{Eff;Wigner,SI,Goldstone}_{E-AHM;\phi} = L^{Kinetic}_{E-AHM;\phi} \nonumber \\
&& \quad \quad -V^{Wigner,SI,Goldstone}_{E-AHM;\phi} +{\cal O}^{Ignore}_{E-AHM;\phi}  \nonumber \\
&&L^{Kinetic}_{E-AHM;\phi} \nonumber \\
&&\quad \quad =\half \Big( \Gamma^{Extended}_{0,2}(;p,-p) - \Gamma^{Extended}_{0,2}(;00) \Big) h^2 \nonumber \\
&&\quad \quad +\half \Big( \Gamma^{Extended}_{0,2}(;q,-q)- \Gamma^{Extended}_{0,2}(;00)  \Big) { \pi}^2 \nonumber \\
&&V^{Wigner,SI,Goldstone}_{E-AHM;\phi} = \mpisq \Big[ \frac{h^2 + { \pi}^2}{2} +\HVEV h \Big] \nonumber \\
&& \quad \quad +\lambda_{\phi}^2 \Big[ \frac{h^2 + { \pi}^2}{2} +\HVEV h \Big]^2 
\end{eqnarray} 
with finite non-trivial wavefunction renormalization 
\be
\label{WavefunctionB}
\Gamma^{Extended}_{0,2}(;q,-q)-\Gamma^{Extended}_{0,2}(;00) \sim q^2 \,.
\ee

The $\phi$-sector effective Lagrangian (\ref{LEffectiveE-AHM}) for the extended-AHM has in-sufficient boundary conditions to distinguish between the 3 modes of the 
BRST-invariant Lagrangian $L_{E-AHM}$.\footnote{
	It is instructive, and dangerous and famously worrisome,  to 
	ignore vacuum energy and re-write the potenial in (\ref{LEffectiveE-AHM}) as:
	\begin{eqnarray}
	\label{VViolatesGoldstoneTheoremE-AHM}
	V^{Eff;Wigner;SI;Goldstone}_{E-AHM;\phi} &=& {\lambda_\phi^2}\left[ \phi^\dagger \phi - \half \left(\HVEV^2 - { \frac{\mpisq}{\lambda_\phi^2} } \right) \right]^2 \quad \quad
	\end{eqnarray} 
	using $\frac{h^2 + { \pi}^2}{2} +\HVEV h = \phi^\dagger \phi - \half \HVEV^2$.

	A so-called BEH fine-tuning (FT) problem arises when one {\em mistakenly} minimizes 
	$V^{Eff;Wigner;SI;Goldstone}_{E-AHM;\phi} $ in (\ref{VViolatesGoldstoneTheoremE-AHM}), 
	while ignoring the crucial constraint  by the Goldstone theorem  (see Subsection \ref{LagrangianE-AHM};ii).
	The resultant incorrect un-physical minimum
	$ \big< H \big>_{FT}^2 = \Big( \HVEV^2 - \frac{\mpisq} {\lambda_\phi^2}  \Big)  $ 
	does not distinguish properly between the 3 modes (\ref{WignerSIGoldstonePotentialsE-AHM}) of the BRST-invariant Lagrangian $L_{E-AHM}$.
	At issue is the fine-tuning of renormalized
	\begin{eqnarray}
	\label{FTExtendeAHM}
	\mpisq &=& \mu_{\phi;Bare}^2 +C_\Lambda \Lambda^2+C_{BEH} m_{BEH}^2 +\delta \mpisq \nonumber \\
	&+&C_{Heavy} M_{Heavy}^2+C_{Heavy;ln}M_{Heavy}^2 \ln{(M_{Heavy}^2)}  \nonumber \\
	&+&C_{Heavy;\Lambda} M_{Heavy}^2\ln{(\Lambda^2)} +\lambda_\phi^2 \HVEV^2
	\end{eqnarray}
	where the $C$'s are constants. It is fashionable to simply drop the UVQD term $C_\Lambda \Lambda^2$ in (\ref{FTExtendeAHM}), and argue 
	that it is somehow an artifact of dimensional regularization (DR), even though M.J. G. Veltman \cite{Veltman1981} showed that UVQD {\em do} appear at 1-loop 
	in the SM, and are properly handled by DR's poles at dimension $Dim=2$.  We call  this ``the Dim-Reg Herring." 
	We will keep UVQD. 
\newline \indent
	{\bf Wigner mode} where $\HVEV =0$:
	\begin{eqnarray}
	\label{WignerDimRegFTE-AHM}
	m_{BEH}^2 &=& m_{\pi}^2 \sim \Lambda^2, M_{Heavy}^2 \gg m_{Weak}^2
	\end{eqnarray} 
	During renormalization of a  tree-level weak-scale BEH mass-squared $m_{BEH;Bare}^2 \sim m_{Weak}^2$, relevant operators originating in quantum loops appear to ``naturally" force the renormalized value 
	up to the heavy scale $m^2_{BEH}\sim M_{Heavy}^2$. 
	Extended-AHM Wigner mode is therefore quantum-loop unstable, because the heavy scale cannot decouple from the weak scale!
	Eqn. (\ref{WignerDimRegFTE-AHM}) is the basis of the so-called ``BEH fine-tuning problem", and the motivation for much BSM physics.
	By GEN standards, weak-scale $m_{BEH}^2$ in Wigner mode is indeed fine-tuned. 
	\newline \indent
{\bf The Scale-Invariant point's} FT properties  are beyond the scope of this paper.
	\newline \indent
{\bf Spontaneously broken Goldstone  mode}, where $\HVEV \neq0$: 
	In obedience to the Goldstone theorem  
	(\ref{TMatrixGoldstoneTheoremE-AHM},\ref{TMatrixGoldstoneTheoremExtended}) below, 
	the bare counter-term $\mu_{\phi;Bare}^{2}$ 
	in (\ref{FTExtendeAHM}) is defined by 
	\begin{eqnarray}
	\label{GoldstoneFTE-AHM}
	-\HVEV \Gamma_{0,2}(;00)=-\HVEV T_{0,2}(;00)=-\HVEV\mpisq = 0
	\end{eqnarray} 
	We show below that, for constant $\theta$, the zero-value in  (\ref{GoldstoneFTE-AHM})  is protected by the NGB shift symmetry 
	\begin{eqnarray}
	\label{BEHFTShiftSymmetry}
	{\tilde \pi}\to {\tilde \pi}+\HVEV \theta
	\end{eqnarray}
\newline \indent
	In addition, minimization of (\ref{VViolatesGoldstoneTheoremE-AHM})
	violates stationarity of the true minimum at $\HVEV$\cite{ItzyksonZuber}, and destroys the theory's renormalizability and unitarity, which require that dimensionless wavefunction  renormalization 
	$\HVEV_{Bare}=\Big[ Z^{\phi}_{E-AHM}\Big] ^{1/2}\HVEV$ 
	contain no relevant operators \cite{Lynn2011,Bjorken1965,ItzyksonZuber}. The crucial observation is that, in obedience to the Goldstone theorem, 
	$Renormalized (\HVEV_{Bare}^2) \neq  \HVEV_{FT}^2$
	in Lorenz gauge SSB extended-AHM.
\newline \indent
If one persisted in regarding 
	$\mu_{\phi;Bare}^{2}$ or $\HVEV^2$ as FT, 
	they would have to be FT to $\sim 10^{\infty}$ to satisfy 
	(\ref{GoldstoneFTE-AHM}), not just a mere $\sim 10^{34}$. That of course is the nature of symmetry: e.g. (\ref{BEHFTShiftSymmetry}).  
	SSB  Goldstone mode extended-AHM is Goldstone Exceptionally Natural, not fine-tuned.
}
The effective potential $V^{Wigner,SI,Goldstone}_{E-AHM;\phi}$ in (\ref{LEffectiveE-AHM}) becomes in various limits: 
Extended-AHM Wigner mode $(m_A^2=0;\HVEV =0;\mpisq =m_{BEH}^2\neq 0)$; 
Extended-AHM Scale-Invariant point $(m_A^2=0;\HVEV =0;\mpisq = m_{BEH}^2= 0)$; 
or Extended-AHM Goldstone mode $(m_A^2\neq0;\HVEV \neq 0;\mpisq = 0; m_{BEH}^2\neq 0)$;
\begin{eqnarray}
\label{WignerSIGoldstonePotentialsE-AHM}
V^{Wigner}_{E-AHM;\phi} &=& \mpisq \Big[ \frac{h^2 + { \pi}^2}{2} \Big] + \lambda_{\phi}^2 \Big[ \frac{h^2 + { \pi}^2}{2}  \Big]^2 \nonumber \\
V^{ScaleInvariant}_{E-AHM;\phi} &=& \lambda_{\phi}^2 \Big[ \frac{h^2 + { \pi}^2}{2}  \Big]^2 \\
V^{Goldstone}_{E-AHM;\phi} &=& \lambda_{\phi}^2 \Big[ \frac{h^2 + { \pi}^2}{2} +\HVEV h \Big]^2 \nonumber
\end{eqnarray}

But (\ref{LEffectiveE-AHM})  has exhausted the constraints (i.e. on the allowed terms in the $\phi$-sector effective extended-AHM Lagrangian) 
due to those $U(1)_Y$ WTIs which govern 1-$\phi$-I connected amputated Green's functions $\Gamma^{Extended}_{N,M}$. 

{\bf ii) 1-$\phi$-R connected amputated $\phi$-sector T-Matrix elements $T^{Extended}_{N,M}$:}

In order to provide such boundary conditions (i.e. which distinguish between the effective potentials in
(\ref{WignerSIGoldstonePotentialsE-AHM})), we turn to the off-shell  T-Matrix
and strict obedience to the wisdom of K. Symanzik's edict  at the top of Subsection \ref{TMatrixAHM}:
 {\bf ``... unless otherwise constrained by Ward identies"}. We can further constrain the allowed terms in the $\phi$-sector effective extended-AHM Lagrangian with those $U(1)_Y$ Ward-Takahashi identities  which govern 1-$\phi$-R T-Matrix elements.

In Appendix \ref{DerivationWTIE-AHM}, we derive 3 such identities governing 1-$\phi$-R connected amputated T-Matrix elements 
$T_{N,M}^{Extended}$ in the $\phi$-sector of the extended-AHM.

\begin{itemize}

\item  Adler self-consistency conditions  (originally written for the {\em global} $SU(2)_L\times SU(2)_R$ \GMLfull model with PCAC\cite{Adler1965,AdlerDashen1968}) constrain  the extended-AHM {\em gauge theory's} effective $\phi$-sector Lagrangian in 
Lorenz gauge (\ref{ExtendedAdlerSelfConsistency})
\begin{eqnarray}
\label{AdlerSelfConsistencyE-AHM} 
&&\HVEV T^{Extended}_{N,M+1}(p_1...p_N;0q_1...q_M)\nonumber \\
&& \quad \quad \times (2\pi)^4\delta^4 \Big(\sum_{n=1}^N p_n +\sum_{m=1}^M q_m \Big) \Big\vert^{p_1^2 =p_2^2...=p_N^2=m_{BEH}^2}_{q_1^2 =q_2^2...=q_M^2=0}  \nonumber \\
&& \quad \quad =0 
\end{eqnarray}
The extended-AHM T-matrix vanishes as one of the pion momenta goes to zero (i.e. 1-soft-pion theorems), provided all other physical scalar particles are on mass-shell. 
Eqn. (\ref{AdlerSelfConsistencyE-AHM}) also ``asserts the absence of infrared (IR) divergences in the ($\phi$-sector extended-AHM) Goldstone mode 
(in Lorenz gauge). Although individual Feynman diagrams are IR divergent, those IR divergent parts cancel exactly in each order of perturbation theory. Furthermore, the Goldstone mode amplitude must vanish in the 1-soft-pion limit \cite{Lee1970}".

\item The  $N=0,M=1$ case of (\ref{AdlerSelfConsistencyE-AHM}) is the Goldstone theorem (\ref{TMatrixGoldstoneTheoremExtended}) itself \cite{Lee1970,LSS-3Short}:  
\begin{eqnarray}
\label{TMatrixGoldstoneTheoremE-AHM}
\HVEV T^{Extended}_{0,2}(;00) =0 \nonumber \\
\HVEV \Gamma^{Extended}_{0,2}(;00) =0
\end{eqnarray}

\item Define $T_{N,M+1}^{Extended;External}$ as the 1-$ \phi$-R $\phi$-sector T-Matrix  {\em with one soft $\pi (q_\mu =0)$ attached to an external-leg} 
as in  Figure \ref{fig:LeeFig10} .
\begin{figure}
\centering
\includegraphics[width=1\hsize,trim={0cm 5cm 0cm 5.5cm},clip]{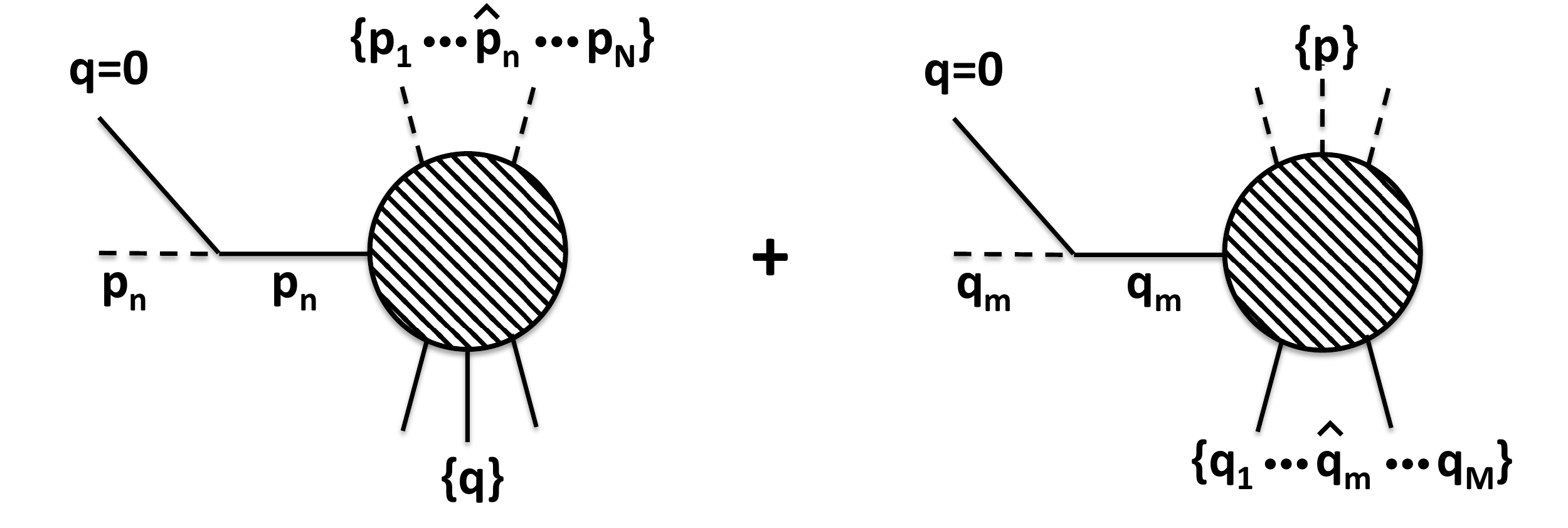}
\caption{
\label{fig:LeeFig10} 
$T_{N,M+1}^{Extended;External}$:
Hashed circles are 1-$\phi$-R $T^{Extended}_{N,M}$, solid lines $\pi$, dashed lines $h$.
One (zero-momentum) soft pion is attached to an external leg in all possible ways.
$T^{Extended}_{N,M}$ is 1-$A^\mu$-R by cutting an $A^\mu$ line,
and also 1-$\Phi$-R by cutting a $\Phi$ line.
Fig. \ref{fig:LeeFig10}  is the extended-AHM analogy of B.W. Lee's  Figure 10 \cite{Lee1970}.
The same graph topologies, but without internal Beyond-AHM $\Phi,\psi$ heavy matter, are used in the proof of (\ref{InternalTMatrix}) for the (unextended) AHM.
}
\end{figure} 
Now separate
\begin{eqnarray}
\label{DefineInternalTMatrixE-AHM}
&&T^{Extended}_{N,M+1}(p_1...p_N;0q_1...q_M) \nonumber \\
&&\quad \quad =T_{N,M+1}^{Extended;External}(p_1...p_N;0q_1...q_M) \nonumber \\
&&\quad \quad +T_{N,M+1}^{Extended;Internal}(p_1...p_N;0q_1...q_M)  
\end{eqnarray}
Appendix \ref{DerivationWTIE-AHM} (\ref{InternalTMatrixExtended})  proves that 
\begin{eqnarray}
\label{InternalTMatrixE-AHM}
&&\HVEV T_{N,M+1}^{Extended;Internal}(p_1...p_N;0q_1...q_M) \nonumber \\
&&\quad \quad =\sum_{m=1}^M T^{Extended}_{N+1,M-1}(q_mp_1...p_N;q_1....{\widehat{q_m}}...q_M)  \nonumber \\
&&\quad \quad -\sum_{n=1}^N T^{Extended}_{N-1,M+1}(p_1...{\widehat{p_n}}...p_N;p_nq_1...q_M) 
\end{eqnarray}

\end{itemize}

The $U(1)_Y$ WTIs  (\ref{GreensWTIPrimeExtended},\ref{ExtendedGreensFWTI}) governing  1-$\phi$-I connected amputated  Greens functions 
$\Gamma^{Extended}_{N,M}$ are solutions to (\ref{InternalTMatrixE-AHM},\ref{InternalTMatrixExtended}).

We re-write the Extended-AHM effective $\phi$-sector Lagrangian (\ref{LEffectiveE-AHM}) but now include the constraint from the Goldstone theorem (\ref{TMatrixGoldstoneTheoremE-AHM},\ref{TMatrixGoldstoneTheoremExtended}):
\begin{eqnarray}
\label{LEffectiveGoldstoneTheoremE-AHM}
L^{Eff;Goldstone}_{E-AHM;\phi} &=& L^{Kinetic}_{E-AHM;\phi}+{\cal O}_{Ignore}^{E-AHM} \nonumber \\
&-&V^{Eff;Goldstone}_{E-AHM;\phi}  \nonumber \\
V^{Eff;Goldstone}_{E-AHM;\phi} &=&\lambda_{\phi}^2  \left[ {\frac{h^2+{\vec \pi}^2}{2}} +\HVEV h\right] ^2 \quad \quad
\end{eqnarray} 
and wavefunction renormalization
\begin{eqnarray}
\label{GoldstoneWavefunctionE-AHM}
&&\Gamma^{Extended}_{0,2}(;q,-q)-\Gamma^{Extended}_{0,2}(;00) \nonumber \\
&&\quad \quad = q^2 +{\cal O}_{Ignore}^{E-AHM}
\end{eqnarray}

A crucial effect of the Goldstone theorem,
together with the  $N=0, M=1$ Ward-Takahashi Greens function identity (\ref{GreensWTIPrimeExtended}), is to automatically eliminate tadpoles in (\ref{LEffectiveGoldstoneTheoremE-AHM})
\begin{eqnarray}
\label{ZeroTadpolesE-AHM}
\Gamma^{Extended}_{1,0}(0;) &=& \HVEV \Gamma^{Extended}_{0,2}(;00) =0
\end{eqnarray}
so that separate tadpole renormalization is un-necessary.

We form the effective Goldstone mode $\phi$-sector Lagrangian in {\bf coordinate space}
\footnote{
It is not lost on the authors that, since we derived it from {\bf connected} amputated Greens functions (where all vacuum energy and disconnected vacuum bubbles are absorbed into an overall phase, which cancels exactly in the S-matrix \cite{Bjorken1965,ItzyksonZuber}), the vacuum energy in  $V^{Eff;Goldstone}_{E-AHM;\phi}$ in (\ref{GoldstoneLagrangianE-AHM}) is exactly zero.
} 
\begin{eqnarray}
\label{GoldstoneLagrangianE-AHM}
L_{E-AHM;\phi}^{Eff;Goldstone} &=& \vert \partial_{\mu}\phi \vert ^2 -V_{E-AHM;\phi}^{Eff;Goldstone}  \nonumber \\
&+&{\cal O}^{Ignore}_{E-AHM;\phi} \nonumber \\
V_{E-AHM;\phi}^{Eff;Goldstone} &=& \lambda_{\phi}^2  \Big[ {\frac{h^2+{\pi}^2}{2}} +\HVEV h\Big] ^2
\end{eqnarray} 

Eqn.  (\ref{GoldstoneLagrangianE-AHM}) is the $\phi$-sector effective  Lagrangian of the  {\em spontaneously broken} 
extended-AHM in 
Lorenz gauge:

\begin{itemize}

\item It obeys the Goldstone theorem (\ref{TMatrixGoldstoneTheoremE-AHM},\ref{TMatrixGoldstoneTheoremExtended}) and all other $U(1)_Y$ WTI (\ref{GreensWTIPrimeExtended},\ref{AdlerSelfConsistencyE-AHM},\ref{TMatrixGoldstoneTheoremE-AHM},\ref{InternalTMatrixE-AHM},\ref{ExtendedAdlerSelfConsistency},\ref{TMatrixGoldstoneTheoremExtended},\ref{InternalTMatrixExtended},\ref{ExtendedGreensFWTI}); 
\item It is minimized at $(H=\HVEV, {\pi}=0)$, and obeys stationarity  \cite{ItzyksonZuber} of that true minimum; 
\item It preserves the theory's renormalizability and unitarity, which require that wavefunction  renormalization, $\HVEV_{Bare}
=\Big[ Z^{\phi}_{E-AHM}\Big]^{1/2}\HVEV$ \cite{LSS-2,Bjorken1965,ItzyksonZuber}, forbid {\em any} relevant operator corrections to $\HVEV$;

\item It includes all divergent
${\cal O}(\Lambda^2),{\cal O}(\ln \Lambda^2)$ and finite terms which arise, 
to all perturbative loop-orders in the full 
$U(1)_Y$ theory, due to virtual transverse gauge bosons, AHM scalars, ghosts, new scalars, and new fermions 
$A^\mu;h,\pi;{\bar \omega},\omega;\Phi;\psi;$ respectively. 

\item  {\bf The Goldstone theorem
(\ref{TMatrixGoldstoneTheoremE-AHM},\ref{TMatrixGoldstoneTheoremExtended})
has caused all relevant operators in (\ref{GoldstoneLagrangianE-AHM}) to vanish!}

\end{itemize}

{\bf iii) Decoupling of heavy matter representations:} We take 
all of the new scalars $\Phi$ and fermions $\psi$ to be very heavy.
For Beyond-AHM scalar(s)
\begin{eqnarray}
\label{GlobalInvariantBeyondAHMPhi}
&&L_{BeyondAHM;\Phi}^{GlobalInvariant} = \Big\vert  \partial_\mu \Phi \Big\vert ^2 -V_{\Phi} -V_{\phi \Phi} \nonumber \\
&&V_{\Phi} = M_\Phi^2\Big( \Phi^\dagger \Phi \Big) + \lambda_\Phi^2\Big( \Phi^\dagger \Phi \Big) ^2  \nonumber \\
&&V_{\phi \Phi} = \lambda_{\phi \Phi}^2 \Big( \phi^\dagger \phi \Big)  \Big( \Phi^\dagger \Phi \Big)
\end{eqnarray}
we take
\begin{eqnarray}
\label{HeavyScaleScalar}
M_\Phi^2&\sim& M_{Heavy}^2 \nonumber \\
&\gg& \Big( \big\vert q^2 \big\vert,m_A^2, m_{BEH}^2 \Big) \sim m_{Weak}^2 \sim (100GeV)^2 \quad \quad
\end{eqnarray}
with $q_\mu$ typical for a studied low-energy process. For pedagogical simplicity, we have chosen a single $\Phi$ representation with $U(1)_Y$ hypercharge $Y_\Phi =Y_\phi = -1$, but the analysis is easily extended \cite{LSS-3Short} to other and multiple $U(1)_Y$ Beyond-AHM representations.

For Beyond-AHM fermion(s)
\begin{eqnarray}
\label{GlobalInvariantBeyondAHMpsi}
&&L_{BeyondAHM;\psi}^{GlobalInvariant} =  i{\bar \psi}_L \gamma^\mu \partial_\mu \psi_L +  i{\bar \psi}_R \gamma^\mu \partial_\mu \psi_R  \nonumber  \\
&& \quad \quad +L_{BeyondAHM;\psi}^{Yukawa}+L_{BeyondAHM;\psi}^{Majorana} \\
&&L_{BeyondAHM;\psi}^{Yukawa}=
	-\Big( y_{\phi \psi}{\bar \psi}_L \phi \psi_R 
		+ y_{\phi \psi}^\star {\bar \psi}_R \phi^\dagger \psi_L 
		\Big)  \nonumber \\
&& \quad \quad - \Big(  
	y_{\Phi \psi}{\bar \psi}_L \Phi \psi_R + 
	y_{\Phi \psi}^\star {\bar \psi}_R \Phi^\dagger \psi_L \Big) \nonumber \\
&&L^{Majorana}_{BeyondAHM;\psi} = -\half M_{\psi_L} \Big( {\psi_L^{Weyl} }{\psi_L^{Weyl}}+{\bar \psi}_L^{Weyl} {\bar \psi}_L^{Weyl}\Big) \nonumber \\ 
&& \quad \quad  -\half M_{\psi_R} \Big( {\psi_R^{Weyl} }{\psi_R^{Weyl}}+{\bar \psi}_R^{Weyl} {\bar \psi}_R^{Weyl}\Big) \nonumber
\end{eqnarray}
with fermion $U(1)_Y$ hypercharges chosen so that the {\em axial anomaly is zero}.
To remain perturbative, we keep the Yukawa couplings  
$y_{\phi \psi},y_{\Phi \psi} {\buildrel  {<} \over {\sim}} 1$, 
but take the Majorana masses-squared
\begin{eqnarray}
\label{HeavyScaleNu}
M_{\psi_L}^2,M_{\psi_R}^2&\sim& M_{Heavy}^2 \nonumber \\
&\gg& \Big( \big\vert k^2 \big\vert,m_A^2, m_{BEH}^2 \Big)  \sim m_{Weak}^2
\end{eqnarray}
We keep all Yukawas and masses real for pedagogical simplicity.

Some comments are in order:
\begin{itemize}
\item We have ignored finite  ${\cal O}_{E-AHM;\phi}^{1/M_{Heavy}^2;Irrelevant}$ 
which decouple and vanish as $m_{Weak}^2/ M_{Heavy}^2 \to 0$.

\item Among the terms included in (\ref{GoldstoneLagrangianE-AHM}) are {\em finite} relevant operators dependent on the heavy matter representations:
\begin{eqnarray}
\label{HeavyRelevantOperators}
&&M_{Heavy}^2, M_{Heavy}^2 \ln{ \big(M_{Heavy}^2\big)}, \nonumber \\
&& \quad \quad M_{Heavy}^2 \ln{ \big(m_{Weak}^2\big)}, m_{Weak}^2 \ln{ \big(M_{Heavy}^2\big)}  \quad 
\end{eqnarray}
but they have vanished because of the Goldstone theorem (\ref{TMatrixGoldstoneTheoremE-AHM},\ref{TMatrixGoldstoneTheoremExtended})! That fact is one of the central results of this paper.

\item Marginal operators $\sim \ln{ \big(M_{Heavy}^2\big)}$ have been absorbed in  (\ref{GoldstoneLagrangianE-AHM}): i.e. in the renormalization of gauge-independent observables (i.e. the quartic-coupling constant
$\lambda_\phi^2$ and the BEH VEV $\HVEV$), and in un-observable wavefunction renormalization (\ref{GoldstoneWavefunctionE-AHM}).
\end{itemize}

No trace of $M_{Heavy}$-scale $\Phi,\psi$ survives in (\ref{GoldstoneLagrangianE-AHM})! All the heavy Beyond-AHM matter representations have completely decoupled.

{\bf iv) 1st decoupling theorem: SSB $\phi$-sector connected amputated 1-$\phi$-R T-Matrices.}
\footnote{
We take ${\cal O}^{1/\Lambda^2;Irrelevant}_{E-AHM;\phi} \to0$ so to un-encumber our notation. 
}
\begin{eqnarray}
\label{SSBTMatrixDecouplingTheorem}
T_{N,M}^{Extended} \quad
 {\buildrel  {{m_{Weak}^2}/{M_{Heavy}^2} \to 0} \over {=\joinrel=\joinrel=\joinrel=\joinrel=\joinrel=\joinrel=\joinrel=\joinrel=\joinrel=\joinrel\Longrightarrow}} 
\quad T_{N,M} 
\end{eqnarray}
become equal in the limit ${{m_{Weak}^2}/{M_{Heavy}^2} \to 0}$.

{\bf v) 2nd decoupling theorem: SSB $\phi$-sector connected amputated 1-$\phi$-I Green's functions.}
\begin{eqnarray}
\label{SSBGreensFunctionDecouplingTheorem}
\Gamma_{N,M}^{Extended} \quad
 {\buildrel  {{m_{Weak}^2}/{M_{Heavy}^2} \to 0} \over {=\joinrel=\joinrel=\joinrel=\joinrel=\joinrel=\joinrel=\joinrel=\joinrel=\joinrel=\joinrel\Longrightarrow}} 
\quad \Gamma_{N,M}
\end{eqnarray}
become equal in the limit ${{m_{Weak}^2}/{M_{Heavy}^2} \to 0}$.

{\bf vi) 3rd decoupling theorem: SSB $\phi$-sector BEH pole-mass-squared.}
The  $N=1,M=1$ connected amputated Green's function $U(1)_Y$ WTI (\ref{GreensWTIPrimeExtended}), augmented by the Goldstone theorem (\ref{TMatrixGoldstoneTheoremE-AHM}) reads
\begin{eqnarray}
\label{BEHMassE-AHM}
\Gamma^{Extended}_{2,0}(00;)&=&\HVEV\Gamma^{Extended}_{1,2}(0;00) \nonumber \\
&=&-2\lambda^2_\phi \HVEV^2 \nonumber \\
\lim_{\HVEV \to 0}\Gamma^{Extended}_{2,0}(00;)&=&0 
\end{eqnarray} 
shows that {\bf the BEH pole-mass-squared arises entirely from SSB.} Define
\begin{eqnarray}
\label{E-AHMBEHPropagator}
\Delta^{BEH}_{E-AHM}(q^2) &=& 
\frac{1}{q^2-m^2_{BEH;Pole} + i\epsilon} \nonumber \\
&+& \int dm^2 \frac{\rho^{BEH}_{E-AHM}(m^2)}{q^2-m^2 + i\epsilon} \qquad \qquad
\end{eqnarray}

$m^2_{BEH;Pole}$ is the gauge-independent observable BEH resonance pole-mass-squared. 
We now show that it is not-FT.
Since, in analogy with
 (\ref{SpectralDensityPropagators}), the spectral density $\rho^{BEH}_{E-AHM}(M_{Heavy}^2) \sim 1/M_{Heavy}^2$  is not FT
\begin{eqnarray}
\label{E-AHMBEHPropagatorAtZero}
&& \rho^{BEH}_{E-AHM}(m^2) = \rho^{BEH}_{AHM}(m^2)  + {\cal O}^{1/M_{Heavy}^2;Irrelevant}_{E-AHM;\phi} \nonumber \\
&&\Gamma^{Extended}_{2,0}(00;)\equiv\Big[ \Delta^{BEH}_{E-AHM}(0) \Big]^{-1}\nonumber \\ 
&&\quad \quad =-2\lambda_\phi^2 \HVEV^2 \nonumber \\
&&\quad \quad =-m^2_{BEH;Pole} \Big[ 1+  m^2_{BEH;Pole} \int dm^2 \frac{\rho^{BEH}_{AHM}(m^2)}{m^2 - i\epsilon} \Big]^{-1} \nonumber \\
&&\quad \quad + {\cal O}^{1/M_{Heavy}^2;Irrelevant}_{E-AHM;\phi} \qquad \qquad
\end{eqnarray}
and we have
\begin{eqnarray}
\label{GaugeIndependentBEHPoleMass}
&&m^2_{BEH;Pole} \nonumber = 2\lambda_\phi^2 \HVEV^2\Big[ 1- 2\lambda_\phi^2 \HVEV^2 \int dm^2 \frac{\rho^{BEH}_{AHM}(m^2)}{m^2 - i\epsilon} \Big]^{-1} \nonumber \\
&&\qquad + {\cal O}^{1/M_{Heavy}^2;Irrelevant}_{E-AHM;\phi} 
\end{eqnarray}

Because $\lambda^2_\phi , Z^{\phi}_{ExrendedAHM}$ are dimensionless, $\lambda^2_\phi$ and
\begin{eqnarray}
\label{BEHVEVE-AHM}
\HVEV=\Big[Z^{\phi}_{ExrendedAHM}\Big]^{-\half}\HVEV_{Bare}
\end{eqnarray}
absorb no relevant operators 
and are therefore not FT,  Eqn.  (\ref{GaugeIndependentBEHPoleMass}) shows that  the  BEH pole-mass-squared $m_{BEH;Pole}^2$ also absorbes no relevant 
operators, and is also not fine-tuned. 

No trace of $M_{Heavy}$-scale $\Phi,\psi$ survives in (\ref{GaugeIndependentBEHPoleMass})! All the heavy Beyond-AHM matter representations have completely decoupled, and the BEH-pole masses-squared
\begin{eqnarray}
\label{SSBBEHMassDecouplingTheorem}
m^{2;E-AHM}_{BEH;Pole}  \quad
 {\buildrel  {{m_{Weak}^2}/{M_{Heavy}^2} \to 0} \over {=\joinrel=\joinrel=\joinrel=\joinrel=\joinrel=\joinrel=\joinrel=\joinrel=\joinrel=\joinrel\Longrightarrow}} 
\quad m^{2;AHM}_{BEH;Pole} \quad \quad
\end{eqnarray}
become equal in the limit ${{m_{Weak}^2}/{M_{Heavy}^2} \to 0}$. 
We call (\ref{SSBBEHMassDecouplingTheorem}) the {\bf ``SSB BEH-Mass Decoupling Theorem"}.

{\bf vii)}
By dimensional analysis, heavy $\Phi,\psi$ decouple from the $\pi$ spectral functions
\begin{eqnarray}
\label{E-AHMBEHPropagatorAtZeroB}
&& \Delta^{\pi;Spectral}_{E-AHM}(q^2) = \Delta^{\pi;Spectral}_{AHM}(q^2) + {\cal O}\Big(1/M_{Heavy}^2\Big) \quad \quad
\end{eqnarray}

The  SSB extended-AHM $\phi$-sector  is therefore Goldstone Exceptionally Natural, with far more powerful suppression of fine-tuning than G. 't Hooft's no-FT criteria would demand.

\subsection{Decoupling of gauge singlet $M^2_S \gg m_{Weak}^2$ real scalar field $S$ with discrete $Z_2$ symmetry and $\SVEV=0$}
\label{HeavyScalar}

For the heavy scalar we consider  a $U(1)_Y$ gauge singlet  real scalar $S$, with  ($S\to-S$) $Z_2$ symmetry, 
$M_S^2\gg m_{Weak}^2$, and $\SVEV =0$. 
We add to the renormalized theory
\begin{eqnarray}
\label{SingletScalarLagrangian}
&&L_S=\half(\partial_{\mu}S)^2 -V_{\phi S} \nonumber \\
&&V_{\phi S} = \half M_S^2 S^2 + \frac {\lambda_S^2}{4} S^4 + \half \lambda_{\phi S}^2 S^2 \left[ \phi^\dagger\phi -\half \HVEV^2 \right] \nonumber \\
&&\phi^\dagger\phi -\half \HVEV^2=\frac{h^2 +\pi^2}{2}+\HVEV h
\end{eqnarray}
Since $S$ is a gauge singlet, it is also a rigid/global singlet. Its $U(1)_Y$ hypercharge,  transformation and current 
\begin{eqnarray}
\label{MajoranaCurrent}
Y_S=0;\quad \delta S(t,{\vec y})&=&0 \nonumber \\
{J}^{\mu;S}_{BeyondAHM}&=&0
\end{eqnarray}
therefore satisfy all of the de-coupling criteria in Appendix \ref{DerivationWTIE-AHM}:

\begin{itemize}

\item Since it is massive, $S$ cannot carry information to the surface $z^{3-surface}\to \infty$ of the (all-space-time) 4-volume $\int d^4 z$, and so satisfies (\ref{ExtendedSurfaceIntegral});

\item The equal-time commutators satisfy (\ref{EqTimeCommE-AHM})
\begin{eqnarray}
\label{ExtendedEqTimeCommAHMA}
 \delta(z_0-y_0)\left[ {J}^{0;S}_{BeyondAHM}(z),H(y)\right] &=&0 \nonumber \\
 \delta(z_0-y_0)\left[ {J}^{0;S}_{BeyondAHM}(z),\pi(y)\right] &=&0
\end{eqnarray}

\item The classical equation of motion  
\begin{eqnarray}
\label{DivergenceCurrentAHMSA}
&&\partial_{\mu} \Big( { J}^{\mu;S}_{BeyondAHM} + { J}^{\mu}_{AHM} \Big)  \\
&&\qquad \qquad =\partial_{\mu}   { J}^{\mu}_{AHM} = -  e\HVEV H \partial_{\beta}{A}^{\beta}   \nonumber
\end{eqnarray}
restores conservation of the rigid/global $U(1)_Y$ extended current for $\phi$-sector physical states, and satisfies (\ref{ExtendedQuantumCurrentConservation})
\begin{eqnarray}
\label{ExtendedQuantumCurrentConservationSA}
&&\Big< 0\vert T\Big[  \partial_{\mu} \Big( { J}^{\mu;S}_{BeyondAHM} + { J}^{\mu}_{AHM} \Big)(z)  \nonumber \\
&&\quad \quad \times h(x_1)...h(x_N) \pi(y_1)...\pi(y_M)\Big]\vert 0\Big>_{Connected} \nonumber \\
&&\quad \quad=0
\end{eqnarray}

\item The zero VEV $\big< S \big>=0$ satisfies (\ref{E-AHMVEV});

\end{itemize}

The $U(1)_Y$ WTI  governing  the extended $\phi$-sector transition matrix $T_{N,M}^{Extended;S}$ are therefore true, namely: the extended Adler self-consistency conditions (\ref{AdlerSelfConsistencyE-AHM},\ref{ExtendedAdlerSelfConsistency}), together with their proof of infra-red finiteness in the presence of massless NGB;
the extended Goldstone theorem (\ref{TMatrixGoldstoneTheoremE-AHM},\ref{TMatrixGoldstoneTheoremExtended}); 
the extended 1-soft-$\pi$ theorems (\ref{InternalTMatrixE-AHM},\ref{InternalTMatrixExtended});
The extended $U(1)_Y$ WTI  (\ref{GreensWTIPrimeExtended},\ref{ExtendedGreensFWTI}) governing  connected amputated $\phi$-sector Green's functions $\Gamma_{N,M}^{Extended;S}$ are also true.

The 3 decoupling theorems 
(\ref{SSBTMatrixDecouplingTheorem},\ref{SSBGreensFunctionDecouplingTheorem},\ref{SSBBEHMassDecouplingTheorem})
therefore follow, so that no trace of the $M_S^2 \sim M^2_{Heavy}$ scalar $S$ survives the 
${m_{Weak}^2}/{M_{Heavy}^2} \to 0$ limit: 
i.e. it has completely decoupled! 
The $\phi$-sector connected amputated T-Matrices and Green's functions, and the BEH pole masses-squared 
\begin{eqnarray}
\label{PhiSectorDecouplingTheorem}
T_{N,M}^{Extended;S} \quad
& {\buildrel  {{m_{Weak}^2}/{M_S^2} \to 0} \over {=\joinrel=\joinrel=\joinrel=\joinrel=\joinrel=\joinrel=\joinrel=\joinrel=\joinrel=\joinrel
\Longrightarrow}} &
\quad T_{N,M} \\
\Gamma_{N,M}^{Extended;S} \quad
& {\buildrel  {{m_{Weak}^2}/{M_S^2} \to 0} \over {=\joinrel=\joinrel=\joinrel=\joinrel=\joinrel=\joinrel=\joinrel=\joinrel=\joinrel=\joinrel \Longrightarrow}} &
\quad \Gamma_{N,M} \nonumber  \\
m^{2;E-AHM;S}_{BEH;Pole;\phi}  \quad 
&{\buildrel  {{m_{Weak}^2}/{M_{Heavy}^2} \to 0} \over {=\joinrel=\joinrel=\joinrel=\joinrel=\joinrel=\joinrel=\joinrel=\joinrel=\joinrel=\joinrel
\Longrightarrow}} &
\quad m^{2;AHM}_{BEH;Pole;\phi} \nonumber
\end{eqnarray}
become equal in the limit ${{m_{Weak}^2}/{M_{Heavy}^2} \to 0}$. 
\footnote{$m^{2;AHM}_{BEH;Pole}$ decoupling is in exact disagreement with \cite{deGouvea:2014xba,Espinosa,Farina:2013mla}.
}

\subsection{The lightest generation of Standard Model quarks and leptons, augmented by a right-handed neutrino $\nu_R$ with 
Dirac mass $m_\nu^{Dirac}$: Gauged hypercharge and global colors}
\label{SMQuarksLeptons}

These 16 spin $S=\half$ fermions: 
$u^{c}_L, d^{c}_L$; $u^{c}_R$; $d^{c}_R$; $e_L,\nu_L$; $e_R$; $\nu_R$;
with global $SU(3)$ colors $c=$red, white, blue, and gauged $U(1)_Y$ hypercharge,
are regarded here as extended-AHM matter representations.
Baryon and lepton-number conserving {\em  Dirac} masses-squared arise entirely from SSB and are very light: $m_{Quark}^2,m_{Lepton}^2  \ll
 m_{Weak}^2$.  
The so-extended $U(1)_Y$ AHM gauge theory has zero axial-anomaly because quark/lepton  AHM quantum numbers are chosen to be their SM hypercharges
(including $Y_{\nu_R}=0$).

{\bf i) Beyond-AHM Dirac quarks:}
\begin{eqnarray}
\label{BeyondAHMQuark}
&&L_{BeyondAHM;q}^{GlobalInvariant} =  L_{BeyondAHM;q}^{Kinetic} +L_{BeyondAHM;q}^{Yukawa}\quad \quad \\
&&L_{BeyondAHM;q}^{Kinetic}= i \sum_{color}^{r,w,b} \sum^{u,d}_{flavor}  \Big( {\bar q}_L^c \gamma^\mu \partial_\mu q_L^c 
+  {\bar q}_R^c \gamma^\mu \partial_\mu q_R^c \Big) \nonumber \\
&&L_{BeyondAHM;q}^{Yukawa}= - \sum_{color}^{r,w,b} \sum^{u,d}_{flavor} y_q  \Big( {\bar q}_L^c \phi q_R^c 
+  {\bar q}_R^c \phi^\dagger q_L^c \Big) \nonumber 
\end{eqnarray}

The $U(1)_Y$ quark current  and transformation properties are
\begin{eqnarray}
\label{BeyondAHMQuarkCurrent}
J_{BeyondAHM;q}^{\mu;Dirac}&=& -\sum_{color}^{r,w,b} \sum^{u,d}_{flavor} \nonumber \\
&\times&  \Big( Y_{q_L}{\bar q}_L^c \gamma^\mu q_L^c 
+  Y_{q_R}{\bar q}_R^c \gamma^\mu  q_R^c \Big) \quad \nonumber \\
\delta q^c_L(t,{\vec x}) &=& -iY_{q_L} q^c_L(t,{\vec x})\theta \nonumber \\
\delta q^c_R(t,{\vec x}) &=& -iY_{q_R} q^c_R(t,{\vec x})\theta \nonumber \\
Y_{u_L}=\frac{1}{3};Y_{d_L}&=&\frac{1}{3};Y_{u_R}=\frac{4}{3};Y_{d_R}=-\frac{2}{3};
\end{eqnarray}

{\bf ii) Beyond-AHM Dirac leptons:}
\begin{eqnarray}
\label{BeyondAHMpsi}
&&L_{BeyondAHM;l}^{GlobalInvariant} =  L_{BeyondAHM;l}^{Kinetic} +L_{BeyondAHM;l}^{Yukawa}\quad \quad \\
&&L_{BeyondAHM;l}^{Kinetic}= i \sum^{\nu,e}_{flavor}  \Big( {\bar l}_L \gamma^\mu \partial_\mu l_L 
+  {\bar l}_R \gamma^\mu \partial_\mu l_R \Big) \nonumber \\
&&L_{BeyondAHM;l}^{Yukawa}= - \sum^{\nu,e}_{flavor} y_l  \Big( {\bar l}_L \phi l_R 
+  {\bar l}_R \phi^\dagger l_L \Big) \nonumber 
\end{eqnarray}

The lepton $U(1)_Y$  current and transformation properties are
\begin{eqnarray}
\label{BeyondAHMLeptonCurrent}
J_{BeyondAHM;l}^{\mu;Dirac}&=&  - \sum^{\nu,e}_{flavor}  \Big( Y_{l_L}{\bar l}_L \gamma^\mu l_L
+  Y_{l_R}{\bar l}_R \gamma^\mu  l_R \Big) \quad \nonumber \\
\delta l_L(t,{\vec x}) &=& -iY_{l_L} l_L(t,{\vec x})\theta \nonumber \\
\delta l_R(t,{\vec x}) &=& -iY_{l_R} l_R(t,{\vec x})\theta \nonumber \\
Y_{\nu_L}=-1;Y_{e_L}&=&-1;Y_{\nu_R}=0;Y_{e_R}=-2; 
\end{eqnarray}

With these Standard Model quark and lepton hypercharges $Y_i$, our $U(1)_Y$ WTI have zero axial anomaly. 

We now prove applicability of  our $U(1)_Y$ WTI: i.e. for connected amputated $\phi$-sector Greens functions 
$\Gamma_{N,M}^{Extended}$ and T-Matrix elements $T_{N,M}^{Extended}$.

\begin{itemize}

\item The equal-time quantum commutators satisfy (\ref{EqTimeCommE-AHM})
\begin{eqnarray}
\label{ExtendedEqTimeCommAHMB}
 \delta(z_0-y_0)\left[ J_{BeyondAHM;q}^{0;Dirac}(z),H(y)\right] &=&0 \nonumber \\
 \delta(z_0-y_0)\left[ J_{BeyondAHM;q}^{0;Dirac},\pi(y)\right] &=&0 \nonumber \\
\delta(z_0-y_0)\left[ J_{BeyondAHM;l}^{0;Dirac}(z),H(y)\right] &=&0 \nonumber \\
 \delta(z_0-y_0)\left[ J_{BeyondAHM;l}^{0;Dirac},\pi(y)\right] &=&0
\end{eqnarray}

\item The classical equation of motion  
\begin{eqnarray}
\label{DivergenceCurrentAHMSB}
&&\partial_{\mu} \Big( J_{BeyondAHM;l}^{\mu;Dirac}+ J_{BeyondAHM;q}^{\mu;Dirac} + { J}^{\mu}_{AHM} \Big) \nonumber \\
&&\quad \quad = -  e\HVEV H \partial_{\beta}{A}^{\beta}   
\end{eqnarray}
restores conservation of the rigid/global $U(1)_Y$ extended current for $\phi$-sector physical states, and satisfies (\ref{ExtendedQuantumCurrentConservation})
\begin{eqnarray}
\label{ExtendedQuantumCurrentConservationSB}
&&\Big< 0\vert T\Big[  \partial_{\mu} \Big( J_{BeyondAHM;l}^{\mu;Dirac}+ J_{BeyondAHM;q}^{\mu;Dirac} + { J}^{\mu}_{AHM}\Big)(z)  \nonumber \\
&&\quad \quad \times h(x_1)...h(x_N) \pi_{t_1}(y_1)...\pi_{t_M}(y_M)\Big]\vert 0\Big>_{Connected} \nonumber \\
&& \quad \quad  =0 
\end{eqnarray}

\item  {\bf Dirac quark surface terms:} Since they are here taken massive $m_u=\frac{1}{\sqrt 2}y_u\HVEV$ and $m_d=\frac{1}{\sqrt 2}y_d\HVEV$ (i.e. in deference to the SM with its massive strongly-interacting hadronic pion), and we need only {\em connected} graphs, the light quarks $u,d$ cannot carry information to the surface $z^{3-surface}\to \infty$ of the (all-space-time) 4-volume $\int d^4 z$, and so satisfy (\ref{ExtendedSurfaceIntegral}). 
In contrast, massless quarks {\em could} carry (on the light-cone) $U(1)_Y$ information  to the $z^{3-surface}\to \infty$, and would violate  (\ref{ExtendedSurfaceIntegral}),
and so destroy  the spirit, results and essence of our $U(1)_Y$-WTI-based no-FT and heavy particle decoupling results here in Section \ref{E-AHM}.
Still, from the harm they do our global $U(1)_Y$ WTIs, we worry that the infra-red structure of massless quarks might also harm $U(1)_Y$ {\em local} Slavnov-Taylor identities.
\footnote{
It is amusing to elevate $U(1)_Y$ WTIs to a ``Principle of Nature", so as to give them predictive power. Imagine we impose, on quark-extended AHM, an extra gauge/local $SU(3)_{Color}$, an extra  rigid/global $SU(2)_L$ on left-handed $(u_L,d_L)$ quarks, and an extra rigid/global $SU(2)_R$ on right-handed $(u_R,d_R)$ quarks. The structure group would be
\begin{eqnarray}
\label{AHMChiralPerturbationTheory}
U(1)_{Y;Local}&\otimes& SU(3)_{Color;Local} \nonumber \\
&\otimes& SU(2)_{L;Global} \otimes SU(2)_{R;Global}
\end{eqnarray}
The result would add to the quark-extended AHM a Chiral Perturbation Theory 
($\chi PT$) \cite{Georgi2009}
with massive neutrons and protons, and 3  massless hadronic pions: there is no QED electric charge in this toy model. Now {\em impose} our 
SSB $U(1)_Y$ WTIs, which require and demand non-zero quark masses 
$m_u = m_d \neq 0$. The result is $\chi PT$, but now {\em broken} (in just the right way, after further breaking $m_u \neq m_d$) to give strong-interaction {\em hadronic pions} their masses.
Would we then claim that spontaneouly broken $U(1)_Y$ WTIs {\em predict} the breaking of hadronic $SU(2)_L\times SU(2)_R$ $\chi PT$?
}

\item  {\bf Charged Dirac lepton surface terms:} Since it is massive $m_e=\frac{1}{\sqrt 2}y_e\HVEV$, and we need only connected graphs, the electron $e$ cannot carry information to the surface $z^{3-surface}\to \infty$ of the (all-space-time) 4-volume $\int d^4 z$, and so satisfies (\ref{ExtendedSurfaceIntegral});

\item  {\bf Dirac neutrino surface terms:} Since  it is here taken massive $m^{Dirac}_{\nu}=\frac{1}{\sqrt 2}y_{\nu}\HVEV$ 
(i.e. in deference to observed SM neutrino mixing), and we need only {\em connected} graphs, the light 
neutrino $\nu_L$ cannot carry information to the surface $z^{3-surface}\to \infty$ of the (all-space-time) 
4-volume $\int d^4 z$, and so satisfies (\ref{ExtendedSurfaceIntegral}). 
In contrast, a massless neutrino {\em could} carry (on the light-cone) $U(1)_Y$ information  to the $z^{3-surface}\to \infty$, and would violate  (\ref{ExtendedSurfaceIntegral}),
and so destroy  the spirit, results and essence of our $U(1)_Y$-WTI-based no-FT and heavy particle decoupling results here in Section \ref{E-AHM}.
Still, from the harm it does our global $U(1)_Y$ WTIs, we worry that the infra-red structure of a massless neutrino might also harm $U(1)_Y$ {\em local} Slavnov-Taylor identities.
\footnote{
Imagine we are able to extend this work to the Standard Model itself \cite{LSS-4Proof,LSS-4}! With its local/gauge group 
$SU(3)_{Color} \times SU(2)_L \times U(1)_Y$, we would build 3 sets of rigid/global WTIs: unbroken $SU(3)_{Color}$; unbroken electromagnetic  $U(1)_{QED}$; and spontaneously broken $SU(2)_L$.  
It is then amusing to elevate such rigid/global  WTIs to a ``Principle of Nature", so as to give them predictive power for 
actual experiments and observations. 
The $SU(3)_{Color}$ and $U(1)_{QED}$ WTIs are {\bf unbroken vector-current IDs}, and will not yield information analogous with that of SSB extended-AHM here.
But the axial-vector current inside the SSB $SU(2)_L$ WTIs will require and demand a non-zero SSB Dirac mass for each and every one of the the weak-interaction eigenstates 
$m^{Dirac}_{\nu_e}, m^{Dirac}_{\nu_\mu}, m^{Dirac}_{\nu_\tau} \neq 0$. The observable  PNMS mixing matrix would then rotate those to mass-eigenstates  $m^{Dirac}_{\nu_1}, m^{Dirac}_{\nu_2}, m^{Dirac}_{\nu_3} \neq 0$.
Would we then claim that spontaneouly broken $SU(2)_L$ WTIs {\em predict} neutrino oscillations?
To make possible connection with Nature, 
although current experimental neutrino  mixing data cannot rule out an exactly-zero mass for the lightest neutrino \cite{KarolLang},
the mathematical self-consistency of $SU(2)_L$ WTIs  would!
}

\end{itemize}

Having satisfied all of the criteria in Appendix \ref{DerivationWTIE-AHM}, the $U(1)_Y$ WTI   governing  the extended  connected amputated $\phi$-sector $T_{N,M}^{Extended;q;l}$ are therefore true, namely: the extended Adler self-consistency conditions (\ref{AdlerSelfConsistencyE-AHM},\ref{ExtendedAdlerSelfConsistency}), together with their proof of infra-red finiteness in the presence of massless NGB;
the extended Goldstone theorem (\ref{TMatrixGoldstoneTheoremE-AHM},\ref{TMatrixGoldstoneTheoremExtended});
the extended 1-soft-$\pi$ theorems (\ref{InternalTMatrixE-AHM},\ref{InternalTMatrixExtended}); 
The extended $U(1)_Y$ WTI  (\ref{GreensWTIPrimeExtended},\ref{ExtendedGreensFWTI}) governing  connected amputated $\phi$-sector Green's functions $\Gamma_{N,M}^{Extended;q;l}$ are also true.

\subsection{$\nu$AHM: (in practice) decoupling of gauge singlet right-handed Type I See-saw Majorana \\ neutrino $\nu_R$ with $M_{\nu_R}^2\gg m_{BEH}^2$}
\label{HeavyNeutrino}

For the heavy fermion we consider a $U(1)_Y$ gauge-singlet right-handed Majorana neutrino $\nu_R$, with $M_{\nu_R}^2\gg m_{Weak}^2$, involved in a Type 1 See-Saw with a
left-handed neutrino $\nu_L$, 
Yukawa coupling $y_{\nu}$ and resulting Dirac mass $m_\nu^{Dirac}=y_{\nu}\HVEV /\sqrt{2}$. 

We add to the renormalized theory in Subsection \ref{SMQuarksLeptons} a Majorana mass
\begin{eqnarray}
\label{NeutrinoMajoranaMass}
L^{Majorana}_{\nu_R}&=& -\half M_{\nu_R} \Big( {\nu_R^{Weyl} }{\nu_R^{Weyl}}+{\bar \nu}_R^{Weyl} {\bar \nu}_R^{Weyl}\Big) \qquad 
\end{eqnarray}
Since $\nu_R$ is a gauge singlet, it is also a rigid/global singlet. Its hypercharge, $U(1)_Y$ transformation and current 
\begin{eqnarray}
\label{MajoranaCurrentB}
Y_{\nu_R}=0; \quad \delta \nu_R(t,{\vec y})&=&0 \nonumber \\
{J}^{\mu;Majorana}_{BeyondAHM;\nu_R}&=&0
\end{eqnarray}
therefore satisfy all of the de-coupling criteria in Appendix \ref{DerivationWTIE-AHM}:

\begin{itemize}

\item Since it has a Dirac mass, $\nu_L$ cannot carry information to the surface $z^{3-surface}\to \infty$ of the (all-space-time) 4-volume $\int d^4 z$, and so satisfies (\ref{ExtendedSurfaceIntegral});

\item The equal-time quantum commutators satisfy (\ref{EqTimeCommE-AHM})
\begin{eqnarray}
\label{ExtendedEqTimeCommAHMC}
 \delta(z_0-y_0)\left[ {J}^{0;Majorana}_{BeyondAHM;\nu_R}(z),H(y)\right] &=&0 \nonumber \\
 \delta(z_0-y_0)\left[ {J}^{0;Majorana}_{BeyondAHM;\nu_R}(z),\pi(y)\right] &=&0
\end{eqnarray}

\item The classical equation of motion  
\begin{eqnarray}
\label{DivergenceCurrentAHMSC}
&&\partial_{\mu} \Big( {J}^{\mu;Majorana}_{BeyondAHM;\nu_R} +J_{BeyondAHM;l}^{\mu;Dirac} \nonumber \\
&&\qquad \qquad + J_{BeyondAHM;q}^{\mu;Dirac}  + { J}^{\mu}_{AHM} \Big) \nonumber \\
&&\qquad \qquad=\partial_{\mu} \Big( J_{BeyondAHM;l}^{\mu;Dirac}+ J_{BeyondAHM;q}^{\mu;Dirac} + { J}^{\mu}_{AHM} \Big) \nonumber \\
&&\qquad \qquad= -  e\HVEV H \partial_{\beta}{A}^{\beta}   
\end{eqnarray}
restores conservation of the rigid/global $U(1)_Y$ extended current for $\phi$-sector physical states, and satisfies (\ref{ExtendedQuantumCurrentConservation})
\begin{eqnarray}
\label{ExtendedQuantumCurrentConservationSC}
&&\Big< 0\vert T\Big[  \partial_{\mu} \Big( {J}^{\mu;Majorana}_{BeyondAHM;\nu_R} +J_{BeyondAHM;l}^{\mu;Dirac}\\
&&\quad \quad +J_{BeyondAHM;q}^{\mu;Dirac}  + { J}^{\mu}_{AHM}\Big)(z)  \nonumber \\
&&\quad \quad \times h(x_1)...h(x_N) \pi_{t_1}(y_1)...\pi_{t_M}(y_M)\Big]\vert 0\Big>_{Connected} \nonumber \\
&&\quad \quad =0 \nonumber
\end{eqnarray}

\end{itemize}

Having satisfied all of the criteria in Appendix \ref{DerivationWTIE-AHM}, the $U(1)_Y$ WTI   governing  the extended  connected amputated $\phi$-sector $T_{N,M}^{Extended;q;l}$ are therefore true, namely: the extended Adler self-consistency conditions (\ref{AdlerSelfConsistencyE-AHM},\ref{ExtendedAdlerSelfConsistency}), together with their proof of infra-red finiteness in the presence of massless NGB;
the extended Goldstone theorem (\ref{TMatrixGoldstoneTheoremE-AHM},\ref{TMatrixGoldstoneTheoremExtended});
the extended 1-soft-$\pi$ theorems (\ref{InternalTMatrixE-AHM},\ref{InternalTMatrixExtended}); 
The extended $U(1)_Y$ WTI  (\ref{GreensWTIPrimeExtended},\ref{ExtendedGreensFWTI}) governing  connected amputated $\phi$-sector Green's functions $\Gamma_{N,M}^{Extended;q;l}$ are also true.

The 3 decoupling theorems   (\ref{SSBTMatrixDecouplingTheorem},\ref{SSBGreensFunctionDecouplingTheorem},\ref{SSBBEHMassDecouplingTheorem}) therefore follow, but there is a ``non-decoupling subtlety." Recall that the vanishing of the $\nu_L$ surface terms requires a non-zero neutrino Dirac mass
\begin{eqnarray}
\label{DiracNeutrinoMass}
m_\nu^{Dirac}=\frac{1}{\sqrt 2}y_\nu\HVEV \neq 0
\end{eqnarray}
The light and heavy Type I See-saw $\nu$ masses are 
\begin{eqnarray}
\label{TotalNeutrinoMass}
m_{Light} &\sim& m_\nu^{2;Dirac} / M_{\nu_R} \nonumber \\
m_{Heavy} &\sim& M_{\nu_R} 
\end{eqnarray}
But, in obedience to our proof of $U(1)_Y$ WTI, $m_{Light}$ must  not vanish. Therefore Type I See-saw $\nu$s 
do not allow the  
$M_{\nu_R}\to \infty$ limit!
For the decoupling theorems, we instead imagine huge,  but {\em finite}, $M_{\nu_R}$ with
\begin{eqnarray}
\label{NeutrinoMassLimit}
1 \gg m_\nu^{2;Dirac}/M_{\nu_R}^2 \neq 0
\end{eqnarray}

No {\em practical} trace of the $M_{\nu_R}^2 \sim M^2_{Heavy}$ 
right-handed neutrino $\nu_R$ survives 
$1 \gg {m_{Weak}^2}/{M_{Heavy}^2} \neq 0$: 
i.e. it has become practically invisible! 
The $\phi$-sector connected amputated T-Matrices and Green's functions, and the BEH pole masses-squared 
\begin{eqnarray}
\label{DecouplingTheorem}
T_{N,M}^{Extended;u,d;e,\nu_L,\nu_R} 
& {\buildrel  {1\gg {m_{Weak}^2}/{M_{\nu_R}^2} \neq 0} \over {=\joinrel=\joinrel=\joinrel=\joinrel=\joinrel=\joinrel=\joinrel=\joinrel=\joinrel=\joinrel=\joinrel=\joinrel=\joinrel=\joinrel
\Longrightarrow}} & \quad \quad  \\
& T_{N,M}^{Extended;u,d;e,\nu_L,\nu_{R;Invisible}} & \nonumber \\
\Gamma_{N,M}^{Extended;u,d;e,\nu_L,\nu_R}
& {\buildrel  {1\gg {m_{Weak}^2}/{M_{\nu_R}^2} \neq 0} \over {=\joinrel=\joinrel=\joinrel=\joinrel=\joinrel=\joinrel=\joinrel=\joinrel=\joinrel=\joinrel=\joinrel=\joinrel=\joinrel=\joinrel
\Longrightarrow}} & \nonumber \\
& \Gamma_{N,M}^{Extended;u,d;e,\nu_L,\nu_{R;Invisible}}  & \nonumber \\
m^{2;Extended;u,d;e,\nu_L,\nu_R}_{BEH;Pole;\phi}  \quad 
& {\buildrel  {1\gg {m_{Weak}^2}/{M_{\nu_R}^2} \neq 0} \over {=\joinrel=\joinrel=\joinrel=\joinrel=\joinrel=\joinrel=\joinrel=\joinrel=\joinrel=\joinrel=\joinrel=\joinrel=\joinrel=\joinrel
\Longrightarrow}} & 
\nonumber \\
& m^{2;Extended;u,d;e,\nu_L,\nu_{R;Invisible}}_{BEH;Pole;\phi} & \nonumber
\end{eqnarray}
become, to high approximation,  {\em equal in practice}, for $1 \gg {{m_{Weak}^2}/{M_{Heavy}^2} \neq 0}$.
\footnote{$m^{2;AHM}_{BEH;Pole}$ decoupling is in exact disagreement with \cite{deGouvea:2014xba,Espinosa,Farina:2013mla}.
} 

Still, our $U(1)_Y$ WTIs insist that {\em in principle} a very heavy Majorana mass $M_{\nu_R}$ cannot completely decouple, and may still have some measureable or observational effect.

\section{SSB extended-AHM's physical particle spectrum  excludes the NGB $\tilde \pi$: Decoupling of heavy particles;}
\label{ParticlePhysicsAHM}

G.S. Guralnik, C.R. Hagan and T.W.B. Kibble \cite{Guralnik1964} first showed  in the spontaneously broken Abelian Higgs model that, although there are no massless particles in the $(A^0=0,{\vec \nabla}\cdot {\vec  A} =0)$ ``radiation gauge",  there is a Goldstone theorem, and a true massless NGB,   in the covariant $\partial_\mu A^\mu =0$ Lorenz gauge. T.W.B. Kibble then showed \cite{Kibble1967} that the results of experimental measurements are nevertheless the same in radiation and Lorenz gauges, 
and that the spectrum and dynamics of the {\em observable particle states} are gauge-independent. 

\subsection{SSB extended-AHM's physical particle spectrum  excludes the NGB $\tilde \pi$ whose S-Matrix elements all vanish \cite{LSS-3Short}}
\label{NGBDisappearsExtendeAHM}

We repeat here (and include further detail) the discussion from the companion Letter \cite{LSS-3Short} in order that this paper be pedagogically self-contained, clear and complete.

The BRST-invariant Lagrangian for the extended-AHM in Lorenz gauge is
\begin{eqnarray}
\label{LagrangianE-AHMLorenzA}
&&L_{E-AHM}^{Lorenz}=L_{AHM}^{Lorenz} \nonumber  \\
&& \quad \quad +L_{BeyondAHM;\Phi}^{GaugeInvariant}+L_{BeyondAHM;\psi}^{GaugeInvariant}
\end{eqnarray}
with $L_{AHM}^{Lorenz}$ in (\ref{LagrangianAHM}), $L_{BeyondAHM;\Phi}^{GaugeInvariant}$ in  (\ref{GlobalInvariantBeyondAHMPhi}),
and $L_{BeyondAHM;\psi}^{GaugeInvariant}$ in (\ref{GlobalInvariantBeyondAHMpsi}).

{\bf i) Lagrangian governing dynamics of observable particles:} We now identify the {\em observable particle spectrum} of Lorenz gauge extended-AHM by re-writing (\ref{LagrangianE-AHMLorenzA})
in terms of a new gauge field
\begin{eqnarray}
\label{ProcaField}
B_\mu&\equiv& A_\mu +\frac{1}{e\HVEV}\partial_\mu {\tilde \pi}
\end{eqnarray}
and transforming to the Kibble representation \cite{Ramond2004}
\begin{itemize}

\item Gauge field
\begin{eqnarray}
\label{BFieldStrength}
A_{\mu\nu}&\equiv&\partial_\mu A_\nu - \partial_\nu A_\mu \nonumber \\
&=&\partial_\mu B_\nu - \partial_\nu B_\mu \equiv B_{\mu\nu}
\end{eqnarray}

\item AHM scalar
\begin{eqnarray}
\label{KibbleRepresentationAHM}
\tilde \pi &=& \HVEV \vartheta \nonumber \\
\phi &=& \frac{1}{\sqrt 2}{\tilde H}e^{-iY_\phi \vartheta}; \quad {\tilde H}={\tilde h}+\HVEV \nonumber \\
D_\mu \phi &=& \frac{1}{\sqrt 2}\Big[ \partial_\mu -ieY_\phi A_\mu \Big] {\tilde H}e^{-iY_\phi \vartheta} \nonumber \\ 
&=&\frac{1}{\sqrt 2}\Big[ \partial_\mu {\tilde H}-ieY_\phi {\tilde H}\Big(A_\mu +\frac{1}{e}\partial_\mu\vartheta \Big)\Big] e^{-iY_\phi \vartheta} \nonumber \\
&=&\frac{1}{\sqrt 2}\Big[ \partial_\mu {\tilde H}-ieY_\phi {\tilde H} B_\mu \Big] e^{-iY_\phi \vartheta}  
\end{eqnarray}

\item Beyond-AHM scalar
\begin{eqnarray}
\label{KibbleScalarsBeyondAHM}
\Phi &=& {\tilde \Phi}e^{-iY_\Phi \vartheta} \nonumber \\
\big< {\tilde \Phi} \big> &=& 0 \nonumber \\
D_\mu \Phi &=& \Big[ \partial_\mu -ieY_\Phi A_\mu \Big] {\tilde \Phi}e^{-iY_\Phi \vartheta} \nonumber \\ 
&=&\Big[ \partial_\mu {\tilde \Phi}-ieY_\Phi {\tilde \Phi}\Big(A_\mu +\frac{1}{e}\partial_\mu\vartheta \Big)\Big] e^{-iY_\Phi \vartheta} \nonumber \\
&=&\Big[ \partial_\mu {\tilde \Phi}-ieY_\Phi {\tilde \Phi} B_\mu \Big] e^{-iY_\Phi \vartheta} 
\end{eqnarray}

\item Beyond-AHM fermion(s)
\begin{eqnarray}
\label{KibbleFermionsBeyondAHM}
\psi &=& {\tilde \psi}e^{-iY_\psi \vartheta} \nonumber \\
D_\mu \psi &=& \Big[ \partial_\mu -ieY_\psi A_\mu \Big] {\tilde \psi}e^{-iY_\psi \vartheta} \nonumber \\ 
&=&\Big[ \partial_\mu {\tilde \psi}-ieY_\psi {\tilde \psi} \Big( A_\mu +\frac{1}{e}\partial_\mu \vartheta \Big) \Big] e^{-iY_\Phi \vartheta} \nonumber \\
&=&\Big[ \partial_\mu {\tilde \psi}-ieY_\psi {\tilde \psi} B_\mu \Big] e^{-iY_\Phi \vartheta}
\end{eqnarray}

\end{itemize}

The extended-AHM Lagrangian, which governs the spectrum and dynamics of {\em particle physics} is
\begin{eqnarray}
\label{LagrangianE-AHMParticlePhysics}
&&L_{E-AHM}^{ParticlePhysics} \Big(  B_\mu; {\tilde H} ; {\tilde \Phi}; {\tilde \psi}\Big) \nonumber \\
&& \quad =L_{AHM;{\tilde H},B_\mu}^{Lorenz} \Big(  B_\mu; {\tilde H;{\bar \omega},\omega} \Big)  \nonumber \\
&& \quad +L_{BeyondAHM; {\tilde \Phi} }^{GaugeInvariant}+L_{BeyondAHM;{\tilde \psi} }^{GaugeInvariant} 
\end{eqnarray}
where the spin $S=1$ field $B_\mu$ 
\begin{eqnarray}
\label{GaugeInvariantLagrangianAHMKibble}
&&L_{AHM}^{Lorenz} \Big(  B_\mu; {\tilde H};  {\bar \omega},\omega \Big) = L^{GaugeInvariant}_{AHM;{\tilde H},B_\mu}  \nonumber \\
&&\quad +L_{AHM;B_\mu}^{GaugeFix;Lorenz} +L_{AHM;B_\mu}^{Ghost;Lorenz} \nonumber \\
&&L^{GaugeInvariant}_{AHM;{\tilde H},B_\mu} = -\frac{1}{4}B_{\mu \nu}B^{\mu \nu} \nonumber +\half e^2Y_\phi^2 \HVEV^2 B_\mu B^\mu \nonumber \\
&& \quad + \half\Big(\partial_\mu {\tilde H}\Big)^2+\half e^2Y_\phi^2 \Big( {\tilde H}^2 -\HVEV^2 \Big) B_\mu B^\mu  -V_{AHM} \nonumber \\
&&L_{AHM;B_\mu}^{GaugeFix;Lorenz}=-\lim_{\xi \to 0}\frac{1}{2\xi}\Big( \partial_\mu B^\mu \Big)^2\nonumber \\
&&L_{AHM;B_\mu}^{Ghost;Lorenz}=- {\bar \omega}\partial^2 \omega  \nonumber \\
&&V_{AHM}=\frac{1}{4}\lambda^2_\phi \Big( {\tilde H}^2-\HVEV^2 \Big) 
\end{eqnarray}

For the Beyond-AHM scalar
\begin{eqnarray}
\label{BeyondAHMPhi}
&&L_{BeyondAHM;{\tilde \Phi}}^{GaugeInvariant} = \Big\vert  D_\mu {\tilde \Phi} \Big\vert ^2 -V_{{\tilde \Phi}}-V_{ {\tilde \phi} {\tilde \Phi}}   \\
&& \quad \quad D_\mu {\tilde \Phi} = \Big[ \partial_\mu-ieY_\Phi B_\mu \Big] {\tilde \Phi} \nonumber \\ 
&&V_{\tilde \Phi} = M_\Phi^2\Big( {\tilde \Phi}^\dagger {\tilde \Phi} \Big) + \lambda_\Phi^2\Big( {\tilde \Phi}^\dagger {\tilde \Phi} \Big) ^2 \nonumber \\
&&V_{ {\tilde \phi} {\tilde \Phi}}=\half \lambda_{\phi \Phi}^2 \Big({\tilde H}^2 \Big)  \Big( \Phi^\dagger \Phi \Big) \nonumber
\end{eqnarray}
while for Beyond-AHM fermions
\begin{eqnarray}
\label{BeyondAHMpsiB}
&&L_{BeyondAHM;{\tilde \psi}}^{GaugeInvariant} =  i{\bar {\tilde \psi} }_L D_\mu {\tilde \psi}_L 
+  i{\bar {\tilde \psi}}_R D_\mu {\tilde \psi}_R  \nonumber  \\
&& \quad \quad +L_{BeyondAHM;{\tilde \psi}}^{Yukawa}+L_{BeyondAHM;{\tilde \psi}}^{Majorana} \\
&& \quad \quad D_\mu {\tilde \psi} _L = \Big[ \partial_\mu -ieY_{\psi_L} B_\mu \Big] {{\tilde \psi}} _L\nonumber \\ 
&& \quad \quad D_\mu {\tilde \psi} _R = \Big[ \partial_\mu -ieY_{\psi_R} B_\mu \Big] {\tilde \psi} _R\nonumber \\ 
&&L_{BeyondAHM;{\tilde \psi}}^{Yukawa}=-\frac{1}{\sqrt 2}y_{\phi \psi} \Big(  {\bar {\tilde \psi}}_L  {\tilde \psi}_R 
+ {\bar {\tilde \psi}}_R  {\tilde \psi}_L \Big) {\tilde H} \nonumber \\
&& \quad \quad -y_{\Phi \psi} \Big(  {\bar {\tilde \psi} }_L {\tilde \Phi} {\tilde \psi}_R 
+ {\bar {\tilde \psi}}_R {\tilde \Phi}^\dagger {\tilde \psi}_L \Big) \nonumber \\
&&L^{Majorana}_{BeyondAHM; {\tilde \psi} } = -\half M_{{\psi}_L} \Big( {{\tilde \psi}_L^{Weyl} }{{\tilde \psi}_L^{Weyl}}
+{\bar {\tilde \psi}}_L^{Weyl} {\bar {\tilde \psi}}_L^{Weyl}\Big) \nonumber \\ 
&& \quad \quad  -\half M_{\psi_R} \Big( {{\tilde \psi}_R^{Weyl} }{{\tilde \psi}_R^{Weyl}}
+{\bar {\tilde \psi}}_R^{Weyl} {\bar {\tilde \psi}}_R^{Weyl}\Big) \nonumber 
\end{eqnarray}

{\bf ii) 4th decoupling theorem: extended-AHM $B_\mu$ pole-mass 
\cite{AppelquistCarrazone}.} 
The $B_\mu$ mass-squared in (\ref{GaugeInvariantLagrangianAHMKibble}) arises entirely from SSB
\begin{eqnarray}
\label{ProcaMass}
m_B^2 = m_A^2 = e^2 \HVEV ^2
\end{eqnarray}
Dimensional analysis 
shows that the contribution of a state  of mass/energy 
$\sim M_{Heavy}$  to the spectral function $\Delta_{E-AHM}^{B;Spectral}$ gives  terms $\sim 1/{M_{Heavy}^2}$, so that 
\begin{eqnarray}
\label{VectorPropagatorExtendedDecoupled}
&&\Delta_{E-AHM}^{B}(q^2) = \Delta^{B}_{AHM}(q^2) + {\cal O}\Big(1/M_{Heavy}^2 \Big) \nonumber \\
&&\Delta^{B}_{AHM}(q^2) =\frac{1}{q^2-m_{B;Pole}^2 + i\epsilon} + \int dm^2 \frac{\rho^{B}_{AHM}(m^2)}{q^2-m^2 + i\epsilon} \nonumber \\
&&Z^{B}_{E-AHM} =  Z^{B}_{AHM} + {\cal O}\Big(1/M_{Heavy}^2 \Big) 
\end{eqnarray}
Therefore the gauge-independent $B_\mu$ observable pole-mass-squared
\begin{eqnarray}
\label{BPoleMassExtended}
&&\Big[ \Delta^{B}_{E-AHM}(0) \Big]^{-1} =-m_{B}^2=-e^2\HVEV^2 \nonumber \\
&& m^2_{B;Pole} = e^2 \HVEV^2\Big[ 1- e^2 \HVEV^2 \int dm^2 \frac{{\tilde \rho}^{B}_{AHM}(m^2)}{m^2 - i\epsilon} \Big]^{-1} \nonumber \\
&&\qquad \qquad+ {\cal O}\Big(1/M_{Heavy}^2 \Big) 
\end{eqnarray}
(with Kibble representation ${\tilde \rho}^{B}_{AHM}$) are Goldstone Exceptionally Natural, not fine-tuned.

{\bf iii) Decoupling of NGB $\tilde \pi$, particle spectrum and dynamics \cite{LSS-3Short}:} It is crucial for SSB gauge theories \cite{Guralnik1964,Kibble1967} to remember the additional gauge-fixing term inside (\ref{LagrangianE-AHMLorenzA})
\begin{eqnarray}
\label{LagrangianE-AHMLorenzB}
&&L_{E-AHM}^{Lorenz} = L_{E-AHM}^{ParticlePhysics} \nonumber \\
&& \quad \quad -\lim_{\xi \to 0} \frac{1}{2\xi}\Big( \frac{1}{e\HVEV}\partial^2{\tilde \pi} \Big) \Big( \frac{1}{e\HVEV}\partial^2{\tilde \pi} - 2\partial_\mu B^\mu\Big) \quad \quad 
\end{eqnarray}

The Lagrangian (\ref{LagrangianE-AHMLorenzB}) is guarranteed to generate all of the results in Sections \ref{AbelianHiggsModel} 
and \ref{E-AHM}, and Appendices \ref{DerivationWTIAHM} and \ref{DerivationWTIE-AHM}. 
In practice, this is done via the manifestly renormalizeable extended-AHM Lagrangian 
(\ref{LagrangianE-AHMLorenzB}).

G. Guralnik, T. Hagan and T.W.B. Kibble \cite{Guralnik1964}, and T.W.B. Kibble \cite{Kibble1967}, showed that, in the Kibble representation in Lorenz gauge, the $U(1)_Y$ AHM quantum states factorize. In the analogous $U(1)_Y$ extended-AHM, and in the
$m_{Weak}^2/M_{Heavy}^2 \to 0$ limit 
the  analogous $U(1)_Y$ E-AHM also factorizes \cite{LSS-3Short}
\begin{eqnarray}
\label{E-AHMQuantumStates}
&&\Big\vert \Psi \Big( A^\mu;\phi;{\bar \omega},\omega;\Phi;\psi \Big) \Big> \to  
	\Big\vert \Psi^{Particles} \Big( B^\mu;{\tilde H} \Big) \Big> \\
&&\quad \times \Big\vert \Psi^{Ghost} \Big( {\bar \omega},\omega \Big) \Big>
	\Big\vert \Psi^{Goldstone} \Big( {\tilde \pi} \Big) \Big>
	\Big\vert \Psi^{B-AHM} \Big( {\tilde \Phi};  {\tilde \psi}\Big) \Big>\nonumber
\end{eqnarray}

With $\partial^2\omega =0; \partial^2{\bar \omega} =0$,
the ghosts $\omega$ and $\bar \omega$ are free and massless 
and de-couple in Lorenz gauge. 
Furthermore,
the final gauge-fixing condition $\partial^2  {\tilde \pi} =0$ in (\ref{LagrangianE-AHMLorenzB})  
forces the NGB $\tilde \pi$ to also be a free massless particle,
which completely decouples from, and disappears from,  
the observable particle spectrum and  its dynamics \cite{Guralnik1964,Kibble1967}, 
and whose states factorize as in (\ref{E-AHMQuantumStates}). 

In the $m_{Weak}^2/M_{Heavy}^2 \to 0$ limit, all physical measurements and observations 
are then entirely predicted by the {\bf AHM} Lagrangian (\ref{GaugeInvariantLagrangianAHMKibble}) and its states in (\ref{E-AHMQuantumStates})
\begin{eqnarray}
\label{ParticleDynamics}
&&L_{AHM;B_\mu}^{Lorenz}\Big( {\tilde H};B_\mu; {\bar \omega},\omega \Big); \nonumber \\
&&\Big\vert \Psi^{ParticlePhysics} \Big( B^\mu;{\tilde H} ;{\bar \omega},\omega \Big) \Big>
\end{eqnarray}

What has become of our SSB $U(1)_Y$ Ward-Takahashi identities?
Although the NGB ${\tilde \pi}$ has de-coupled, it still governs the SSB dynamics and particle spectrum of (\ref{ParticleDynamics}) \cite{LSS-3Short}:
it is simply {\em hidden} from explicit view.
But that decoupling NGB still causes powerful {\bf hidden constraints} on (\ref{ParticleDynamics}) to arise from its {\bf hidden shift symmetry} \cite{LSS-3Short}
\begin{eqnarray}
\label{HiddenShiftSymmetry}
{\tilde \pi}\to {\tilde \pi} +\HVEV \theta \nonumber
\end{eqnarray}
for constant $\theta$.

Our SSB $U(1)_Y$ WTIs, and all of the results of Section \ref{AbelianHiggsModel}, Section \ref{E-AHM}, 
Appendix \ref{DerivationWTIAHM} and Appendix \ref{DerivationWTIE-AHM}
are also hidden but still in force: connected amputated Greens functions $\Gamma_{N,M}$ (\ref{GreensWTIPrimeExtended},\ref{ExtendedGreensFWTI}); connected amputated  T-Matrix elements $T_{N,M}$ (\ref{InternalTMatrixE-AHM},\ref{InternalTMatrixExtended}); 
Adler self-consistency conditions (\ref{AdlerSelfConsistencyE-AHM},\ref{ExtendedAdlerSelfConsistency}) together with their proof of IR finiteness;
Goldstone theorem  (\ref{TMatrixGoldstoneTheoremE-AHM},\ref{TMatrixGoldstoneTheoremExtended});
1-soft-$\pi$ theorems (\ref{InternalTMatrixE-AHM},\ref{ExtendedAdlerSelfConsistency},\ref{InternalTMatrixExtended}); 
decoupling theorems for Green's functions and T-Matrix elements (\ref{SSBTMatrixDecouplingTheorem},\ref{SSBGreensFunctionDecouplingTheorem});
and the decoupling theorem for the BEH pole-mass-squared $m_{BEH;Pole}^2$ (\ref{GaugeIndependentBEHPoleMass}).
These still govern the SSB dynamics and 
particle spectrum of (\ref{ParticleDynamics}): they are simply hidden from explicit view.
We call them {\bf ``Hidden $U(1)_Y$ Ward-Takahashi identities of the SSB Abelian Higgs model"}.

\subsection{SSB causes decoupling of heavy $M_{Heavy}^2 \gg m_{Weak}^2$ particles. This  fact is hidden, from the  observable particle spectrum of the $U(1)_Y$ extended-AHM and its dynamics, by the decoupling of the NGB $\tilde \pi$}
\label{DecouplingInGaugedE-AHM}

We now take
all of the new scalars $\tilde \Phi$ and fermions $\tilde \psi$ in the extended-AHM  
to be very heavy, and are only interested in low-energy processes:
\begin{eqnarray}
\label{TildeHeavy}
&&M^2_{\tilde \Phi}, M^2_{\tilde \psi} \sim M_{Heavy}^2 \gg m_{Weak}^2 \nonumber \\
&&q_\mu {\buildrel {<} \over {\sim}} m_{Weak}
\end{eqnarray} 
where $q_\mu$ is a typical momentum transfer.
In the limit
$m_{Weak}^2 / M_{Heavy}^2 \to 0$
the effective Lagrangian of the {\em spontaneously broken} extended-AHM gauge theory \cite{AppelquistCarrazone}
\begin{eqnarray}
\label{EffLagrangianLimit}
&&L^{Eff;SSB}_{E-AHM} \Big( k_\mu;B_\nu;{\tilde H};{\tilde \Phi};{\tilde \psi} \Big) \nonumber \\
&&\quad \rightarrow L^{Eff;SSB}_{AHM} \Big( k_\mu;B_\nu;{\tilde H} \Big) + {\cal O}\Big( m_{Weak}^2 / M_{Heavy}^2\Big) \quad \quad
\end{eqnarray}
{\bf with the possible exception} that the dimension $Dim=2$ operator $\propto \mu^2_\phi$ in 
$V_{AHM}$ in (\ref{LagrangianAHM}).
\begin{eqnarray}
\label{GaugeInvariantAHMPotential}
&&V_{AHM}=\mu_\phi^2 \Big(  \phi^\dagger \phi \Big)  + \lambda^2_\phi \Big(  \phi^\dagger \phi  \Big)^2
\end{eqnarray}
has caused a fine-tuning problem, and {\bf raised the BEH massed-squared to the heavy scale}: $m_{BEH}^2 \sim M_{Heavy}^2$.
We now show that this is not the case.

{\bf i) 5th decoupling theorem: SSB Abelian Higgs model:} 
Eqn. (\ref{GaugeInvariantAHMPotential}) lies entirely within the $\phi$-sector of the extended theory, and is therefore subject to all of the results of Sections \ref{AbelianHiggsModel} and \ref{E-AHM}, and Appendices \ref{DerivationWTIAHM} and \ref{DerivationWTIE-AHM}.
Therefore we know, instead, that
{\bf the BEH pole-mass-squared (\ref{GaugeIndependentBEHPoleMass}) arises entirely from SSB and (un-extended) AHM decays.} We also know that
\begin{eqnarray}
\label{E-AHMEffectivePotential}
&&V_{E-AHM}^{Eff}=\lambda^2_\phi \Big(  \phi^\dagger \phi -\half \HVEV^2 \Big)^2 +{\cal O}^{Ignore}_{E-AHM} \nonumber \\ \nonumber \\
&&\qquad ={\lambda^2_\phi} \Big(  {\tilde \phi}^\dagger {\tilde \phi} -\half \HVEV^2 \Big)^2 + {\cal O}^{Ignore}_{E-AHM} \nonumber \\
&&\qquad =\frac{{\lambda^2_\phi}}{4} \Big(  {\tilde H}^2 - \HVEV^2 \Big)^2 + {\cal O}^{Ignore}_{E-AHM} \nonumber \\
&&\qquad =\frac{{\lambda^2_\phi}}{4} \Big(  {\tilde h}^2 + 2 \HVEV {\tilde h} \Big)^2 + {\cal O}^{Ignore}_{E-AHM} 
\end{eqnarray}

\begin{itemize}
\item In (\ref{GaugeIndependentBEHPoleMass},\ref{E-AHMEffectivePotential}) finite  ${\cal O}_{E-AHM;\phi}^{1/M_{Heavy}^2;Irrelevant}$ 
decouple and vanish as $m_{Weak}^2/ M_{Heavy}^2 \to 0$.

\item Among the terms included in (\ref{E-AHMEffectivePotential}) are {\em finite} relevant operators dependent on the heavy matter representations:
\begin{eqnarray}
\label{HeavyRelevantOperatorsB}
&&M_{Heavy}^2, M_{Heavy}^2 \ln{ \big(M_{Heavy}^2\big)}, \nonumber \\
&& \quad \quad M_{Heavy}^2 \ln{ \big(m_{Weak}^2\big)}, m_{Weak}^2 \ln{ \big(M_{Heavy}^2\big)}  \quad 
\end{eqnarray}
but the Goldstone theorem (\ref{TMatrixGoldstoneTheoremE-AHM}) has made them vanish! That fact is a central point of this paper.

\item Marginal operators $\sim \ln{ \big(M_{Heavy}^2\big)}$ have been absorbed in  (\ref{E-AHMEffectivePotential}): i.e. in the renormalization of gauge-independent observables (i.e. the quartic-coupling constant
$\lambda_\phi^2$ and the BEH VEV $\HVEV$), and in the un-observable wavefunction renormalization 
$Z_{E-AHM}^\phi$ 
(\ref{GoldstoneWavefunctionE-AHM}).
\end{itemize}

Therefore, no trace of $M_{Heavy}$-scale $\Phi,\psi$ survives in (\ref{GaugeIndependentBEHPoleMass},\ref{E-AHMEffectivePotential})! All the heavy Beyond-AHM matter representations have completely decoupled, and the two SSB gauge theories
\begin{eqnarray}
\label{AHMGaugeTheoryDecouplingTheorem}
E-AHM\quad  {\buildrel  {{m_{Weak}^2}/{M_{Heavy}^2} \to 0} \over {=\joinrel=\joinrel=\joinrel=\joinrel=\joinrel=\joinrel=\joinrel=\joinrel=\joinrel=\joinrel\Longrightarrow}} \quad AHM \quad \quad
\end{eqnarray}
become equivalent in the limit ${{m_{Weak}^2}/{M_{Heavy}^2} \to 0}$, 
a central result of this paper.

{\bf ii)  Gauge-independence of our results \cite{LSS-3Short}:} 
S.-H. Henry Tye and Y. Vtorov-Karevsky \cite{Tye1996} proved that, displayed in the Kibble representation,
 $\lambda_\phi^2, \HVEV^2$ and  the AHM effective potential,
\begin{eqnarray}
\label{E-AHMEffectivePotentialSHHTye}
V_{AHM}^{Eff}&=&\frac{{\lambda^2_\phi}}{4} \Big(  {\tilde h}^2 + 2 \HVEV {\tilde h} \Big)^2 + {\cal O}^{Ignore}_{AHM} \quad \quad
\end{eqnarray}
are all-loop-orders  gauge-independent. The renormalized gauge-coupling-constant-squared at zero momentum $e^2\equiv e^2(0)$ is gauge-independent. With our 5 decoupling theorems 
(\ref{SSBTMatrixDecouplingTheorem},\ref{SSBGreensFunctionDecouplingTheorem},\ref{GaugeIndependentBEHPoleMass},\ref{BPoleMassExtended},\ref{AHMGaugeTheoryDecouplingTheorem}), 
so are 
$\lambda_\phi^2, \HVEV^2$ and $V_{E-AHM}^{Eff}$ 
in (\ref{E-AHMEffectivePotential}), 
  and the $B_\mu$ pole-mass-squared  (\ref{BPoleMassExtended}). These all appear in the {\em decoupled} particle physics
 (\ref{ParticleDynamics}) of extended-AHM.

{\bf iii) Observable gauge-independent BEH mass:} Slavnov-Taylor identities guarrantee that the on-shell T-Matrix-element definition of the experimentally observable BEH mass-squared
\begin{eqnarray}
\label{ObservableBEHMass}
T^{extended}_{2,0}(p,-p;)\vert _{p^2= m^2_{BEH;Experimental}}=0 \quad \quad
\end{eqnarray}
is  gauge-independent to all-loop-orders. But in Lorenz gauge we have the specific instance
\begin{eqnarray}
\label{BEHPropagatorAtPoleMass}
&&T^{extended}_{2,0}(p,-p;)\vert _{p^2=m^2_{BEH;Pole}}^{KibbleRepresentation} \nonumber \\
&& \quad \quad \equiv \Big[ \Delta^{BEH}_{E-AHM} \big(p^2=m^2_{BEH;Pole} \big)\Big]^{-1} \nonumber \\
&&  \quad \quad =\Big[ \frac{1}{p^2-m_{BEH;Pole}^2 + i\epsilon}\nonumber \\
&& \qquad \qquad \qquad+ \int dm^2 \frac{{\tilde \rho}^{BEH}_{AHM}(m^2)}{p^2-m^2 + i\epsilon} \Big]^{-1}_{p^2=m^2_{BEH;Pole}} \nonumber \\
&&\quad \quad =0
\end{eqnarray}
with ${\tilde \rho}^{BEH}_{AHM}$ in the Kibble representation.
Therefore the experimentally observable BEH mass, in extended-AHM, is the BEH pole-mass
\begin{eqnarray}
\label{BEHPoleEqualExperimentalMass}
&& m^2_{BEH;Experimental}=m^2_{BEH;Pole} \nonumber \\
&& \quad \quad = 2\lambda_\phi^2 \HVEV^2\Big[ 1- 2\lambda_\phi^2 \HVEV^2 \int dm^2 \frac{{\tilde \rho}^{BEH}_{AHM}(m^2)}{m^2 - i\epsilon} \Big]^{-1} \nonumber \\
&&\qquad \qquad+ {\cal O}^{1/M_{Heavy}^2;Irrelevant}_{E-AHM;\phi} 
\end{eqnarray}
and is gauge-independent to all-loop-orders.

\section{BWL \& GDS: Notre vision $\acute{\mathrm \bf A}$ travers le prisme de la rigueur math$\acute{\mathrm \bf E}$matique q$\acute{\mathrm \bf U}$ imposait Raymond Stora
}
\label{RigueurMathematiqueExigeante}

Raymond Stora regarded Vintage-QFT as incomplete, fuzzy in its definitions, and primitive in technology. For example, he worried 
whether the off-shell T-Matrix could be mathematically rigourously  defined {\em to exist}  in 
Lorenz gauge: e.g. without running into some infra-red (IR) sublety.
This, even though we prove here the IR finiteness of the $\phi$-sector on-shell T-Matrix. 

Although he agreed on the correctness of the results presented here, Raymond might complain that we fall short of a 
strict mathematically rigourous proof (i.e. according to his  exacting mathematical standards).
He reminded us that much has been learned about QFT, via modern path integrals, in the recent $\sim 45$ years.
In the time up to his passing, he was intent on improving this work by focussing on 3 issues: 
\begin{itemize}

\item Properly defining and proving the 
Lorenz gauge results presented here, but with modern path integrals; 

\item Tracking our central results  (i.e. no-FT and heavy-particle decoupling)  directly to SSB,  via BRST methods,  in an arbitrary manifestly IR finite 't Hooft $R_\xi$ gauge: i.e.  proving {\em to his satisfaction} that they are not an artifact of 
Lorenz gauge;

\item  Tracking our central  results  
directly to those Slavnov-Taylor IDS 
governing the SSB Goldstone mode of the BRST-invariant extended-AHM Lagrangian.
\end{itemize}

Any errors, wrong-headedness, mis-understanding, or mis-representation appearing in  this paper are solely the fault of BWL and GDS.

\section{Conclusion: Historically, complete decoupling of heavy invisible particles is the usual physics experience}
\label{Conclusions}

We showed, in Sections \ref{E-AHM}, \ref{ParticlePhysicsAHM} and Appendix \ref{DerivationWTIE-AHM}, that the low-energy weak-scale effective SSB extended-AHM Lagrangian is protected, by new  ``hidden" rigid/global SSB $U(1)_Y$ WTIs and a hidden Goldstone theorem, against contributions from certain heavy 
$M_{Heavy}^2 \gg m_{Weak}^2$ Beyond-AHM particles. 
Five decoupling theorems 
(\ref{SSBTMatrixDecouplingTheorem},\ref{SSBGreensFunctionDecouplingTheorem},\ref{GaugeIndependentBEHPoleMass},\ref{BPoleMassExtended},\ref{AHMGaugeTheoryDecouplingTheorem})
govern certain heavy particles $\Phi,\psi$. Renormalized gauge-independent observable $(\HVEV^2, m_{BEH;Pole}^2)$ are therefore not fine-tuned, but instead
Goldstone Exceptionally Natural, with far more powerful suppression of fine-tuning than G. 't Hooft's  naturalness criteria \cite{tHooft1980} would demand.

What is remarkable is that heavy particle decoupling and Goldstone Exceptional Naturalness are obscured/hidden from the physical particle spectrum (\ref{ParticleDynamics}) and its dynamics. The decoupling of the NGB $\tilde \pi$ has famously spared the AHM an observable massless particle  \cite{Guralnik1964,Englert1964,Higgs1964}. But it has also hidden, from that physical particle spectrum and  dynamics, our $U(1)_Y$ WTI
 (\ref{GreensWTIPrimeExtended}, \ref{AdlerSelfConsistencyE-AHM}, \ref{TMatrixGoldstoneTheoremE-AHM},
\ref{InternalTMatrixE-AHM}, \ref{ExtendedAdlerSelfConsistency}, \ref{TMatrixGoldstoneTheoremExtended},  \ref{InternalTMatrixExtended},  \ref{ExtendedGreensFWTI}), 
and their severe constraints on the effective low-energy extended-AHM Lagrangian.  In particular, the {\bf weak-scale} extended-AHM SSB gauge theory has a {\bf hidden $U(1)_Y$ shift symmetry},
for constant $\theta$
\begin{eqnarray}
\label{HiddenshiftSymmetryPrime}
{\tilde \pi} \to {\tilde \pi} +\HVEV \theta
\end{eqnarray}
which has protected it from any
Brout-Englert-Higgs  fine-tuning problem, 
 and  caused the  complete
\footnote{
Modulo special cases: e.g. heavy Majorana $\nu_R$ in Subsection \ref{HeavyNeutrino}, and possibly 
${\cal O}_{E-AHM\phi}^{Dim \leq 4;NonAnalytic;Heavy}$ in (\ref{IgnoreE-AHMOperators}).
}
 decoupling of certain heavy $M^2_{Heavy} \gg m_{Weak}^2$ $U(1)_Y$ matter-particles.


But such heavy-particle decoupling is historically (i.e. except  for high-precision electro-weak S,T and U \cite{Kennedy1988, 
PeskinTakeuchi, Ramond2004}) the usual physics experience, at each energy scale, as experiments probed smaller and smaller distances. After all, Willis Lamb did 
not need to know the top quark or BEH mass in order to interpret theoretically the experimentally observed ${\cal O}(m_e \alpha^5 \ln \alpha)$ splitting in the spectrum of hydrogen. 

Such heavy-particle decoupling may be the reason why the Standard Model \cite{LSS-4,LSS-4Proof}, viewed as an effective low-energy weak-scale theory, is the most experimentally and observationally successfull 
and accurate theory of Nature known to humans, i.e. when augmented by classical General Relativity and neutrino mixing: that ``Core  Theory"
\cite{WilcekCoreTheory} has no known experimental or observational counter-examples.

\smallskip

{\bf Acknowledgements:}
BWL thanks: 
Jon Butterworth and the Department of Physics and Astronomy at University College London for support as a UCL Honorary Senior Research Associate;
Albrecht Karle and U Wisconsin at Madison for hospitality during the 2014-2015 academic year;
Chris Pope, the George P. and Cynthia W. Mitchell Center for Fundamental Physics and Astronomy, and Texas A\&M University,  for support and hospitality during the 
2010-2011 academic year, where this work began.
 GDS is partially supported by CWRU grant DOE-SC0009946.  

\appendix
\section{$U(1)_Y$ Ward-Takahashi identities in the SSB Abelian Higgs Model}
\label{DerivationWTIAHM}

In this Appendix \ref{DerivationWTIAHM}, we present the full detailed derivations of our $U(1)_Y$ WTI for the SSB AHM. For pedagogical completeness, we reproduce all details of the entire argument.
\begin{itemize}
\item This paper is interested in building the effective $\phi$ potential, and showing no-fine-tuning of the BEH mass.

\item A companion Letter \cite{LSS-3Short} focusses instead on the WTI themselves, showing that AHM {\bf physics} has more symmetry than the AHM effective Lagrangian.
\end{itemize}

We focus on the rigid/global  current ${J}^{\mu}_{AHM}$ of the Abelian Higgs model, the spontaneously broken gauge theory of a complex scalar $\phi = \frac{1}{{\sqrt 2}}\big( H+i\pi \big)= \frac{1}{{\sqrt 2}}{\tilde H}e^{i{\tilde \pi}/\HVEV}$, 
and a massive $U(1)_Y$ gauge field $A_\mu$. 
\begin{eqnarray}
\label{AHMCurrent}
 {J}^{\mu}_{AHM}&=& \pi \partial^\mu H-H\partial^\mu \pi-eA^\mu\Big(\pi^2 + H^2 \Big) \quad \quad
\end{eqnarray}
The classical equations of motion reveal the crucial fact: due to gauge-fixing terms in the BRST-invariant Lagrangian, the 
classical axial-vector current 
(\ref{AHMCurrent}) is 
not conserved. In 
Lorenz gauge  
\begin{eqnarray}
\label{DivergenceAHMCurrent}
\partial_{\mu} {J}^{\mu}_{AHM}&=& -H m_A F_A   \nonumber \\
m_A &=& e\HVEV \nonumber \\
F_A &=& \partial_{\beta}{A}^{\beta} 
\end{eqnarray}
with $F_A$ the gauge fixing function.  Still, the {\em physical states} $A_\mu;h,\pi$ of the theory (but not the BRST-invariant Lagrangian) obey $F_A=0$.
In 
Lorenz gauge, $A_\mu$ is transverse and $\tilde \pi$ is a massless Nambu-Goldstone Boson (NGB).

The purpose of this Appendix \ref{DerivationWTIAHM} is to derive a tower of quantum $U(1)_Y$ Ward-Takahashi identities, which exhausts the information content of (\ref{DivergenceAHMCurrent}), and severely constrains the dynamics (i.e. the connected time-orderd products) of the physical states 
of the {\em spontaneously broken} Abelian Higgs model. 

{\bf 1) We study a total differential of a certain {\em connected} time-ordered product}
\begin{eqnarray}
\label{TotalDerivativeAHM}
&&\partial_{\mu} \Big< 0\vert T\Big[ {J}^{\mu}_{AHM}(z)  \\
&&\quad \quad \times h(x_1)...h(x_N) \pi_(y_1)...\pi(y_M)\Big]\vert 0\Big>_{Connected} \nonumber
\end{eqnarray}
written in terms of the {\em physical states} of the complex scalar $\phi$.
Here we have N external renormalized scalars $h=H-\HVEV$ (coordinates x, momenta p), 
and M external ($CP=-1$) renormalized pseudo-scalars ${ \pi}$ (coordinates y, momenta q). 

{\bf 2) Conservation of the global $U(1)_Y$ current for the {\em physical states}:} Strict quantum constraints are imposed, which force the relativistically-covariant theory of gauge bosons to propagate {\it only} its true number of quantum spin $S=1$ degrees of freedom: these constraints are implemented by use of spin $S=0$ fermionic Fadeev-Popov ghosts $({\bar \omega},\omega)$ and, in Lorenz gauge,  $S=0$ massless $ \pi$.  Physical states and their connected time-ordered products, but not the BRST-invariant Lagrangian,  obey \cite{tHooft1971} the gauge-fixing condition ${F}_A = \partial_{\beta}{A}^{\beta}=0$ in 
Lorenz gauge
\begin{eqnarray}
\label{GaugeConditions}
&&\big< 0\vert T\Big[ \Big( \partial_{\beta}{A}^{\beta}(z)\Big) \nonumber \\
&&\quad \quad \times h(x_1)...h(x_N)\pi(y_1)...\pi(y_M)\Big]\vert 0\big>_{Connected} \nonumber \\
&&\quad \quad =0
\end{eqnarray}
which restores conservation of the rigid/global $U(1)_Y$ current for physical states
\begin{eqnarray}
\label{QuantumCurrentConservation}
&&\Big< 0\vert T\Big[ \Big( \partial_{\mu}{J}^{\mu}_{AHM}(z) \Big) \\
&&\quad \quad \times h(x_1)...h(x_N) \pi(y_1)...\pi(y_M)\Big]\vert 0\Big>_{Connected}  \nonumber \\
&& \quad \quad =0
\end{eqnarray}
It is in this ``time-ordered-product" sense that the ``physical" rigid global $U(1)_Y$ current ${J}^{\mu}_{AHM}$ is ``conserved", and
it is this conserved current  which generates 2 towers of quantum $U(1)_Y$ WTI. These WTI severely constrain the dynamics of the $\phi$-sector.

{\bf 3) Vintage QFT and canonical quantization:} Equal-time commutators are imposed on the exact renormalized fields, yielding equal-time quantum commutators at space-time points $y, z$.
\begin{eqnarray}
\label{EqTimeCommAHM}
 \delta(z_0-y_0)\left[ {J}^0_{AHM}(z),H(y)\right] &=& -i{\pi}(x)\delta^4(z-y) \nonumber \\
 \delta(z_0-y_0)\left[ {J}^0_{AHM}(z),\pi(y)\right] &=& iH(y)\delta^4(z-y) \nonumber \\
 \delta(z_0-y_0)\left[ {J}^0_{AHM}(z),A^\mu (y)\right] &=& 0 \nonumber \\
 \delta(z_0-y_0)\left[ {J}^0_{AHM}(z),\omega (y)\right] &=& 0 \nonumber \\
 \delta(z_0-y_0)\left[ {J}^0_{AHM}(z),{\bar \omega} (y)\right] &=& 0 
\end{eqnarray} 
Field normalization follows from the non-trivial commutators
\begin{eqnarray}
\label{ZeroEqTimeCommAHM}
  \delta(z_0-y_0)\left[ \partial^0 H(z),H(y)\right] &=& -i\delta^4(z-y) \nonumber \\
 \delta(z_0-y_0)\left[ \partial^0 \pi (z),\pi(y)\right] &=& -i\delta^4(z-y) 
\end{eqnarray}

{\bf 4) Certain surface integrals vanish:} 
As appropriate to our study of the massless $\pi$, we use pion pole dominance to derive 1-soft-pion theorems, and form the surface integral
\begin{eqnarray}
\label{SurfacePionPoleDominance}
&&\lim_{k_\lambda \to 0} \int d^4z e^{ikz} \partial_{\mu} \Big< 0\vert T\Big[ \Big({J}^{\mu}_{AHM}+\HVEV \partial^\mu \pi\Big)(z)  \nonumber \\
&&\quad \quad \times h(x_1)...h(x_N) \pi_(y_1)...\pi(y_M)\Big]\vert 0\Big>_{Connected} \nonumber \\
&& =\int d^4z \partial_{\mu} \Big< 0\vert T\Big[ \Big({J}^{\mu}_{AHM}+\HVEV \partial^\mu \pi\Big)(z)  \nonumber \\
&&\quad \quad \times h(x_1)...h(x_N) \pi_(y_1)...\pi(y_M)\Big]\vert 0\Big>_{Connected} \nonumber \\
&& =\int_{3-surface\to\infty}  d^3z \quad {\widehat {z}_\mu}^{3-surface} \nonumber \\
&&\quad \quad \times \Big< 0\vert T\Big[ \Big({J}^{\mu}_{AHM}+\HVEV \partial^\mu \pi\Big)(z)  \nonumber \\
&&\quad \quad \times h(x_1)...h(x_N) \pi_(y_1)...\pi(y_M)\Big]\vert 0\Big>_{Connected} \nonumber \\
&&\quad \quad =0
\end{eqnarray}
where we have used Stokes theorem, and $ {\widehat {z}_\mu}^{3-surface}$ is a unit vector normal to the $3-surface$. The time-ordered-product constrains the $3-surface$ to lie on, or inside, the light-cone. 

At a given point on the surface of a large enough 4-volume $\int d^4z$ (i.e. the volume of all space-time): all fields are asymptotic in-states and out-states, properly quantized as free fields, with each field species orthogonal to the others,
and  they are evaluated at equal times, making time-ordering un-necessary at $(z^{3-surface}\to \infty)$. 
Input the global AHM current (\ref{AHMCurrent}) to  (\ref{SurfacePionPoleDominance}), using $\partial_\mu \HVEV =0$
\begin{eqnarray}
\label{SurfaceIntegral}
&& \int_{3-surface\to\infty}  d^3z \quad {\widehat {z}_\mu}^{3-surface} \Big< 0\vert T\Big[  \nonumber \\
&&\quad \quad \times \Big(\pi \partial^\mu h-h\partial^\mu \pi-eA^\mu(\pi^2 + H^2 )\Big)(z)  \nonumber \\
&&\quad \quad \times h(x_1)...h(x_N) \pi_(y_1)...\pi(y_M)\Big]\vert 0\Big>_{Connected} \nonumber \\
&&\quad \quad =0
\end{eqnarray}

The surface integral (\ref{SurfaceIntegral}) vanishes because both $(h, A^\mu)$ are massive in the spontaneously broken $U(1)_Y$ AHM, with $(m_{BEH}^2\neq 0, m_A^2=e^2\HVEV^2)$ respectively. Propagators connecting $(h, A^\mu)$, from points on $z^{3-surface}\to \infty$ to the localized interaction points $(x_1...x_N;y_1...y_M)$, must stay inside the light-cone, die off exponentially with mass,
and are incapable of carrying information that far.

It is very important for ``pion-pole-dominance"  and this paper, that {\em this argument fails} for the remaining term in $J^\mu_{AHM}$ in (\ref{AHMCurrent}):
\begin{eqnarray}
\label{NGBSurfaceIntegral}
&& \int_{2-Surface\to\infty}  d^2z \quad {\widehat {z}_\mu}^{2-surface}  \nonumber \\
&&\quad \quad \times \Big< 0\vert T\Big[  \Big(-\HVEV \partial^\mu \pi(z)\Big) \nonumber \\
&&\quad \quad \times h(x_1)...h(x_N) \pi_(y_1)...\pi(y_M)\Big]\vert 0\Big>_{Connected} \nonumber \\
&&\quad \quad \neq 0
\end{eqnarray}
$\pi$ is massless in the SSB AHM, capable of carrying (along the light-cone) long-ranged pseudo-scalar forces out to the $2$-surface  $(z^{2-surface}\to \infty)$: i.e. the very ends of the light-cone (but not inside it).
That massless-ness is the basis of our pion-pole-dominance-based $U(1)_Y$ WTIs, which give 1-soft-pion theorems 
(\ref{SoftPionLimitPropagator}), infra-red finiteness for $\mpisq =0$ (\ref{AdlerSelfConsistency}),  and a Goldstone theorem ({\ref{TMatrixGoldstoneTheorem}).

{\bf 5) Master equation:}
Using (\ref{QuantumCurrentConservation},\ref{ZeroEqTimeCommAHM}) in (\ref{TotalDerivativeAHM}) to form the right-hand-side, and (\ref{SurfaceIntegral}) in (\ref{TotalDerivativeAHM}) to form the left-hand-side, we write the master equation
\begin{eqnarray}
\label{MasterEquation}
&&-\HVEV\partial_{\mu}^z \big< 0\vert T\Big[  \big(\partial^\mu\pi(z)\big)  \nonumber \\
&& \quad \times h(x_1)...h(x_N) \pi_(y_1)...\pi(y_M)\Big]\vert 0\big>_{Connected} \nonumber \\
&&\quad = \sum_{m=1}^M \quad  i\delta^4(z-y_m) \big< 0\vert T\Big[ h(z) h(x_1)...h(x_N) \nonumber \\
&&\quad \quad \quad \quad \times  \pi(y_1)...{\widehat {\pi (y_m)}}...\pi(y_M)\Big]\vert 0\big>_{Connected} \nonumber \\
&&\quad - \sum_{n=1}^N \quad  i\delta^4(z-x_n) \big< 0\vert T\Big[ h(x_1)...{\widehat {h(x_n)}}...h(x_N) \nonumber \\
&&\quad \quad \quad \quad \times  \pi(z){\pi}(y_1)...\pi(y_M)\big]\vert 0\big>_{Connected}
\end{eqnarray}
where the ``hatted" fields ${\widehat {h(x_n)}}$ and ${\widehat {\pi (y_m)}}$ are to be removed. We have also thrown away a sum of $M$ terms, proportional to $\HVEV$,
which corresponds entirely to disconnected graphs.

{\bf 6) $\phi$-sector connected amplitudes:} Connected momentum-space amplitudes, with $N$ external BEHs and $M$ external $\pi$s, are defined in terms of $\phi$-sector connected time-ordered products
\begin{eqnarray}
\label{Amplitudes} 
&&iG_{N,M}(p_1...p_N;q_1...q_M)(2\pi)^4\delta^4 \Big(\sum_{n=1}^N p_n +\sum_{m=1}^M q_m \Big) \nonumber \\
&& \quad =\prod_{n=1}^N\int d^4x_n e^{ip_nx_n} \prod_{m=1}^M\int d^4y_m e^{iq_my_m}  \\
&&\quad \times \big< 0\vert T\Big[ h(x_1)...h(x_N) \pi_(y_1)...\pi(y_M)\Big]\vert 0\big>_{Connected} \nonumber
\end{eqnarray}

The master eqn. (\ref{MasterEquation}) can then be re-written
\begin{eqnarray}
\label{AmplitudeIdentity} 
&&i\HVEV k^2 G_{N,M+1}(p_1...p_N;kq_1...q_M) \\
&&\quad =\sum_{n=1}^N G_{N-1,M+1}(p_1...{\widehat {p_n}}...p_N;(k+p_n)q_1...q_M) \nonumber \\
&&\quad -\sum_{m=1}^M G_{N+1,M-1}((k+q_m)p_1...p_N;q_1...{\widehat {q_m}}...q_M) \nonumber 
\end{eqnarray}
with the ``hatted" momenta $({\widehat {p_n}},{\widehat {q_m}})$ removed  in (\ref{AmplitudeIdentity}), and an overall momentum conservation factor of $(2\pi)^4\delta^4 \Big(k+\sum_{n=1}^N p_n +\sum_{m=1}^M q_m \Big)$. 

{\bf 7) $\phi$-propagators:} Special cases of (\ref{Amplitudes}) are the BEH and $\pi$ propagators
\begin{eqnarray}
\label{ConnectedAmplitudePropagators} 
iG_{2,0}(p_1,-p_1;)&=&i\int \frac{d^4p_2}{(2\pi)^4}  G_{2,0}(p_1,p_2;)\nonumber \\
&=&\int d^4x_1 e^{ip_1x_1} \big< 0\vert T \Big[ h(x_1)h(0)\Big]\vert 0 \big> \nonumber \\
&\equiv& i\Delta_{BEH}(p_1^2) \nonumber \\
iG_{0,2}(;q_1,-q_1)&=&i\int \frac{d^4q_2}{(2\pi)^4}  G_{0,2}(;q_1,q_2) \nonumber \\
&=&\int d^4y_1 e^{iq_1y_1}\big< 0\vert T \Big[ \pi (y_1)\pi(0)\Big]\vert 0\big> \nonumber \\
&\equiv& i\Delta_{\pi}(q_1^2)
\end{eqnarray}

{\bf 8) $\phi$-sector connected amputated 1-$(h,\pi)$-Reducible (1-$\phi$-R) transition matrix (T-matrix):} With an overall momentum conservation factor 
$(2\pi)^4\delta^4 \Big(\sum_{n=1}^N p_n +\sum_{m=1}^M q_m \Big)$, the $\phi$-sector connected amplitudes are related to $\phi$-sector connected amputated T-matrix elements
\begin{eqnarray}
\label{TMatrix} 
&&G_{N,M}(p_1...p_N;q_1...q_M)  \\
&&\equiv\prod_{n=1}^N\Big[i\Delta_{BEH}(p_n^2)\Big] \prod_{m=1}^M\Big[i\Delta_{\pi}(q_m^2)\Big] T_{N,M}(p_1...p_N;q_1...q_M) \nonumber 
\end{eqnarray}
so that the master equation (\ref{MasterEquation}) can be written
\begin{eqnarray}
\label{TMatrixIdentity} 
&&i\HVEV k^2 \Big[i\Delta_\pi(k^2)\Big]T_{N,M+1}(p_1...p_N;kq_1...q_M) \nonumber \\
&&\quad =\sum_{n=1}^N T_{N-1,M+1}(p_1...{\widehat {p_n}}...p_N;(k+p_n)q_1...q_M) \nonumber \\
&&\quad \times \Big[i\Delta_\pi((k+p_n)^2)\Big] \Big[i\Delta_{BEH}(p_n^2)\Big]^{-1} \nonumber \\
&&\quad -\sum_{m=1}^M T_{N+1,M-1}((k+q_m)p_1...p_N;q_1...{\widehat {q_m}}...q_M) \nonumber \\
&&\quad \times \Big[i\Delta_{BEH}((k+q_m)^2)\Big] \Big[i\Delta_{\pi}(q_m^2)\Big]^{-1} 
\end{eqnarray}
with the ``hatted" momenta $({\widehat {p_n}},{\widehat {q_m}})$ removed  in (\ref{TMatrixIdentity}), and an overall momentum conservation factor of $(2\pi)^4\delta^4 \Big(k+\sum_{n=1}^N p_n +\sum_{m=1}^M q_m \Big)$.

{\bf 9) ``Pion pole dominance" and ``1-soft-pion theorems" for the T-matrix:} Consider the ``soft-pion limit" 
\begin{eqnarray}
\label{SoftPionLimitPropagator}
\lim_{k_\mu \to 0} k^2\Delta_\pi (k^2) =1 
\end{eqnarray}
where the $\pi$ is hypothesized to be all-loop-orders massless, and written in the K$\ddot a$ll$\acute e$n-Lehmann representation \cite{Bjorken1965} with spectral density $\rho^{\pi}_{AHM}$
\begin{eqnarray}
\label{NGBPropagator}
\Delta_{\pi}(k^2) &=& \frac{1}{k^2
+ i\epsilon} + \int dm^2 \frac{\rho^{\pi}_{AHM}(m^2)}{k^2-m^2 + i\epsilon}
\end{eqnarray} 
The master equation (\ref{MasterEquation}) then becomes
\begin{eqnarray}
\label{SoftPionTMatrixID} 
&&-\HVEV T_{N,M+1}(p_1...p_N;0q_1...q_M) \nonumber \\
&&\quad =\sum_{n=1}^N T_{N-1,M+1}(p_1...{\widehat {p_n}}...p_N;p_nq_1...q_M) \nonumber \\
&&\quad \times \Big[i\Delta_\pi(p_n^2)\Big] \Big[i\Delta_{BEH}(p_n^2)\Big]^{-1} \nonumber \\
&&\quad -\sum_{m=1}^M T_{N+1,M-1}(q_m p_1...p_N;q_1...{\widehat {q_m}}...q_M) \nonumber \\
&&\quad \times \Big[i\Delta_{BEH}(q_m^2)\Big] \Big[i\Delta_{\pi}(q_m^2)\Big]^{-1} 
\end{eqnarray}
in the 1-soft-pion limit.
As usual the ``hatted" momenta $({\widehat {p_n}},{\widehat {q_m}})$ and associated fields are removed  in (\ref{SoftPionTMatrixID}), and an overall momentum conservation factor  $(2\pi)^4\delta^4 \Big(\sum_{n=1}^N p_n +\sum_{m=1}^M q_m \Big)$ applied.

The set of 1-soft-pion theorems (\ref{SoftPionTMatrixID}) have the form 
\begin{eqnarray}
\HVEV T_{N,M+1}\sim T_{N-1,M+1} - T_{N+1,M-1}
\end{eqnarray}
relating, by the addition of a zero-momentum pion, an $N+M+1$-point function to $N+M$-point functions.

{\bf 10) The Adler self-consistency relations (but now for a gauge theory} rather than global $SU(2)_L \times SU(2)_R$ \cite{Adler1965,AdlerDashen1968} are gotten by putting the remainder of the (\ref{SoftPionTMatrixID}) particles on mass-shell
\begin{eqnarray}
\label{AdlerSelfConsistency} 
&&\HVEV T_{N,M+1}(p_1...p_N;0q_1...q_M)\nonumber \\
&& \quad \quad \times (2\pi)^4\delta^4 \Big(\sum_{n=1}^N p_n +\sum_{m=1}^M q_m \Big) \Big\vert^{p_1^2 =p_2^2...=p_N^2=m_{BEH}^2}_{q_1^2 =q_2^2...=q_M^2=0}  \nonumber \\
&& \quad \quad =0 
\end{eqnarray}
which guarrantees the infra-red (IR) finiteness of the $\phi$-sector on-shell T-matrix in the SSB AHM gauge theory in 
Lorenz gauge, with massless $\pi$ in the 1-soft-pion limit.
These ``1-soft-pion" theorems \cite{Adler1965,AdlerDashen1968} force the T-matrix to vanish as one of the pion momenta goes to zero, 
provided all other physical scalar particles are on mass-shell. 
Eqn.
(\ref{AdlerSelfConsistency})  asserts the absence of infrared divergences 
in the physical-scalar sector in Goldstone mode.
``Although individual Feynman diagrams may be IR divergent, 
those IR divergent parts cancel exactly in each order of perturbation theory. 
Furthermore, the Goldstone mode amplitude must vanish in the soft-pion limit \cite{Lee1970}".

{\bf 11) 1-$(h,\pi)$ Reducibility (1-$\phi$-R) and 1-$(h,\pi)$ Irreducibility (1-$\phi$-I):} With some exceptions, the $\phi$-sector connected amputated transition matrix $T_{N,M}$ can be cut apart by cutting an internal $h$ or $\pi$ line, and are designated 1-$\phi$-R. In contrast, the $\phi$-sector connected amputated Green's functions $\Gamma_{N,M}$ are defined to be 1-$\phi$-I: i.e. they cannot be cut apart by cutting an internal $h$ or $\pi$ line.
\begin{eqnarray}
\label{1SPReducibility}
T_{N,M} = \Gamma_{N,M} + (1-\phi-R)
\end{eqnarray}

Both $T_{N,M}$ and $\Gamma_{N,M}$ are 1-$(A_\mu)$-Reducible (1-$A^\mu$-R): i.e. they can be cut apart by cutting an internal transverse-vector $A_\mu$ gauge-particle line.

{\bf 12) $\phi$-sector 2-point functions, propagators and a 3-point vertex:} The special 2-point functions $T_{0,2}(;q,-q)$ and
$T_{2,0}(p,-p;)$, and the 3-point vertex  $T_{1,2}(q;0,-q)$, are 1-$\phi$-I (i.e. they are not 1-$\phi$-R), and are therefore equal to the corresponding 1-$\phi$-I connected amputated Green's functions. The 2-point functions
\begin{eqnarray}
\label{TMatrix2PointA}
T_{2,0}(p,-p;)&=&\Gamma_{2,0}(p,-p;)=\big[\Delta_{BEH}(p^2)\big]^{-1} \nonumber \\
T_{0,2}(;q,-q)&=&\Gamma_{0,2}(;q,-q)=\big[\Delta_{\pi}(q^2)\big]^{-1}  
\end{eqnarray}
are related to the $(1h,2\pi)$ 3-point $h\pi^2$ vertex 
\begin{eqnarray}
\label{3PointVertex}
T_{1,2}(p;q,-p-q) = \Gamma_{1,2}(p;q,-p-q) 
\end{eqnarray}
by a 1-soft-pion theorem (\ref{SoftPionTMatrixID})
\begin{eqnarray}
\label{TMatrix2and3Point}
&&\HVEV T_{1,2}(q;0,-q) -T_{2,0}(q,-q;)+T_{0,2}(;q,-q) \nonumber \\
&&\quad \quad =\HVEV T_{1,2}(q;0,-q) -\big[\Delta_{BEH}(q^2)\big]^{-1} +\big[\Delta_{\pi}(q^2)\big]^{-1} \nonumber \\
&&\quad \quad =\HVEV \Gamma_{1,2}(q;0,-q) -\Gamma_{2,0}(q,-q;)+\Gamma_{0,2}(;q,-q) \nonumber \\
&&\quad \quad =\HVEV \Gamma_{1,2}(q;0,-q) -\big[\Delta_{BEH}(q^2)\big]^{-1} +\big[\Delta_{\pi}(q^2)\big]^{-1} \nonumber \\
&&\quad \quad =0 
\end{eqnarray}

{\bf 13) The Goldstone theorem, in the spontaneously broken AHM in 
Lorenz gauge,} is a special case of that SSB gauge theory's Adler self-consistency relations (\ref{AdlerSelfConsistency})
\begin{eqnarray}
\label{TMatrixGoldstoneTheorem} 
\HVEV T_{0,2}(;00)&=&0 \nonumber \\
\HVEV \Gamma_{0,2}(;00)&=&0 \nonumber \\
\HVEV \big[ \Delta_\pi (0) \big]^{-1} &=&0
\end{eqnarray}
proving that $\pi$ is massless. That all-loop-orders renormalized massless-ness is protected/guarranteed by the global $U(1)_Y$ symmetry of the {\em physical states} of the gauge theory after spontaneous symmetry breaking.

{\bf 14) $T_{N,M+1}^{External}$ $\phi$-sector T-Matrix with one soft $\pi (q_\mu =0)$ attached to an external-leg:} 
Figure \ref{fig:LeeFig10}  shows that
\begin{eqnarray}
\label{ExternalLegTMatrix}
&&\HVEV T_{N,M+1}^{External}(p_1...p_N;0q_1...q_M) \nonumber \\
&&\quad \quad =\sum_{n=1}^N \Big[ i\HVEV\Gamma_{1,2}(p_n,0,-p_n)\Big] \Big[ i\Delta_\pi (p_n^2)\Big]  \nonumber \\
&&\quad \quad \times T_{N-1,M+1}(p_1...{\widehat{p_n}}...p_N;p_nq_1...q_M) \nonumber \\
&&\quad \quad +\sum_{m=1}^M \Big[ i\HVEV\Gamma_{1,2}(q_m,0,-q_m)\Big] \Big[ i\Delta_{BEH} (q_m^2)\Big]  \nonumber \\
&&\quad \quad \times T_{N+1,M-1}(q_mp_1...p_N;q_1....{\widehat{q_m}}...q_M) \nonumber \\
&&\quad \quad =\sum_{n=1}^N \Big(1-\Big[ i\Delta_\pi (p_n^2)\Big]\Big[ i\Delta_{BEH} (p_n^2)\Big]^{-1} \Big) \nonumber \\
&&\quad \quad \times T_{N-1,M+1}(p_1...{\widehat{p_n}}...p_N;p_nq_1...q_M) \nonumber \\
&&\quad \quad -\sum_{m=1}^M \Big(1-\Big[ i\Delta_{BEH} (q_m^2)\Big]\Big[ i\Delta_\pi (q_m^2)\Big]^{-1} \Big)  \nonumber \\
&&\quad \quad \times T_{N+1,M-1}(q_mp_1...p_N;q_1....{\widehat{q_m}}...q_M)  
\end{eqnarray}
where we used (\ref{TMatrix2and3Point}). Now separate
\begin{eqnarray}
\label{DefineInternalTMatrixA}
&&T_{N,M+1}(p_1...p_N;0q_1...q_M) \nonumber \\
&&\quad \quad =T_{N,M+1}^{External}(p_1...p_N;0q_1...q_M) \nonumber \\
&&\quad \quad +T_{N,M+1}^{Internal}(p_1...p_N;0q_1...q_M)  
\end{eqnarray}
so that 
\begin{eqnarray}
\label{InternalTMatrix}
&&\HVEV T_{N,M+1}^{Internal}(p_1...p_N;0q_1...q_M) \nonumber \\
&&\quad \quad =\sum_{m=1}^M T_{N+1,M-1}(q_mp_1...p_N;q_1....{\widehat{q_m}}...q_M)  \nonumber \\
&&\quad \quad -\sum_{n=1}^N T_{N-1,M+1}(p_1...{\widehat{p_n}}...p_N;p_nq_1...q_M) \quad \quad 
\end{eqnarray}

{\bf 15) Recursive $U(1)_Y$ WTI  for 1-$(h,\pi)$-Irreducible (1-$\phi$-I) connected amputated Green's functions $\Gamma_{N,M}$:} Removing the 1-$(h,\pi)$-Reducible (1-$\phi$-R)  graphs from both sides of (\ref{InternalTMatrix}) yields the recursive identity
\begin{eqnarray}
\label{GreensFWTI}
&&\HVEV \Gamma_{N,M+1}(p_1...p_N;0q_1...q_M) \nonumber \\
&&\quad \quad =\sum_{m=1}^M \Gamma_{N+1,M-1}(q_mp_1...p_N;q_1....{\widehat{q_m}}...q_M)  \nonumber \\
&&\quad \quad -\sum_{n=1}^N \Gamma_{N-1,M+1}(p_1...{\widehat{p_n}}...p_N;p_nq_1...q_M) \quad \quad
\end{eqnarray}

B.W. Lee \cite{Lee1970} gave an inductive proof for the corresponding recursive $SU(2)_L \times SU(2)_R$ WTI in the {\em global}  Gell-Mann Levy model with PCAC \cite{GellMannLevy1960}. 
Specifically, he proved  that, given the global $SU(2)_L \times SU(2)_R$ analogy of (\ref{InternalTMatrix}), the global $SU(2)_L \times SU(2)_R$ analogy of (\ref{GreensFWTI}) follows. This he did by examination of the explicit reducibility/irreducibility of the various  Feynman graphs involved. 

That proof also works for the $U(1)_Y$ SSB AHM, thus establishing our tower of 1-$\phi$-I connected amputated Green's functions' recursive
$U(1)_Y$ WTI (\ref{GreensFWTI}) for a {\em local/gauge} theory. 

Rather than including the lengthy proof here, we paraphrase \cite{Lee1970} as follows: 
(\ref{TMatrix2and3Point}) shows that (\ref{GreensFWTI}) is true for $(N=1,M=1)$. Assume it is true for all $(n,m)$ such that $n+m < N+M$.
Consider (\ref{InternalTMatrix}) for $n=N,m=M$. The two classes of graphs contributing to $T_{N,M+1}^{Internal}(p_1...p_N;0q_1...q_M)$ are displayed in Figure \ref{fig:LeeFig11}. 

\begin{figure}
\centering
\includegraphics[width=1\hsize,trim={0cm 4cm 0cm 3cm},clip]{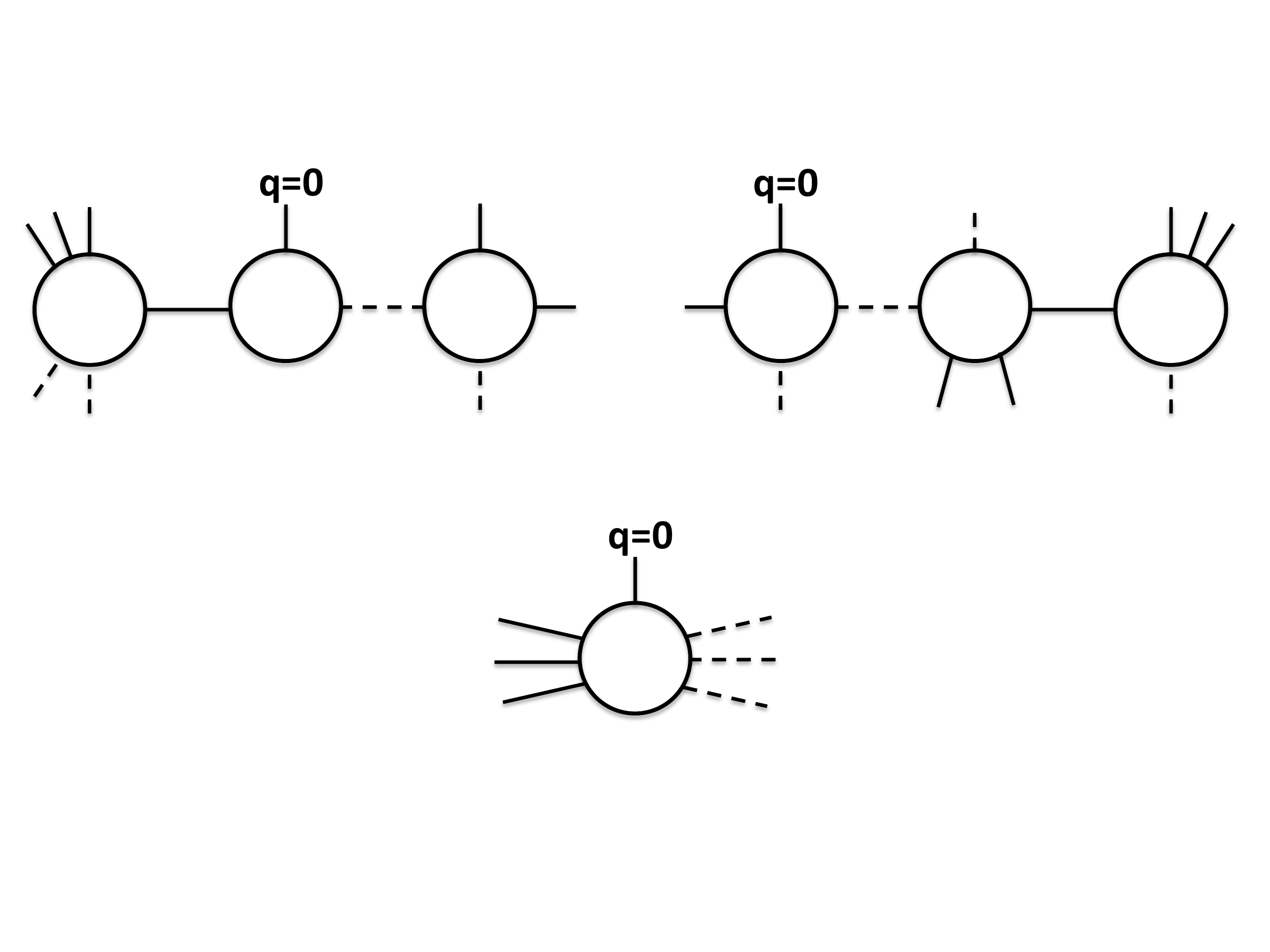}
\caption{
\label{fig:LeeFig11} 
Circles are 1-$\phi$-I $\Gamma^{E-AHM}_{n,m}$, solid lines $\pi$, dashed lines $h$, with $n+m<N+M$.
1 (zero-momentum) soft pion emerges in all possible ways from the  connected amputated Green's functions.
$\Gamma^{E-AHM}_{n,m}$ is 1-$A^\mu$-R by cutting an $A^\mu$ line,
and also 1-$\Phi$-R by cutting a $\Phi$ line.
Fig. \ref{fig:LeeFig11}  is the extended-AHM analogy of B.W. Lee's Figure 11 \cite{Lee1970}.
The same graph topologies, but without internal Beyond-AHM $\Phi,\psi$ heavy matter, are used in the proof of (\ref{GreensFWTI}) for the (unextended) AHM.
}
\end{figure}

The top graphs
 in Figure \ref{fig:LeeFig11} are 1-$\phi$-R. For $(n,m;n+m<N+M)$ we may use (\ref{GreensFWTI}), 
for those 1-$\phi$-I Green's functions $\Gamma_{n,m}$ which contribute to (\ref{InternalTMatrix}),
 to show that the contribution of 1-$\phi$-R graphs to both sides of  (\ref{InternalTMatrix}) are identical.

The bottom graphs in  Figure \ref{fig:LeeFig11} are 1-$\phi$-I and so already obey (\ref{GreensFWTI}). 

{\bf 16) Goldstone theorem makes tadpoles vanish:} 
\begin{eqnarray}
\label{Tadpoles}
&&\big<0\vert h(x=0)\vert0\big>_{Connected} \nonumber \\
&& \qquad \qquad = i \Big[i\Delta_{BEH}(0)\Big]\Gamma_{1,0}(0;) \qquad \quad
\end{eqnarray}
but the $N=0,M=1$ case of (\ref{GreensFWTI}) reads
\begin{eqnarray}
\label{GoldstoneTadpoles}
\Gamma_{1,0}(0;)&=& \HVEV \Gamma_{0,2}(;00) \nonumber \\
&=&0 
\end{eqnarray}
where we used (\ref{TMatrixGoldstoneTheorem}), so that tadpoles all vanish automatically, and separate tadpole renormalization is un-necessary.
Since we can choose the origin of coordinates anywhere we like
\begin{eqnarray}
\label{GoldstoneTadpolesVanishExtended}
\big<0\vert h(x)\vert0\big>_{Connected} &=& 0
\end{eqnarray}

{\bf 17) Renormalized  gauge-independent observable $\HVEV$}.
\begin{eqnarray}
\label{HVEV}
\big<0\vert H(x)\vert0\big>_{Connected} &=&\big<0\vert h(x)\vert0\big>_{Connected} +\HVEV \nonumber \\
&=& \HVEV \nonumber \\
\partial_\mu \HVEV &=&0
\end{eqnarray}

{\bf 18) Benjamin W. Lee's 1970 Cargese summer school lectures' \cite{Lee1970}} proof of $\phi$-sector WTI focusses on the global $SU(2)_L \times SU(2)_R$ Gell-Mann Levy theory and Partially Conserved Axial-vector Currents (PCAC). But it gives a detailed pedagogical account of the appearance of the Goldstone theorem and true massless Nambu-Goldstone bosons in {\em global} theories, and is recommended reading. We include a translation guide in Table 1.

\newpage
\bigskip
\vbox{
\baselineskip=15pt
\halign{
\hfil\sl#&\hfil\quad\it#\hfil&\quad#\hfil\cr
\multispan3\hfil\bf Table 1: Derivation of Ward-Takahashi identities \hfil\cr
{\bf Property}&{\bf This paper}&{\bf B.W.Lee \cite{Lee1970}}\cr
LagrangianInvariant&BRST&global group\cr
structure group&$U(1)_Y$&$SU(2)_L \times SU(2)_R$\cr
local/gauge group&$U(1)_Y$& \cr
rigid/global group&$U(1)_Y$&$SU(2)_L \times SU(2)_R$\cr
global currents&${J}_{AHM}^\mu$&${\vec V}^\mu ;{\vec A}^\mu$ \cr
PCAC&no&yes\cr
current divergence&$-Hm_A\partial_\beta A^\beta$&$0; f_\pi\mpisq{\vec \pi}$\cr
$L_{GaugeFixing}$&$Lorenz$&\cr
gauge&Lorenz&\cr
ghosts ${\bar \omega},\omega$&decouple&\cr
conserved current&physical states&Lagrangian \cr
physical states&$A_\mu,h, \pi , \Phi,\psi$&$s,{\vec \pi}$\cr
interaction&weak&strong \cr
fields&$A_\mu,H, \pi , {\bar \omega},\omega$&$\sigma,{\vec \pi}$\cr
BEH scalar&$h=H-\HVEV$&$s=\sigma-<\sigma>$\cr
VEV&$\HVEV$&$<\sigma>=v=f_\pi$\cr
particles in loops&$Physical\&Ghosts$&$s,{\vec \pi}$\cr
renormalization&all-loop-orderss&all-loop-orders\cr
amplitudes&&G\cr
ConnectedAmplitudes&$G_{N,M}$&H\cr
NoPionPoleSingularity&&$\bar H$\cr
T-Matrix&$T_{N,M}$&$T$\cr
1-$\phi$&$h,{ \pi}$&$s,{\vec \pi}$\cr 
$\phi$-sector $T_{N,M}$&1-$\phi$-R&1-$\phi$-R\cr
$\phi$SectorGreensF's&$\Gamma_{N,M}$&$\Gamma_{N,M}$\cr 
connected $\Gamma_{N,M}$&amputated&amputated\cr
connected $T_{N,M}$&amputated&amputated\cr
$\Gamma_{N,M}$&1-$\phi$-I&1-$\phi$-I\cr
External $\pi (q_\mu =0)$&$T^{External}_{N,M+1}$&$T_1$\cr
Internal $\pi (q_\mu =0)$&$T^{Internal}_{N,M+1}$&$T_2$\cr
BEH propagator&$\Delta_{BEH}$&$\Delta_\sigma$\cr
TransversePropagator&$\Delta_{A}^{\mu \nu}$&\cr
SSB&GoldstoneMode&GoldstoneMode\cr
NGB after SSB&$\tilde \pi$&$\tilde{\vec \pi}$\cr
Pion propagator&$\Delta_{\pi}$&$\delta^{t_i t_j}\Delta_{\pi}$\cr
Goldstone theorem&physical states&GoldstoneMode\cr
}
}
\bigskip

\newpage

\section{$U(1)_Y$ $(h,\pi)$-sector WTIs, which now  include the all-loop-orders contributions of certain additional virtual $U(1)_Y$ matter representations $\Phi,\psi$
in the extended-Abelian Higgs Model (E-AHM)}
\label{DerivationWTIE-AHM}

In this Appendix \ref{DerivationWTIE-AHM}, we present the full detailed derivations of our $U(1)_Y$ WTI for the SSB E-AHM. For pedagogical completeness, we reproduce all details of the entire argument.
\begin{itemize}
\item This paper is interested in building the effective E-AHM $\phi$ potential, no-fine-tuning of the BEH mass, and the decoupling in E-AHM of certain heavy matter particles from the effective low-energy AHM theory.

\item A companion Letter \cite{LSS-3Short} focusses instead on the WTI themselves, showing that E-AHM {\bf physics} has more symmetry than the E-AHM effective Lagrangian.
\end{itemize}

We focus on the rigid/global {\em extended}-AHM current 
\begin{eqnarray}
\label{extendedAHMCurrent}
{ J}^{\mu}_{E-AHM}&=&{J}^{\mu}_{AHM}(A^\mu,\phi) \nonumber \\
&+&{J}^{\mu}_{BeyondAHM}(\Phi,\Psi) \quad \quad
\end{eqnarray}
 of the {\em ``extended Abelian Higgs model"}, the spontaneously broken gauge theory of a complex spin $S=0$ scalar $\phi = \frac{1}{{\sqrt 2}}\big( H+i\pi \big)$, 
a massive $U(1)_Y$ $S=1$ transverse gauge field $A_\mu$, and certain $S=0$ scalars $\Phi$ and $S=\half$  fermions $\psi$
originating in Beyond-AHM  models.

The classical equations of motion reveal that, due to gauge-fixing terms in the BRST-invariant Lagrangian, the 
classical current 
(\ref{extendedAHMCurrent}) is 
not conserved. In 
Lorenz gauge  
\begin{eqnarray}
\label{ExtendedDivergenceAHMCurrent}
\partial_{\mu}  { J}^{\mu}_{E-AHM} &=& -H m_A F_A   \nonumber \\
m_A &=& e\HVEV \nonumber \\
F_A &=& \partial_{\beta}{A}^{\beta} 
\end{eqnarray}
with $F_A$ the gauge fixing function.  

The purpose of this Appendix is to derive a tower of $U(1)_Y$ extended WTIs, which exhausts the information content of (\ref{ExtendedDivergenceAHMCurrent}), and severely constrains the dynamics (i.e. the connected time-ordered products) of the physical states 
of the SSB {\em extended}-AHM.
We make use here of all of the results in Appendix A concerning ${J}^{\mu}_{AHM}$.

{\bf 1) We study a certain total differential of a connected time-ordered product:}
\begin{eqnarray}
\label{ExtendedTotalDerivativeAHM}
&&\partial_{\mu} \Big< 0\vert T\Big[ { J}^{\mu}_{E-AHM}(z)  \\
&&\quad \quad \times h(x_1)...h(x_N) \pi_(y_1)...\pi(y_M)\Big]\vert 0\Big>_{Connected} \nonumber
\end{eqnarray}
written in terms of the {\em physical states} of the complex scalar $\phi$.
Here we have N external renormalized scalars $h=H-\HVEV$ (coordinates x, momenta p), 
and M external ($CP=-1$) renormalized pseudo-scalars ${ \pi}$ (coordinates y, momenta q).

{\bf 2) Conservation of the global $U(1)_Y$ current for the {\em physical states}:} Strict quantum constraints are imposed, which force the relativistically-covariant theory of a massive transverse gauge boson to propagate {\it only} its true number of quantum spin $S=1$ degrees of freedom.   Physical states and their time-ordered products, but not the BRST-invariant Lagrangian,  obey the gauge-fixing condition ${F}_A = \partial_{\beta}{A}^{\beta}=0$ in
Lorenz gauge \cite{tHooft1971}
\begin{eqnarray}
\label{GaugeConditionExtended} 
&&\big< 0\vert T\Big[ \Big( \partial_{\beta}{A}^{\beta}(z)\Big) \nonumber \\
&&\quad \quad  \times h(x_1)...h(x_N)\pi(y_1)...\pi(y_M)\Big]\vert 0\big>_{Connected} \nonumber \\
&&\quad \quad =0
\end{eqnarray}
which restores conservation of the rigid/global $U(1)_Y$ extended current for physical states
\begin{eqnarray}
\label{ExtendedQuantumCurrentConservation}
&&\Big< 0\vert T\Big[ \Big( \partial_{\mu}{J}^{\mu}_{E-AHM}(z) \Big) \nonumber \\
&&\quad \quad \times h(x_1)...h(x_N) \pi(y_1)...\pi(y_M)\Big]\vert 0\Big>_{Connected} \nonumber \\
 &&\quad \quad=0
\end{eqnarray}
It is in this ``time-ordered-product" sense that the rigid global extended $U(1)_Y$ current ${J}^{\mu}_{E-AHM}$ is conserved, and
it is this conserved current  which generates our tower of $U(1)_Y$ extended WTI. These extended WTI severely constrain the dynamics of $\phi$.

{\bf 3) Vintage QFT and canonical quantization:} Equal-time commutators are imposed on the exact renormalized Beyond-AHM fields, yielding equal-time quantum commutators at space-time points $y, z$.
\begin{eqnarray}
\label{EqTimeCommE-AHM}
 \delta(z_0-y_0)\left[ {J}^0_{BeyondAHM}(z),H(y)\right] &=&0 \nonumber \\
 \delta(z_0-y_0)\left[ {J}^0_{BeyondAHM}(z),\pi(y)\right] &=&0
\end{eqnarray} 
Only certain  $U(1)_Y$  matter particles $\Phi, \psi$ obey this condition.

{\bf $\bullet$ Renormalized $\HVEV$ is defined to match the (un-extended) AHM}. 
Our extended $U(1)_Y$ WTI therefore  require that all of the new spin $S=0$ fields in $ {J}^\mu_{BeyondAHM}$ have zero vacuum expectation value (VEV):
\begin{eqnarray}
\label{E-AHMVEV}
\big<  \Phi_{BeyondAHM}\big> =0
\end{eqnarray}
Only certain  $U(1)_Y$  matter particles $\Phi$ obey this condition.

{\bf 4) Certain connected surface integrals must vanish:} 
As appropriate to our study of massless $\pi$, we again use pion pole dominance to derive 1-soft-pion theorems, and require that the {\em connected} surface integral
\begin{eqnarray}
\label{ExtendedSurfaceIntegral}
&&\lim_{k_\lambda \to 0} \int d^4z e^{ikz} \partial_{\mu} \Big< 0\vert T\Big[ \Big({J}^{\mu}_{BeyondAHM} (z)\Big)  \nonumber \\
&&\quad \quad \times h(x_1)...h(x_N) \pi_(y_1)...\pi(y_M)\Big]\vert 0\Big>_{Connected} \nonumber \\
&& =\int d^4z \partial_{\mu} \Big< 0\vert T\Big[ \Big({J}^{\mu}_{BeyondAHM} (z)\Big) \nonumber \\
&&\quad \quad \times h(x_1)...h(x_N) \pi_(y_1)...\pi(y_M)\Big]\vert 0\Big>_{Connected} \nonumber \\
&& =\int_{3-Surface\to\infty}  d^3z \quad {\widehat {z}_\mu}^{3-surface} \nonumber \\
&&\quad \quad \times \Big< 0\vert T\Big[ \Big({J}^{\mu}_{BeyondAHM} (z) \Big) \nonumber \\
&&\quad \quad \times h(x_1)...h(x_N) \pi_(y_1)...\pi(y_M)\Big]\vert 0\Big>_{Connected} \nonumber \\
&&\quad \quad =0
\end{eqnarray}
where we have used Stokes theorem, and $ {\widehat {z}_\mu}^{3-surface}$ is a unit vector normal to the $3-surface$. The time-ordered-product constrains the $3-surface$ to lie on-or-inside the light-cone.

At a given point on the surface of a large enough 4-volume $\int d^4z$ (i.e. the volume of all space-time): all fields are asymptotic in-states and out-states; are properly quantized as free fields; with each field species orthogonal to the others;
and  they are evaluated at equal times, making time-ordering un-necessary at $(z^{3-surface}\to \infty)$. 

Only certain  $U(1)_Y$  {\bf massive} matter particles $\Phi, \psi$ obey this condition.

{\bf 5) Extended master equation:}
Using (\ref{ExtendedQuantumCurrentConservation},\ref{EqTimeCommE-AHM}) in (\ref{ExtendedTotalDerivativeAHM}) to form the right-hand-side, and (\ref{ExtendedSurfaceIntegral}) in (\ref{ExtendedTotalDerivativeAHM}) to form the left-hand-side, we write the {\em extended} master equation, which relates {\em connected} time-ordered products:
\begin{eqnarray}
\label{ExtendedMasterEquation}
&&-\HVEV\partial_{\mu}^z \big< 0\vert T\Big[  \big(\partial^\mu\pi(z)\big)  \nonumber \\
&&\quad \quad \quad \quad \times h(x_1)...h(x_N) \pi_(y_1)...\pi(y_M)\Big]\vert 0\big>_{Connected} \nonumber \\
&&\quad = \sum_{m=1}^M \quad  i\delta^4(z-y_m) \big< 0\vert T\Big[ h(z) h(x_1)...h(x_N) \nonumber \\
&&\quad \quad \quad \quad \times  \pi(y_1)...{\widehat {\pi (y_m)}}...\pi(y_M)\Big]\vert 0\big>_{Connected} \nonumber \\
&&\quad - \sum_{n=1}^N \quad  i\delta^4(z-x_n) \big< 0\vert T\Big[ h(x_1)...{\widehat {h(x_n)}}...h(x_N) \nonumber \\
&&\quad \quad \quad \quad \times  \pi(z){\pi}(y_1)...\pi(y_M)\big]\vert 0\big>_{Connected}
\end{eqnarray}
where the ``hatted" fields ${\widehat {h(x_n)}}$ and ${\widehat {\pi (y_m)}}$ are to be removed. We have also thrown away a sum of $M$ terms, proportional to $\HVEV$,
which corresponds entirely to disconnected graphs.

{\bf $\bullet$ $U(1)_Y$ Ward-Takahashi identities for the $\phi$-sector of the extended-AHM:} The extended master equation (\ref{ExtendedMasterEquation}) governing the $\phi$-sector of the extended-AHM, is idential to the master equation (\ref{MasterEquation}) governing the $\phi$-sector of the (un-extended) AHM. This proves that, for each $U(1)_Y$ WTI which is true in the AHM, an analogous $U(1)_Y$ WTI is true for the extended-AHM.
Appendix A proved $U(1)_Y$ WTI relations among 1-$\phi$-R $\phi$-sector  T-Matrix elements $T_{N,M}$, as well as $U(1)_Y$ WTI relations among 1-$\phi$-I 
$\phi$-sector Green's functions $\Gamma_{N,M}$, in the spontaneously broken AHM.
Analogous $U(1)_Y$ WTI relations among 1-$\phi$-R $\phi$-sector T-Matrix elements $T_{N,M}^{Extended}$, 
as well as analogous $U(1)_Y$ WTI relations among 1-$\phi$-I 
$\phi$-sector  Green's functions $\Gamma_{N,M}^{Extended}$, are therefore here proved true for the spontaneously broken {\em extended}-AHM.

But there is one {\em huge} difference! The renormalization of  our $U(1)_Y$ WTI, governing $\phi$-sector $T_{N,M}^{Extended}$ and $\phi$-sector $\Gamma_{N,M}^{Extended}$, now includes the all-loop-orders contributions of virtual gauge bosons, $\phi$-scalars, ghosts, new Beyond-AHM scalars and new Beyond-AHM fermions: i.e.  $A^\mu; h, \pi; {\bar \omega}, \omega: \Phi; \psi$ respectively. 

{\bf 10) Adler self-consistency relations, but now for the extended-AHM gauge theory:}
\begin{eqnarray}
\label{ExtendedAdlerSelfConsistency} 
&&\HVEV T_{N,M+1}^{Extended}(p_1...p_N;0q_1...q_M)\nonumber \\
&& \quad \quad \times (2\pi)^4\delta^4 \Big(\sum_{n=1}^N p_n +\sum_{m=1}^M q_m \Big) \Big\vert^{p_1^2 =p_2^2...=p_N^2=m_{BEH}^2}_{q_1^2 =q_2^2...=q_M^2=0}  \nonumber \\
&& \quad \quad =0 
\end{eqnarray}
These prove the infra-red (IR) finiteness of the $\phi$-sector on-shell connected T-matrix in the extended-AHM gauge theory, with massless 
$\pi$, in 
Lorenz gauge, in the 1-soft-pion limit.

{\bf 11) 1-$(h,\pi)$ Reducibility (1-$\phi$-R) and 1-$(h,\pi)$ Irreducibility (1-$\phi$-I):} With some exceptions, the extended $\phi$-sector connected amputated T-Matrix elements  $T_{N,M}^{Extended}$ can be cut apart by cutting an internal $h$ or $\pi$ line: they are designated 1-$\phi$-R. In contrast, the extended $\phi$-sector Green's functions $\Gamma_{N,M}^{Extended}$ are defined to be 1-$\phi$-I: ie. they cannot be cut apart by cutting an internal $h$ or $\pi$ line.
\begin{eqnarray}
\label{Extended1SPReducibility}
T_{N,M}^{Extended} = \Gamma_{N,M}^{Extended} + (1-\phi -R)
\end{eqnarray}

As usual, both $T_{N,M}^{Extended}$ and $\Gamma_{N,M}^{Extended}$ are 1-$(A^\mu)$-Reducible (1-$A^\mu$-R): i.e. they can be cut apart by cutting an internal transverse-vector $A^\mu$ gauge-particle line.

But both $T_{N,M}^{Extended}$ and $\Gamma_{N,M}^{Extended}$ are also 1-$\Phi$-Reducible (1-$\Phi$-R): i.e. they can be cut apart by cutting an internal $\Phi$ scalar line.

{\bf 12) $\phi$-sector 2-point functions, propagators and a 3-point vertex:} The 2-point functions
\begin{eqnarray}
\label{TMatrix2PointB}
T_{2,0}^{Extended}(p,-p;)&=&\Gamma_{2,0}^{Extended}(p,-p;)=\big[\Delta_{BEH}(p^2)\big]^{-1} \nonumber \\
T_{0,2}^{Extended}(;q,-q)&=&\Gamma_{0,2}^{Extended}(;q,-q)= \big[\Delta_{\pi}(q^2)\big]^{-1} \qquad \quad
\end{eqnarray}
are related to the $(1h,2\pi)$ 3-point $h\pi^2$ vertex 
\begin{eqnarray}
\label{3PointVertexB}
T_{1,2}^{Extended}(p;q,-p-q) = \Gamma^{Extended}_{1,2}(p;q,-p-q) \quad \quad
\end{eqnarray}
by a 1-soft-pion theorem (\ref{ExtendedGreensFWTI})
\begin{eqnarray}
\label{TMatrix2and3PointExtended}
&&\HVEV T_{1,2}^{Extended}(q;0,-q) \nonumber \\
&&\quad \quad =\big[\Delta_{BEH}(q^2)\big]^{-1} -\big[\Delta_{\pi}(q^2)\big]^{-1} 
\end{eqnarray}

{\bf 13) The Goldstone theorem, in 
Lorenz-gauge-extended-AHM} is the $N=0,M=1$ case 
of (\ref{ExtendedAdlerSelfConsistency}):
\begin{eqnarray}
\label{TMatrixGoldstoneTheoremExtended} 
\HVEV T_{0,2}^{Extended}(;00)&=&0 \nonumber \\
\HVEV \Gamma_{0,2}^{Extended}(;00)&=&0 \nonumber \\
\HVEV \big[ \Delta_\pi (0) \big]^{-1} &=&0
\end{eqnarray}
proves that $\pi$ is still massless in the extended-AHM, whose all-loop-orders renormalized massless-ness is protected/guarranteed by the global $U(1)_Y$ symmetry of the physical states of the extended-AHM gauge theory after SSB.

{\bf 14)} $T_{N,M+1}^{Extended;External}$ are the 1-$\phi$-R $\phi$-sector connected amputated T-Matrix elements, {\bf with one soft $\pi (q_\mu =0)$ attached to an external-leg,} as shown in Figure \ref{fig:LeeFig10}.
With the separation
\begin{eqnarray}
\label{DefineInternalTMatrixB}
&&T^{Extended}_{N,M+1}(p_1...p_N;0q_1...q_M) \nonumber \\
&&\quad \quad =T_{N,M+1}^{Extended;External}(p_1...p_N;0q_1...q_M) \nonumber \\
&&\quad \quad +T_{N,M+1}^{Extended;Internal}(p_1...p_N;0q_1...q_M)  \qquad \qquad
\end{eqnarray}
we have the recursive $U(1)_Y$ T-Matrix WTI
\begin{eqnarray}
\label{InternalTMatrixExtended}
&&\HVEV T_{N,M+1}^{Extended;Internal}(p_1...p_N;0q_1...q_M) \nonumber \\
&&\quad \quad =\sum_{m=1}^M T^{Extended}_{N+1,M-1}(q_mp_1...p_N;q_1....{\widehat{q_m}}...q_M)  \nonumber \\
&&\quad \quad -\sum_{n=1}^N T^{Extended}_{N-1,M+1}(p_1...{\widehat{p_n}}...p_N;p_nq_1...q_M) \quad\quad
\end{eqnarray}

{\bf 15) Recursive $U(1)_Y$ WTIs for 1-$\phi$-I $\phi$-sector connected amputated extended Green's functions $\Gamma_{N,M}^{Extended}$}
are a solution to (\ref{InternalTMatrixExtended})
\begin{eqnarray}
\label{ExtendedGreensFWTI}
&&\HVEV \Gamma_{N,M+1}^{Extended}(p_1...p_N;0q_1...q_M) \nonumber \\
&&\quad \quad =\sum_{m=1}^M \Gamma_{N+1,M-1}^{Extended}(q_mp_1...p_N;q_1....{\widehat{q_m}}...q_M)  \nonumber \\
&&\quad \quad -\sum_{n=1}^N \Gamma_{N-1,M+1}^{Extended}(p_1...{\widehat{p_n}}...p_N;p_nq_1...q_M) \quad\quad
\end{eqnarray}

{\bf 16) Goldstone theorem makes tadpoles vanish:} 
\begin{eqnarray}
\label{TadpolesExtended}
&&\big<0\vert h(x=0)\vert0\big>_{Connected}  \\
&& \qquad \qquad = i \Big[i\Delta_{BEH}(0)\Big]\Gamma^{Extended}_{1,0}(0;) \nonumber
\end{eqnarray}
but the $N=0,M=1$ case of (\ref{ExtendedGreensFWTI}) reads
\begin{eqnarray}
\label{GoldstoneTadpolesExtended}
\Gamma^{Extended}_{1,0}(0;)&=& \HVEV \Gamma^{Extended}_{0,2}(;00) \nonumber \\
&=&0 
\end{eqnarray}
where we have used (\ref{TMatrixGoldstoneTheoremExtended}), so that tadpoles all vanish automatically, and separate tadpole renormalization is un-necessary. Since we can choose the origin of coordinates anywhere we like
\begin{eqnarray}
\label{GoldstoneTadpolesVanishExtendedB}
\big<0\vert h(x)\vert0\big>_{Connected} &=& 0
\end{eqnarray}
\newline
{\bf 17) Renormalized gauge-independent observable $\HVEV$}.
\begin{eqnarray}
\label{HVEVExtended}
\big<0\vert H(x)\vert0\big>_{Connected} &=&\big<0\vert h(x)\vert0\big>_{Connected} +\HVEV \nonumber \\
&=& \HVEV \nonumber \\
\partial_\mu \HVEV &=&0
\end{eqnarray}

 \end{document}